\RequirePackage{lineno}
\documentclass{aastex62}

\usepackage{graphicx}  
\usepackage[page]{appendix}  
\usepackage{rotating}  
\usepackage[normalem]{ulem}  
\usepackage{amsmath}
\usepackage{amssymb}
\usepackage{graphicx}
\usepackage{tabularx}
\usepackage{rotating}
\usepackage{float}
\usepackage[caption=false, position=top]{subfig}

\accepted{to ApJ on March 30, 2019}

\shorttitle{Archival Pulsar Search}
\shortauthors{The VERITAS Collaboration}

\begin{document}

\title{A Search for Pulsed Very High-Energy Gamma Rays from Thirteen Young Pulsars in Archival VERITAS Data}

\correspondingauthor{G. T. Richards}
\email{grichard@udel.edu}

\correspondingauthor{J. Tyler}
\email{jonathan.tyler@mail.mcgill.ca}


\author{A.~Archer}
\affiliation{Department of Physics, Washington University, St. Louis, MO 63130, USA}

\author{W.~Benbow}
\affiliation{Fred Lawrence Whipple Observatory, Harvard-Smithsonian Center for Astrophysics, Amado, AZ 85645, USA}

\author{R.~Bird}
\affiliation{Department of Physics and Astronomy, University of California, Los Angeles, CA 90095, USA}

\author{R.~Brose}
\affiliation{Institute of Physics and Astronomy, University of Potsdam, 14476 Potsdam-Golm, Germany}
\affiliation{DESY, Platanenallee 6, 15738 Zeuthen, Germany}

\author{M.~Buchovecky}
\affiliation{Department of Physics and Astronomy, University of California, Los Angeles, CA 90095, USA}

\author{J.~H.~Buckley}
\affiliation{Department of Physics, Washington University, St. Louis, MO 63130, USA}

\author{A.~J.~Chromey}
\affiliation{Department of Physics and Astronomy, Iowa State University, Ames, IA 50011, USA}

\author{W.~Cui}
\affiliation{Department of Physics and Astronomy, Purdue University, West Lafayette, IN 47907, USA}
\affiliation{Department of Physics and Center for Astrophysics, Tsinghua University, Beijing 100084, China.}

\author{A.~Falcone}
\affiliation{Department of Astronomy and Astrophysics, 525 Davey Lab, Pennsylvania State University, University Park, PA 16802, USA}

\author{Q.~Feng}
\affiliation{Physics Department, McGill University, Montreal, QC H3A 2T8, Canada}

\author{J.~P.~Finley}
\affiliation{Department of Physics and Astronomy, Purdue University, West Lafayette, IN 47907, USA}

\author{L.~Fortson}
\affiliation{School of Physics and Astronomy, University of Minnesota, Minneapolis, MN 55455, USA}

\author{A.~Furniss}
\affiliation{Department of Physics, California State University - East Bay, Hayward, CA 94542, USA}

\author{A.~Gent}
\affiliation{School of Physics and Center for Relativistic Astrophysics, Georgia Institute of Technology, 837 State Street NW, Atlanta, GA 30332-0430}

\author{O.~Gueta}
\affiliation{DESY, Platanenallee 6, 15738 Zeuthen, Germany}

\author{D.~Hanna}
\affiliation{Physics Department, McGill University, Montreal, QC H3A 2T8, Canada}

\author{T.~Hassan}
\affiliation{DESY, Platanenallee 6, 15738 Zeuthen, Germany}

\author{O.~Hervet}
\affiliation{Santa Cruz Institute for Particle Physics and Department of Physics, University of California, Santa Cruz, CA 95064, USA}

\author{J.~Holder}
\affiliation{Department of Physics and Astronomy and the Bartol Research Institute, University of Delaware, Newark, DE 19716, USA}

\author{G.~Hughes}
\affiliation{Fred Lawrence Whipple Observatory, Harvard-Smithsonian Center for Astrophysics, Amado, AZ 85645, USA}

\author{T.~B.~Humensky}
\affiliation{Physics Department, Columbia University, New York, NY 10027, USA}

\author{C.~A.~Johnson}
\affiliation{Santa Cruz Institute for Particle Physics and Department of Physics, University of California, Santa Cruz, CA 95064, USA}

\author{P.~Kaaret}
\affiliation{Department of Physics and Astronomy, University of Iowa, Van Allen Hall, Iowa City, IA 52242, USA}

\author{P.~Kar}
\affiliation{Department of Physics and Astronomy, University of Utah, Salt Lake City, UT 84112, USA}

\author{N.~Kelley-Hoskins}
\affiliation{DESY, Platanenallee 6, 15738 Zeuthen, Germany}

\author{M.~Kertzman}
\affiliation{Department of Physics and Astronomy, DePauw University, Greencastle, IN 46135-0037, USA}

\author{D.~Kieda}
\affiliation{Department of Physics and Astronomy, University of Utah, Salt Lake City, UT 84112, USA}

\author{F.~Krennrich}
\affiliation{Department of Physics and Astronomy, Iowa State University, Ames, IA 50011, USA}

\author{S.~Kumar}
\affiliation{Department of Physics and Astronomy and the Bartol Research Institute, University of Delaware, Newark, DE 19716, USA}

\author{M.~J.~Lang}
\affiliation{School of Physics, National University of Ireland Galway, University Road, Galway, Ireland}

\author{T.~T.Y.~Lin}
\affiliation{Physics Department, McGill University, Montreal, QC H3A 2T8, Canada}

\author{A.~McCann}
\affiliation{Physics Department, McGill University, Montreal, QC H3A 2T8, Canada}

\author{P.~Moriarty}
\affiliation{School of Physics, National University of Ireland Galway, University Road, Galway, Ireland}

\author{R.~Mukherjee}
\affiliation{Department of Physics and Astronomy, Barnard College, Columbia University, NY 10027, USA}

\author{S.~O'Brien}
\affiliation{School of Physics, University College Dublin, Belfield, Dublin 4, Ireland}

\author{R.~A.~Ong}
\affiliation{Department of Physics and Astronomy, University of California, Los Angeles, CA 90095, USA}

\author{A.~N.~Otte}
\affiliation{School of Physics and Center for Relativistic Astrophysics, Georgia Institute of Technology, 837 State Street NW, Atlanta, GA 30332-0430}

\author{D.~Pandel}
\affiliation{Department of Physics, Grand Valley State University, Allendale, MI 49401, USA}

\author{N.~Park}
\affiliation{Enrico Fermi Institute, University of Chicago, Chicago, IL 60637, USA}

\author{A.~Petrashyk}
\affiliation{Physics Department, Columbia University, New York, NY 10027, USA}

\author{M.~Pohl}
\affiliation{Institute of Physics and Astronomy, University of Potsdam, 14476 Potsdam-Golm, Germany}
\affiliation{DESY, Platanenallee 6, 15738 Zeuthen, Germany}

\author{E.~Pueschel}
\affiliation{DESY, Platanenallee 6, 15738 Zeuthen, Germany}

\author{J.~Quinn}
\affiliation{School of Physics, University College Dublin, Belfield, Dublin 4, Ireland}

\author{K.~Ragan}
\affiliation{Physics Department, McGill University, Montreal, QC H3A 2T8, Canada}

\author{G.~T.~Richards}
\affiliation{Department of Physics and Astronomy and the Bartol Research Institute, University of Delaware, Newark, DE 19716, USA}
\affiliation{School of Physics and Center for Relativistic Astrophysics, Georgia Institute of Technology, 837 State Street NW, Atlanta, GA 30332-0430}

\author{E.~Roache}
\affiliation{Fred Lawrence Whipple Observatory, Harvard-Smithsonian Center for Astrophysics, Amado, AZ 85645, USA}

\author{I.~Sadeh}
\affiliation{DESY, Platanenallee 6, 15738 Zeuthen, Germany}

\author{M.~Santander}
\affiliation{Department of Physics and Astronomy, University of Alabama, Tuscaloosa, AL 35487, USA}

\author{S.~S.~Scott}
\affiliation{Santa Cruz Institute for Particle Physics and Department of Physics, University of California, Santa Cruz, CA 95064, USA}

\author{G.~H.~Sembroski}
\affiliation{Department of Physics and Astronomy, Purdue University, West Lafayette, IN 47907, USA}

\author{K.~Shahinyan}
\affiliation{School of Physics and Astronomy, University of Minnesota, Minneapolis, MN 55455, USA}

\author{I.~Sushch}
\affiliation{DESY, Platanenallee 6, 15738 Zeuthen, Germany}

\author{J.~Tyler}
\affiliation{Physics Department, McGill University, Montreal, QC H3A 2T8, Canada}

\author{S.~P.~Wakely}
\affiliation{Enrico Fermi Institute, University of Chicago, Chicago, IL 60637, USA}

\author{A.~Weinstein}
\affiliation{Department of Physics and Astronomy, Iowa State University, Ames, IA 50011, USA}

\author{R.~M.~Wells}
\affiliation{Department of Physics and Astronomy, Iowa State University, Ames, IA 50011, USA}

\author{P.~Wilcox}
\affiliation{Department of Physics and Astronomy, University of Iowa, Van Allen Hall, Iowa City, IA 52242, USA}

\author{A.~Wilhelm}
\affiliation{Institute of Physics and Astronomy, University of Potsdam, 14476 Potsdam-Golm, Germany}
\affiliation{DESY, Platanenallee 6, 15738 Zeuthen, Germany}

\author{D.~A.~Williams}
\affiliation{Santa Cruz Institute for Particle Physics and Department of Physics, University of California, Santa Cruz, CA 95064, USA}

\author{T.~J~Williamson}
\affiliation{Department of Physics and Astronomy and the Bartol Research Institute, University of Delaware, Newark, DE 19716, USA}

\author{B.~Zitzer}
\affiliation{Physics Department, McGill University, Montreal, QC H3A 2T8, Canada}

\begin{abstract}

We conduct a search for periodic emission in the very high-energy gamma-ray band (VHE; E $>$ 100\,GeV) from a total of 13 pulsars in an archival VERITAS data set with a total exposure of over 450 hours.  The set of pulsars includes many of the  brightest young gamma-ray pulsars visible in the Northern Hemisphere.  The data analysis 
resulted in non-detections of pulsed VHE gamma rays from each pulsar.  Upper limits on a potential VHE gamma-ray flux are derived at the 95\% confidence level above three energy thresholds using two methods.  These are the first such searches for pulsed VHE emission from each of the pulsars, and the obtained limits constrain a possible flux component manifesting at VHEs as is seen for the Crab pulsar.

\end{abstract}

\keywords{gamma rays: observations --- 
pulsars: general --- stars: neutron}

\section{Introduction} \label{sec:intro}

While just seven gamma-ray pulsars were known prior to the launch of the {\it Fermi} Large Area Telescope (LAT) in 2008, over 200 gamma-ray pulsars have now been discovered in the high-energy (HE) gamma-ray band above 100\,MeV.\footnote{\detokenize{https://confluence.slac.stanford.edu/display/GLAMCOG/Public+List+of+LAT-Detected+Gamma-Ray+Pulsars}}  However, in the gamma-ray band above 50\,GeV (as reported in the 2FHL catalog~\citep{2016ApJS..222....5A}), only the Vela pulsar remains firmly detected in the LAT data.  Further, in a stacked analysis of 115 gamma-ray pulsars by~\cite{2015ApJ...804...86M}, no significant gamma-ray excess was observed at energies above 50\,GeV.  The shapes of the HE spectra for gamma-ray pulsars are well-characterized by exponential or sub-exponential cutoffs above energies of a few GeV, where the fluxes are observed to fall rapidly.  Natural cutoffs in the gamma-ray spectra at a few GeV are expected in synchro-curvature emission models due to the radiation-reaction limit~\citep{2014ARA&A..52..211C}.  
Synchro-curvature radiation from the outer magnetosphere remains the most accepted explanation found in the literature for the emission of the observed HE gamma-ray radiation due to its ability to reproduce the general features of the observed pulsar light curves and spectra~\citep{2009ApJ...695.1289W, 2013ApJS..208...17A, 2015A&A...575A...3P}. 

Since the detection of the Crab pulsar in VHE gamma rays above 100\,GeV by VERITAS~\citep{2011Sci...334...69V} and MAGIC~\citep{2012A&A...540A..69A}, one of the principal questions in VHE astrophysics has been whether or not the Crab pulsar is the sole VHE-emitting pulsar.  The observed VHE emission from the Crab pulsar was found to be consistent with a pure power law extending from HE gamma rays $>$10\,GeV into the VHE band.  These VHE measurements allowed stringent constraints to be placed on the location of the emission region and radiation mechanism responsible for the observed gamma rays.  Synchro-curvature radiation is not an adequate explanation for the observed radiation from the Crab pulsar into the VHE band due to the absence of the expected spectral cutoff~\citep{2012puas.book.....L}.  The VHE emission is generally accepted to be a result of inverse-Compton (IC) scattering by some population(s) of relativistic electrons (e.g.,~\cite{2012ApJ...754...33L},~\cite{2012Natur.482..507A},~\cite{2015ApJ...811...63H},~\cite{2015MNRAS.449L..51M}), though the proposed models largely diverge in their assumed emission locations and specifics of the IC scattering mechanism.  Furthermore, the VHE spectrum of the Crab pulsar was measured to be consistent with a power law up to 1.5\,TeV by MAGIC~\citep{2016A&A...585A.133A} with no evidence of a spectral cutoff.  At present, it is not clear that any one model is capable of simultaneously reproducing the observed VHE light curve and energy spectrum of the Crab pulsar~\citep{2017EPJWC.13603003Z}.

The Vela pulsar has recently been detected at energies above 50\,GeV by the {\it Fermi}-LAT~\citep{2014ApJ...797L..13L} and up to $\sim$100\,GeV by H.E.S.S. II~\citep{2018A&A...620A..66H}, making it the second pulsar detected from the ground by an imaging atmospheric Cherenkov telescope (IACT).  The 10--100\,GeV spectrum seen by H.E.S.S. II is consistent with a pure power law, though the presence of curvature cannot yet be ruled out given the current uncertainties, leaving open the question of the shape of the spectrum in this energy range.

Both the Crab and Vela pulsars belong to a population of young rotation-powered gamma-ray pulsars, which is a population comprising over half of the total known pulsars in gamma rays.  As two of the brightest young pulsars, the Crab and Vela are the most highly ranked\footnote{\detokenize{http://www.atnf.csiro.au/people/pulsar/psrcat/}} among all known pulsars in terms of the detectability metric $\dot E/d^2$, where $\dot E$ is the spin-down luminosity, and $d$ is the distance.  
However, the second-brightest gamma-ray pulsar (also of relatively high $\dot E/d^2$), Geminga, has been observed in deep exposures by VERITAS and MAGIC, and VHE pulsations were not detected in the searches conducted by either instrument~\citep{2015ApJ...800...61A, 2016A&A...591A.138A}.   Given that the Crab and Vela pulsars are the only known to emit at very high energies, a natural starting point for a search for more pulsars in the VHE band would be to selectively target based on $\dot E/d^2$, despite the non-detection of Geminga.
  
Many gamma-ray pulsars of the highest $\dot E/d^2$  are associated with known Galactic pulsar wind nebulae (PWNe), and PWNe are one of the dominant source classes detected at VHEs by IACT arrays.  Through PWN searches, the major IACT observatories have each accumulated a large amount of data that can be probed for pulsed emission from the pulsar powering the nebula, regardless of whether the nebula is detected.    
Indeed, VERITAS has incidentally observed a number of pulsars while targeting a PWN or other object; 13 such pulsars are listed in Table~\ref{tab:pulsar_properties}, along with some of their properties.  These 13  pulsars are hereafter referred to as the \emph{archival pulsars}.  The list contains eight of the top twelve pulsars in the \textit{Fermi}-LAT second pulsar catalog (2PC) located in the northern sky visible to VERITAS when ranked in $\dot E/d^2$.  Two of the top twelve are the Crab (rank 1) and Geminga (rank 5) pulsars, which have already been subjects of VERITAS observational campaigns~\citep{2011Sci...334...69V, 2015ApJ...800...61A} (though we note that the high ranking for Geminga is principally due to its small distance rather than a particularly high $\dot E$).  

\newcolumntype{C}{>{\centering\arraybackslash}X}
\begin{sidewaystable*}
\centering
\tiny
\caption{The archival pulsar properties.  The right ascension and declination (J2000) values given in columns 2 and 3 used in the analysis are taken from the timing solutions.  Columns 4 and 5 give the pulsar period, $P$, and time derivative of the period, $\dot P$.  The spin-down luminosities ($\dot E$) are given in column 6, and the pulsar distances or distance limits are given in column 7.   Column 8 lists the ranking in $\dot E / d^2$ for the Northern Hemisphere.  Column 9 gives the possible PWN counterparts of the pulsars, and columns 10 and 11 give the VERITAS exposure times and average zenith angles of observations.  Values for $P$, $\dot P$, and $\dot E$ have been taken from the second {\it Fermi}-LAT pulsar catalog (2PC)~\citep{2013ApJS..208...17A} unless otherwise noted.  Where a distance upper limit is quoted, the limit is the distance to the Galaxy's edge as calculated in~\cite{2013ApJS..208...17A}.  The possible PWN counterparts are taken from SIMBAD\footnote{\detokenize{http://simbad.u-strasbg.fr/simbad/}} or TeVCat\footnote{\detokenize{http://tevcat.uchicago.edu/}}.}
\vspace{3mm}
\begin{tabularx}{1.0\textwidth}{CCCCCCp{55mm}Cp{24mm}CC}

  \hline
   Pulsar & R.A. ($^\circ$) & Dec. ($^\circ$) & $P$ (ms) & $\dot P$ ($10^{-15}$) & $\dot E$ ($10^{34}$\,erg\,s$^{-1}$) & \hspace{3mm} Distance (kpc) & $\dot E / d^2$ Rank & Possible Counterpart & VERITAS Exposure (hr) & $\bar \theta_{\textrm{zenith}}$ (degrees) \\ \hline
  
J0007+7303 & 1.7565 & 73.0522 & 315.9 & 357 & 44.8 & $1.4\pm0.3$~\citep{1993AJ....105.1060P} & 9 & CTA 1 & 32.4 & 42 \\
J0205+6449 & 31.4080 & 64.8286 & 65.7 & 190 & 2644 & $1.95\pm0.04$~\citep{2006Sci...311...54X} & 3 & 3C 58 & 22.2 & 35 \\
J0248+6021 & 42.0776 & 60.3597 & 217.1 & 55.0 & 21.2 & $2.0\pm0.2$~{\citep{2011A&A...525A..94T}} & 12 & - & 45.9 & 32 \\
J0357+3205 & 59.4680 & 32.0891 & 444.1 & 13.1 & 0.6 & $0.5^{+0.4}_{-0.2}$~\citep{2013ApJ...765...36M} & 14 & - & 7.92 & 14  \\
J0631+1036 & 97.8657 & 10.6165 & 287.8 & 104 & 17.3 & $1.0\pm0.2$~\citep{1996ApJ...456..305Z} & 10 & - & 2.79 & 26 \\
J0633+0632 & 98.4339 & 6.5418 & 297.4 & 79.6 & 11.9 & $<8.7$ & - & - & 108 & 29 \\
J1907+0602 & 286.9782 & 6.0374 & 106.6 & 86.7 & 282 & $3.2\pm0.3$~\citep{2010ApJ...711...64A} & 8 & MGRO J1908+06 & 39.1 & 28 \\
J1954+2836 & 298.5798 & 28.6013 & 92.7 & 21.2 & 105 & $<18.6$ & - & - & 5.18 & 16 \\
J1958+2846 & 299.6667 & 28.7653 & 290.4 & 212 & 34.2 & $<18.5$ & - & - & 13.9 & 10 \\
J2021+3651 & 305.2726 & 36.8513 & 103.7 & 95.6 & 338 & $1.8^{+1.7}_{-1.4}$~\citep{2015ApJ...802...17K} & 4 & Dragonfly Nebula & 58.2 & 18 \\
J2021+4026 & 305.3781 & 40.4461 & 265.3 & 54.2 & 11.4 & $1.5\pm0.4$~{\citep{1980A&AS...39..133L}} & 13 & $\gamma$ Cygni & 20.6 & 21 \\
J2032+4127 & 308.0548 & 41.4568 & 143.2 & 20.4 & 15~\citep{2017MNRAS.464.1211H} & $1.44\pm0.05$~\citep{2018arXiv180409365G} & 11 & 47.9 & 21 \\
J2229+6114 & 337.2720 & 61.2359 & 51.6 & 77.9 & 2231 & $0.80^{+0.15}_{-0.20}$~\citep{2001ApJ...560..236K} & 2 & Boomerang & 47.2 & 33 \\
 
\hline
\end{tabularx}

\label{tab:pulsar_properties}
\end{sidewaystable*}

Brief synopses of some of the more notable pulsars investigated herein are given in the following.
\begin{itemize}
 \item PSR J0007+7303 is a radio-quiet pulsar~\citep{2004ApJ...612..398H} associated with the supernova remnant CTA 1, which is detected in VHE gamma rays above 500\,GeV~\citep{2013ApJ...764...38A}.  It is bright at GeV energies and has the second-highest spectral cutoff energy (4.7\,GeV) among all young gamma-ray pulsars in the 2PC~\citep{2013ApJS..208...17A}.  Above the spectral cutoff energy, its HE gamma-ray spectrum is consistent with a sub-exponential cutoff~\citep{2016ApJ...831...19L}.
 
 \item PSR J0205+6449 is associated with 3C 58, which is a PWN detected at energies above 400\,GeV~\citep{2014A&A...567L...8A}.  It has the third-highest $\dot E$ of any gamma-ray pulsar in the 2PC~\citep{2013ApJS..208...17A}.
 
 \item PSR J0357+3205 is the second slowest-rotating gamma-ray pulsar known\footnote{\detokenize{https://confluence.slac.stanford.edu/display/GLAMCOG/Public+List+of+LAT-Detected+Gamma-Ray+Pulsars}} and also one of the nearest known pulsars at a distance of $\sim$0.5\,kpc~\citep{2013ApJ...765...36M}.  It is notable for having a very long X-ray emission tail that extends $9^\prime$ behind the pulsar~\citep{2011ApJ...733..104D}.  Its estimated runaway velocity of 1900\,km\,s$^{-1}$ makes it one of the highest-velocity pulsars known~\citep{2005AJ....129.1993M}.

 \item PSR J2021+4026 is a radio-quiet gamma-ray pulsar~\citep{2009Sci...325..840A} located within the radio shell of the supernova remnant G78.2+2.1~\citep{2008A&A...490..197L}.  The remnant has also been detected as an extended source in the VHE band by VERITAS~\citep{2013ApJ...770...93A}.  The flux above 100\,MeV from the pulsar was seen to abruptly decrease by $\sim$20\% within less than one week, which coincided with a decrease in the pulsar period time derivative, $\dot P$.  After spending about three years in this low-flux, low $\dot P$ state, the HE flux and $\dot P$ gradually returned to their previous values over the course of a few months~\citep{2017ApJ...842...53Z}. This is currently the only such observation of intermittent behavior (also called mode switching) seen for a pulsar at gamma-ray energies~\citep{2013ApJ...777L...2A}.  The sudden change in HE flux and $\dot P$ are possibly due to a change in the emission beaming from a shift in the magnetic field structure~\citep{2013ApJ...777L...2A}. 
 
 \item PSR J2032+4127 is a pulsar that was thought to be isolated but was recently found to be in a long-period ($P =$ 45--50\,yr~\citep{2017MNRAS.464.1211H}) binary system~\citep{2015MNRAS.451..581L}, orbiting a 15\,$M_{\odot}$ Be star~\citep{2010AN....331..349H} companion.  The pulsar is spatially coincident with the extended VHE gamma-ray source TeV J2032+4130, which would generally be interpreted as a PWN powered by the pulsar~\citep{2008ApJ...675L..25A, 2014ApJ...783...16A}.  In recent observations conducted by VERITAS and MAGIC in fall 2017, both instruments observed a VHE flux elevated by a factor of $\sim$10 over spring/summer 2017~\citep{2017ATel10971....1M, 2018ApJ...867L..19A}.  VERITAS and MAGIC additionally reported a clear detection of a point source at the pulsar location as the system neared periastron, confirming the presence of a gamma-ray binary.
 
\end{itemize}

Searches have been conducted for pulsed emission from all 13 of the pulsars appearing in archival VERITAS data---the first such VHE searches for each pulsar.  The remainder of this article is structured in the following way: in Section~\ref{sec:archival_analysis}, the \textit{Fermi}-LAT and VERITAS data selection and analysis methods are summarized; Section~\ref{sec:archival_results} gives the results of the searches for VHE pulsed emission; and in Section~\ref{sec:archival_discussion} we discuss these results and offer concluding remarks.

\section{Data Selection and Analysis} \label{sec:archival_analysis}

\subsection{VERITAS}

VERITAS is an array of four 12\,m diameter IACTs located at the Fred Lawrence Whipple Observatory in Southern Arizona (31$^{\circ}$ 40' N, 110$^{\circ}$ 57' W, 1.3\,km a.s.l.). Full-array operations started in spring 2007.  The telescopes are a Davies-Cotton design with reflectors consisting of 345 adjustable hexagonal mirror facets and cameras comprising 499 photomultiplier tubes covering a field of view (FoV) of $\sim$3.5$^{\circ}$.  VERITAS is sensitive to VHE gamma-ray photons in the energy range 85\,GeV to $>$30\,TeV with a sensitivity to detect a 1\% Crab Nebula source in approximately 25 hours.  It has an angular resolution of $\sim0.1^{\circ}$ at 68\% containment averaged over the VERITAS energy range and a pointing-accuracy error of less than 50 arcseconds~\citep{2015ICRC...34..771P}.  The VERITAS data analysis results presented herein are obtained using the general methodology outlined in~\cite{2008ApJ...679.1427A}.

Given that the analysis for each pulsar is performed on archival VERITAS data, the amount of available data for each pulsar varies considerably.  The location of each pulsar is taken to be that provided in the corresponding pulsar timing solution, and these coordinates are given in Table~\ref{tab:pulsar_properties}.  The data selected for analysis satisfy two criteria: a) an archival pulsar is within $1.5^{\circ}$ of the center of the instrument FoV, and b) the date the data were taken falls within the epoch of validity of the corresponding timing solution used to phase fold the photon arrival times.  The timing solutions used for the analyses presented here were obtained from the {\it LAT Gamma-Ray Pulsar Timing Models} public webpage.\footnote{\detokenize{https://confluence.slac.stanford.edu/display/GLAMCOG/LAT+Gamma-ray+Pulsar+Timing+Models}} 
The data undergo a quality-selection process, with the sum of all VERITAS data analyzed here constituting a total exposure time of 451.3\,hr.  The exposure time for each individual pulsar is given in column 2 of Table~\ref{tab:pulsedfluxtable}.  After the VERITAS data are processed through the standard analysis pipeline, the data are phase folded with the appropriate timing solution using the \texttt{Tempo2} pulsar timing software package~\citep{2006MNRAS.369..655H}.  

A total of six periodicity tests are applied to the phase-folded data for each pulsar.  All three sets of the standard VERITAS data selection cuts are used for each individual analysis, where each set was originally generated in a way to optimize sensitivity for power-law spectral shapes varying from soft to hard.  The sets of cuts applied to the data are hereafter referred to as \textit{soft}, \textit{moderate}, and \textit{hard}. The primary effect of applying these three sets of cuts is establishing three different energy thresholds per analysis, which is done in order to minimize a priori assumptions about the energy above which emission may be seen.  The application of soft cuts gives the lowest analysis energy threshold of the three sets, which therefore provides the most sensitivity for potential ``Crab-pulsar-like'' power-law spectral shapes extending from the HE band into the VERITAS energy range, while moderate and hard cuts are better suited for searching for possible additional spectral components manifesting at higher energies in the VHE band due to the higher thresholds.
For each set of cuts, two independent tests for a pulsed signal in the phase-folded VERITAS data are conducted.  The first test uses the a-priori-defined expected signal and background phase regions (henceforth referred to as the ``phase-gate test'') described in Section~\ref{sec:gating}.  The significance is calculated using equation 17 in~\cite{1983ApJ...272..317L} from the unbinned phase data by counting $N_{\textrm{on}}$ and $N_{\textrm{off}}$ in the gated phase regions and determining the ratio of the sizes of the signal- and background-counting regions, $\alpha$.  For the second test, de Jager's $H$ test~\citep{1989A&A...221..180D} is applied to the unbinned phase data, which has the advantage that no a-priori knowledge of expected pulse location(s) in the light curves is needed.  The total number of tests is, therefore, six per pulsar search. 

\subsection{{\it Fermi}-LAT}

The {\it Fermi} Large Area Telescope (LAT) is an electron-positron pair-conversion telescope sensitive to gamma-ray photons with energies between $\sim$20\,MeV and more than 300\,GeV.  It has a FoV of $\sim$2.5\,sr and attains full-sky coverage approximately every three hours.  For a complete description of the instrument, see~\cite{2009ApJ...697.1071A, 2012ApJS..203....4A}.

A total of 7.6\,yr of \textit{Fermi}-LAT data for each pulsar is analyzed.  To generate {\it Fermi}-LAT spectra for each of the archival pulsars, the {\it Fermi}-LAT Science Tools (version \texttt{v10r0p5}) are used with the standard quality cuts for a Galactic point-source analysis as detailed on the public {\it LAT Data Selection Recommendations} webpage.\footnote{\detokenize{https://fermi.gsfc.nasa.gov/ssc/data/analysis/documentation/Cicerone/Cicerone_Data_Exploration/Data_preparation.html}}\footnote{Maximum zenith angle $= 90^{\circ}$; event class $= 128$; IRF name = \texttt{P8R2\_SOURCE\_V6}}
Events with energies between 100\,MeV and 300\,GeV collected within a $20^\circ$ region-of-interest (ROI) of the 3FGL catalog~\citep{2015ApJS..218...23A} location of each pulsar are processed with the maximum-likelihood fitting routine using the Pass 8 instrument response functions.   

The spectral reconstruction methodology used here follows the same steps outlined in the second {\it Fermi}-LAT pulsar catalog~\citep{2013ApJS..208...17A}.  To generate spectral energy distributions of the {\it Fermi}-LAT data for each pulsar, a maximum likelihood analysis is performed in each of 12 logarithmically spaced energy bins spanning the range of 100\,MeV to 300\,GeV.  Spectral models for all sources in the 3FGL catalog in the ROI in addition to the galactic and isotropic diffuse backgrounds (gll\_iem\_v06.fits, iso\_p8r2\_source\_V6\_v06.txt) are included in the likelihood fitting\footnote{We note that the 3FGL model files are based on an analysis of four years of LAT data, while here we use 7.6 years of data.  To assess the impact of new sources that could be detectable in a roughly doubled data set, we reprocessed all of the LAT data with model files derived from the Preliminary LAT 8-Year Point Source List (FL8Y).  We find that the reconstructed LAT spectra using the FL8Y model files are the same within errors as those presented in this study in Section~\ref{sec:archival_results}.}.  The normalization parameters of the galactic and isotropic diffuse models and all sources within a circle of $4^{\circ}$ in radius centered at the pulsar location are left free in the fitting routine, while all other sources have parameters fixed to their 3FGL values. For the computation of the spectral points, in each energy bin the pulsar is modelled as a point source with a simple power-law spectral shape:
\begin{linenomath*}
\begin{equation} \label{eq:powerlaw}
 {dF \over dE} = F_0 {\left(E \over E_0\right)}^{-\gamma},
\end{equation}
\end{linenomath*}
where $F_0$ is the flux normalization, $E_0$ is fixed to 300\,MeV, and $\gamma$ is the spectral index fixed to 2.  The normalization of each pulsar is left as a free parameter in the fit.

\subsection{Phase Gating for the VERITAS Analysis} \label{sec:gating}

The regions in the pulsar light curves where signal and background counting are done, also referred to as ``phase gates,'' were defined prior to the application of tests for periodicity.  In short, the method utilizes the pulsar light curves presented in the 2PC (0.1 $<$ E $<$ 100\,GeV) to define the phase gates for the VERITAS search.  The method aims to define the gates in such a way that would maximize the detection significance of the VERITAS search by simulating potential VHE light curves that might be obtained in a VERITAS analysis and subsequently finding the optimal phase-gate combination that maximizes the search significance.  There are two assumptions invoked in the method: the potential VHE light curve will have the same features of the HE light curve seen in the LAT data (e.g. location and shape of the pulse peaks); and the VHE flux of the pulsar in question is $\sim$1\% of the Crab Nebula flux.  If the first assumption is not true, then the search for pulsed emission will be less sensitive.  For this reason, we also use the {\it H} test to search for pulsed emission. 

The method for defining the phase gates for the search for pulsations from the archival pulsars comprises the following steps:
\begin{enumerate}
 \item Determine the signal and background event rates from VERITAS Crab Nebula data using soft cuts.   These event rates are obtained in a reflected-region analysis~\citep{1994APh.....2..137F}.
 \item Multiply the rates from step 1 by the source exposure time to get an expected $N_{\textrm{on}}$ and $\alpha N_{\textrm{off}}$, then find $N_{\textrm{excess}} = N_{\textrm{on}} - \alpha N_{\textrm{off}}$, where $N_{\textrm{excess}}$ is the excess counts, $N_{\textrm{on}}$ is the number of counts in the signal region, $N_{\textrm{off}}$ is the number in the background region, and $\alpha$ is the ratio of the size of the signal region(s) to the background region.
 \item Scale the Crab Nebula excess counts found in step 2 by 0.01 to mimic a 1\% Crab Nebula source (the Crab pulsar flux is approximately 1\% of the Crab Nebula flux at $\sim$200\,GeV).
 \item Obtain the binned pulse profile from the 2PC for the pulsar in question and subtract the value of the lowest bin in the profile from all bins to remove the unpulsed component.
 \item Normalize the pulse profile from the previous step and multiply each bin by the scaled excess found in step 3.
 \item Add the estimated background expected for the VERITAS observations to the profile by adding $\alpha N_{\textrm{off}}/N_{\textrm{bins}}$ to each bin, where $N_{\textrm{bins}}$ is the total number of bins.
 \item Define the number of signal phase gates as one per peak present in the 2PC light curve and calculate the significance corresponding to all non-overlapping phase-gate bin combinations, including a background gate.  The phase gates selected for the VERITAS analysis are those for which the significance is maximized.  As an example, consider a pulsar displaying two HE gamma-ray peaks in its light curve. Two signal phase gates are defined a priori, each with a unique starting and ending edge, which must be located at a bin edge.  
A background gate is also defined in the same way, and none of these three gates are allowed to overlap.  Significances are then calculated for all possible gate combinations by scanning the space of the six
possible gate edges. The gate combination with the highest significance is used for the analysis.  
For cases where two signal phase gates are initially defined but were found to be contiguous, the number of gates is reduced to one, and the procedure is performed again.
\end{enumerate}

\begin{figure}[H]
 \centering
 \includegraphics[scale=1.05]{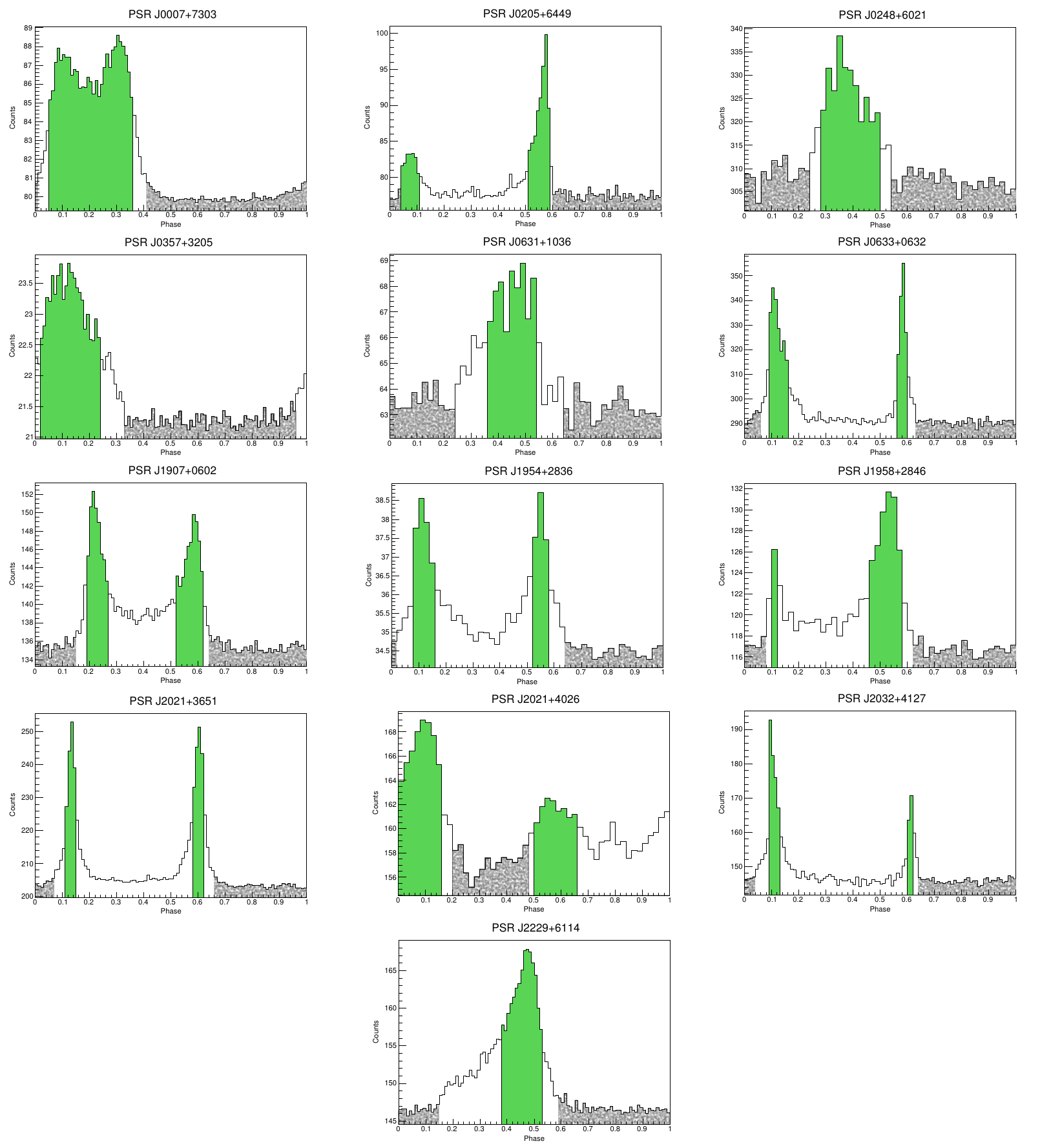}
 \caption{Light curves obtained from the phase-gating procedure (see text) showing the obtained phase-gate definitions for all 13 pulsars appearing in archival VERITAS data.  The signal-counting regions, corresponding to the locations of P1 and P2 (where applicable), are shown in green, and the background-counting region is shown in gray (granulized).}
 \label{fig:phase_gates}
\end{figure}

In some cases where a pulsar shows two distinct, non-overlapping peaks, the significance calculated in this procedure is maximized with only a single signal gate definition.  In order to ensure that a signal gate is defined for each peak, a modified three-pass method is used.  The first pass is simply the procedure described above.  The second pass sets the number of signal phase gates to one and excludes the signal phase gate defined in the first pass from the search region, allowing a signal phase gate to be defined for the peak not found in the first pass.  The third pass sets the number of signal phase gates to two but constrains one of them to be fixed to the signal phase gate determined in pass two.
This modified method was necessary to define gates for the light curves of PSR J0205+6449 and PSR J2021+4026.  The results of the gating procedure are shown in Figure~\ref{fig:phase_gates}, and phase-gate definitions are given in numerical form in Table~\ref{tab:gates}.

The latest publicly available timing solutions from the {\it LAT Gamma-ray Pulsar Timing Models} webpage are used in the VERITAS analysis, which have a longer epoch of validity than those in the 2PC.  The use of these timing solutions to fold the data for the pulsars in most cases introduces appreciable phase offsets with respect to the 2PC light curves, which are calculated and added to the VERITAS event phases allowing the phase gates derived from the 2PC light curves to be used in the analysis.  To determine the phase offsets for the 13 archival pulsars, a {\it Fermi}-LAT data set for each pulsar is phase folded using both the 2PC and the latest timing solutions.  The resulting light curves are then cross-correlated, and the point where the correlation coefficient is maximized is taken to be the offset in phase.  The resulting offset for each pulsar is given in Table~\ref{tab:gates}.


\newcolumntype{C}{>{\centering\arraybackslash}X}
\begin{table*}
\centering
\caption{Phase-gate and phase-offset definitions.  Columns 2 and 3 give the gate definitions for the first peak P1 and the second peak P2 (if present), respectively.  Column 4 gives the background phase-gate definitions.  Column 5 lists the phase offsets between the timing solutions used and those appearing in the 2PC, which were used to calculate the phase gates.}
\vspace{3mm}
\begin{tabularx}{1.0\textwidth}{CCCCC}
  \hline
   Pulsar & P1 & P2 & Background & Phase Offset \\ \hline
  
   J0007+7303 &    0.05--0.36 &     None      &     0.41--0.01 &     $-0.0725$ \\
   J0205+6449 &    0.04--0.11 &     0.51--0.59 &     0.60--0.04 &     $-0.1455$ \\
   J0248+6021 &    0.28--0.50 &     None      &     0.54--0.24 &     0.033  \\
   J0357+3205 &    0.02--0.24 &     None      &     0.33--0.96 &     0.003  \\
   J0631+1036 &    0.36--0.54 &     None      &     0.64--0.24 &     0.023  \\
   J0633+0632 &    0.56--0.60 &     0.09--0.16 &     0.63--0.06 &     0.0145 \\
   J1907+0602 &    0.52--0.62 &     0.19--0.27 &     0.64--0.15 &     0.002  \\
   J1954+2836 &    0.52--0.58 &     0.08--0.16 &     0.64--0.02 &     0.0125 \\
   J1958+2846 &    0.46--0.58 &     0.10--0.12 &     0.62--0.08 &     0.0135 \\
   J2021+3651 &    0.58--0.62 &     0.11--0.15 &     0.66--0.07 &     0.0355 \\
   J2021+4026 &    0.00--0.16 &     0.50--0.66 &     0.20--0.48 &     $-0.0495$ \\
   J2032+4127 &    0.60--0.62 &     0.09--0.13 &     0.64--0.04 &     0.1585 \\
   J2229+6114 &    0.38--0.53 &     None      &     0.59--0.15 &     $-0.0635$ \\
 
\hline
\end{tabularx}

\label{tab:gates}
\end{table*}

\section{Results} \label{sec:archival_results}

None of the six periodicity tests applied to each pulsar data set results in the detection of VHE pulsations.  The distribution of pre-trials significances from the phase-gate test has minimum and maximum values of --1.93$\sigma$ and +1.07$\sigma$, respectively.  The maximum \textit{H} statistic is 10.4, which corresponds to a chance probability of 0.016.  All tests applied to the data therefore do not reveal any evidence for pulsed emission in the VERITAS data.  Significances and \textit{H} statistics for each pulsar are given in Table~\ref{tab:pulsedfluxtable}. 

For each of the six searches for pulsed emission, integral VHE flux upper limits (ULs) from the VERITAS data are computed at the 95\% confidence level.  For the phase-gate test results, the Rolke unbounded method~\citep{2001NIMPA.458..745R} is used to set an upper limit on the excess counts, which is converted into an integral flux UL by dividing by the exposure.  For the results from the \textit{H} test, an integral flux UL is set using the method detailed in~\cite{1994ApJ...436..239D} assuming a single peak with a duty cycle of 10\%.  A spectral index of 3.8 is assumed for all upper limit calculations, which is the same index as seen for the Crab pulsar in~\cite{2011Sci...334...69V}.  We note that assuming a significantly harder ($\gamma = 2$) or softer ($\gamma = 5$) spectral index affects the integral flux ULs at a level of $\sim$25\% on average.  Spectral analysis energy thresholds for the ULs are taken to be the energy corresponding to the peak of the efficiency\footnote{Efficiency in this context is the product of the average effective area vs. energy curve and the assumed spectrum: $dF/dE \propto E^{-\gamma}$, where here we take $\gamma=3.8$.} for each analysis.  
Six 95\% CL integral flux ULs per pulsar are therefore calculated, which are given in Table~\ref{tab:pulsedfluxtable}.  We note that the spectral analysis thresholds vary significantly between pulsars, which is primarily due to the different average zenith angle for each set of observations.         

\newcolumntype{C}{>{\centering\arraybackslash}X}
\begin{sidewaystable*} \scriptsize
\centering
\caption{Results for the 13 pulsars appearing in archival VERITAS data.  Each pulsar has three sets of results, one for each set of cuts applied to the data.  Column 2 lists the exposure time for each pulsar, copied here from Table~\ref{tab:pulsar_properties} for convenience.  Column 3 specifies the set of cuts used in the analysis.  Columns 4 and 5 give the phase-gate test pre-trials significance and \textit{H} statistic, respectively.  Column 6 gives the spectral analysis energy threshold in GeV.  Integral flux upper limits at the 95\% CL above the spectral analysis threshold in Column 6 from the \textit{H} test and Rolke methods are given in columns 7 and 8, respectively.}
\vspace{3mm}
\begin{tabularx}{1.0\textwidth}{CCCCCCCC}
  \hline
  Pulsar & Exposure Time (hr) & Cut Type & Significance & {\it H} Statistic & Spectral Analysis Threshold (GeV)  & {\it H}-Test Flux UL ($10^{-9}$\,m$^{-2}$\,s$^{-1}$) & Rolke Flux UL ($10^{-9}$\,m$^{-2}$\,s$^{-1}$)\\ \hline
 
           &      & soft      & --1.74 & 4.32 & 320  & 16.7 & 1.24 \\
J0007+7303 & 32.4 & moderate  & --0.95 & 2.37 & 460  & 6.20 & 2.48 \\
           &      & hard      & --0.51 & 3.15 & 1100 & 1.38 & 0.767 \\ \hline
           &      & soft      & --1.29 & 1.28 & 240  & 13.7 & 2.77 \\
J0205+6449 & 22.2 & moderate  & --1.11 & 3.29 & 350  & 7.63 & 1.63 \\
           &      & hard      & --1.40 & 3.94 & 500  & 4.12 & 0.575 \\ \hline
           &      & soft      & 0.00   & 3.26 & 220  & 19.4 & 11.0 \\
J0248+6021 & 45.9 & moderate  & 0.85   & 3.69 & 290  & 10.7 & 8.65 \\
           &      & hard      & 0.44   & 1.34 & 600  & 1.9  & 1.72 \\ \hline
           &      & soft      & --0.47 & 0.74 & 140  & 33.6 & 20.9 \\
J0357+3205 & 7.92 & moderate  & --0.17 & 0.32 & 200  & 10.7 & 10.1 \\
           &      & hard      & 0.12   & 2.36 & 380  & 5.26 & 4.01 \\ \hline
           &      & soft      & --1.27 & 3.61 & 150  & 79.4 & 13.9 \\
J0631+1036 & 2.79 & moderate  & 0.81   & 0.56 & 220  & 18.2 & 22.2 \\
           &      & hard      & --1.07 & 1.44 & 460  & 7.44 & 2.44 \\ \hline
           &      & soft      & --1.37 & 3.66 & 180  & 8.92 & 1.00 \\
J0633+0632 & 108  & moderate  & 0.41   & 0.32 & 260  & 1.95 & 1.59 \\
           &      & hard      & 0.70   & 4.80 & 500  & 1.01 & 0.523 \\ \hline
           &      & soft      & --1.49 & 1.60 & 180  & 11.7 & 1.72 \\
J1907+0602 & 39.1 & moderate  & 0.36   & 10.4 & 260  & 7.72 & 3.72 \\
           &      & hard      & --0.15 & 2.60 & 550  & 1.73 & 0.953 \\ \hline
           &      & soft      & 1.07   & 7.01 & 130  & 68.4 & 40.3 \\
J1954+2836 & 5.18 & moderate  & 0.58   & 2.46 & 200  & 19.3 & 14.0 \\
           &      & hard      & --1.50 & 0.60 & 290  & 8.24 & 1.48 \\ \hline
           &      & soft      & --0.70 & 1.62 & 130  & 24.9 & 8.62 \\
J1958+2846 & 13.9 & moderate  & --1.24 & 0.82 & 180  & 9.49 & 2.24 \\
           &      & hard      & --1.54 & 3.00 & 260  & 6.81 & 0.658 \\ \hline
           &      & soft      & --0.56 & 9.46 & 150  & 25.4 & 4.53 \\
J2021+3651 & 58.2 & moderate  & 0.25   & 2.28 & 220  & 7.23 & 2.96 \\
           &      & hard      & 0.95   & 6.46 & 420  & 2.48 & 1.06 \\ \hline
           &      & soft      & 0.18   & 0.73 & 170  & 24.1 & 32.1 \\
J2021+4026 & 20.6 & moderate  & 0.15   & 3.28 & 240  & 15.0 & 13.8 \\
           &      & hard      & --1.93 & 2.42 & 460  & 4.68 & 0.0615 \\ \hline
           &      & soft      & --0.37 & 0.34 & 170  & 10.9 & 4.07 \\
J2032+4127 & 47.9 & moderate  & 0.58   & 4.29 & 220  & 10.4 & 3.56 \\
           &      & hard      & 0.42   & 2.00 & 460  & 2.22 & 0.974 \\ \hline
           &      & soft      & 0.72   & 0.30 & 240  & 8.75 & 9.41 \\
J2229+6114 & 47.2 & moderate  & 0.19   & 0.58 & 320  & 5.28 & 4.07 \\
           &      & hard      & --0.75 & 2.35 & 660  & 1.97 & 0.648 \\ \hline

\hline
\end{tabularx}

\label{tab:pulsedfluxtable}
\end{sidewaystable*}

\begin{figure}[H]
  \centering
  \includegraphics[scale=0.5]{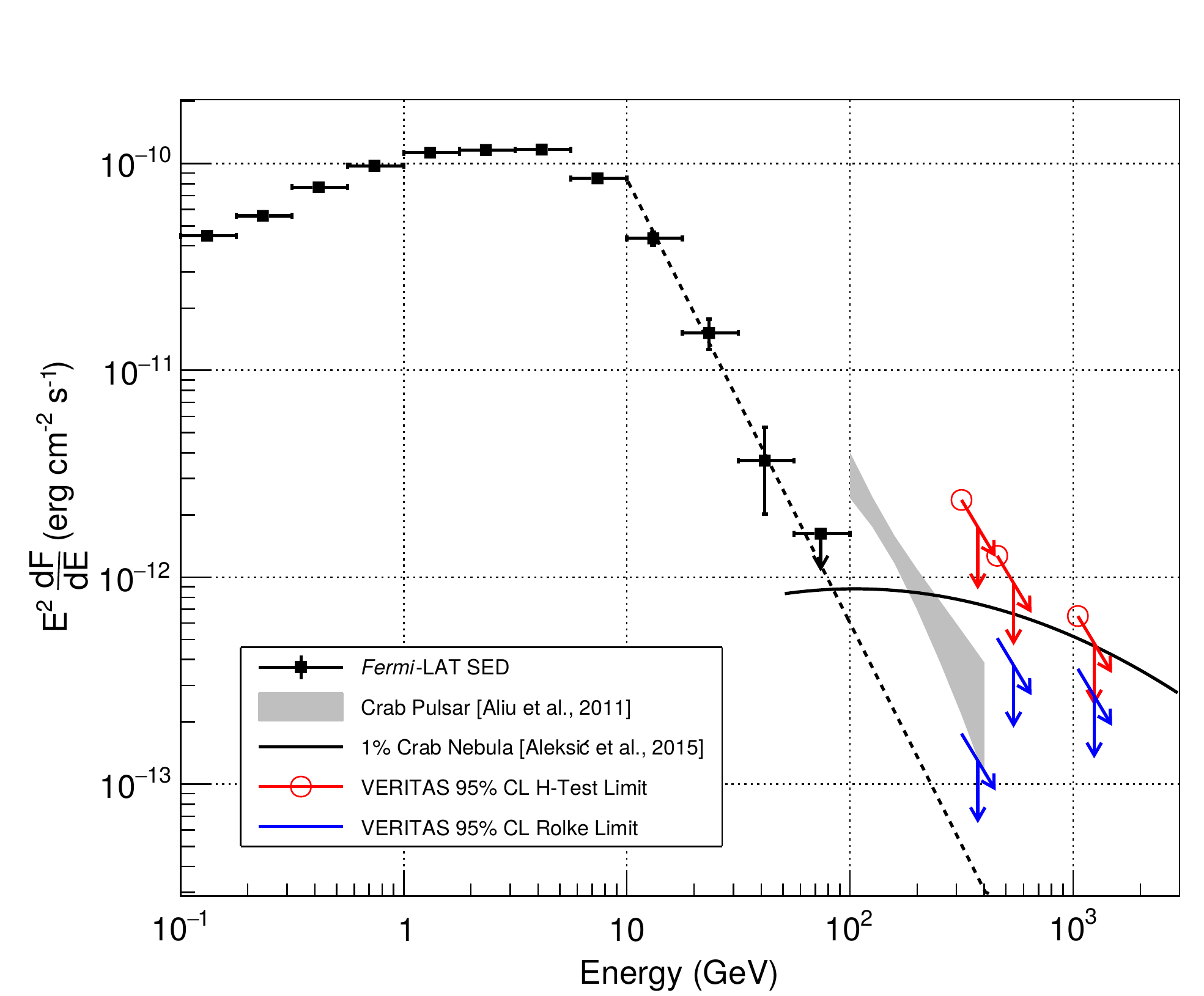}
 \caption{ {\it Fermi}-LAT spectrum of PSR J0007+7303 (black squares) with VERITAS differential flux upper limits from the soft, moderate, and hard-cuts analyses (\emph{H}-Test limits: red arrows, Rolke limits: blue arrows).  A power-law fit to the \textit{Fermi} data above 10\,GeV is given by the black dashed line.}
 \label{fig:PSRJ0007p7303_fermiSED}
\end{figure}
\begin{figure}[H]
  \subfloat[PSR J0205+6449]{\includegraphics[width=0.5\textwidth]{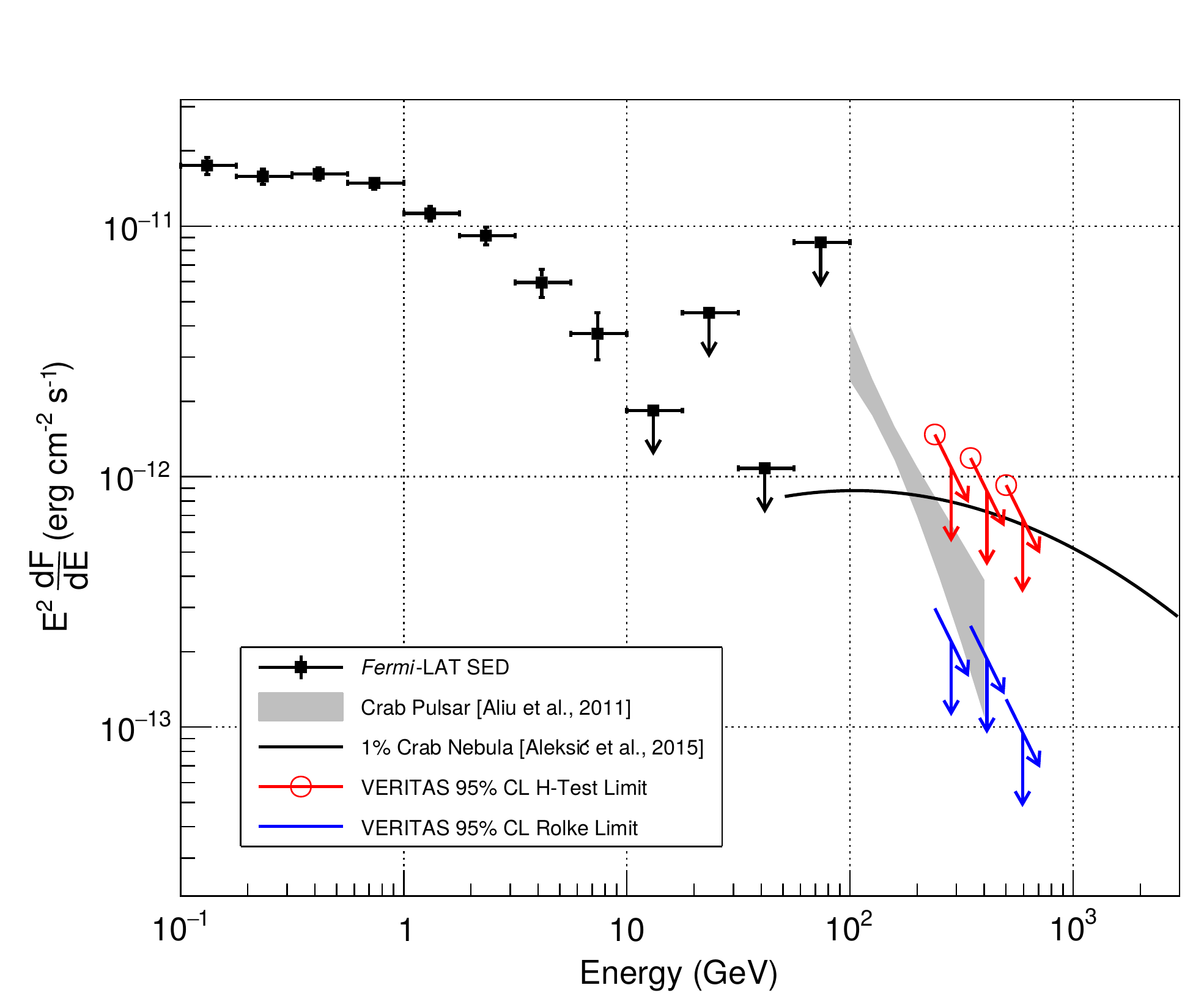}}
    \subfloat[PSR J0248+6021]{\includegraphics[width=0.5\textwidth]{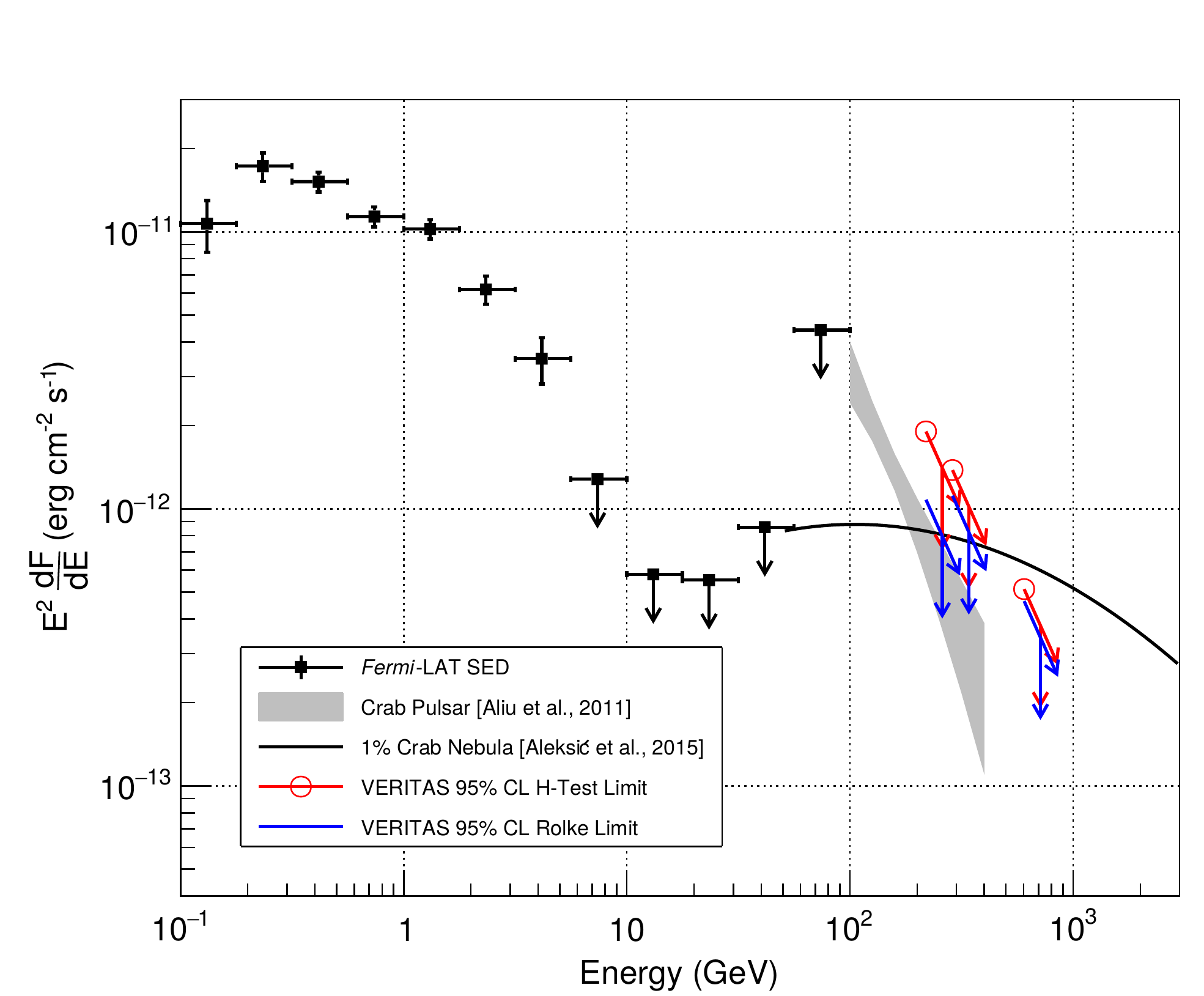}} \\ \\ \\ \\
      \subfloat[PSR J0357+3205]{\includegraphics[width=0.5\textwidth]{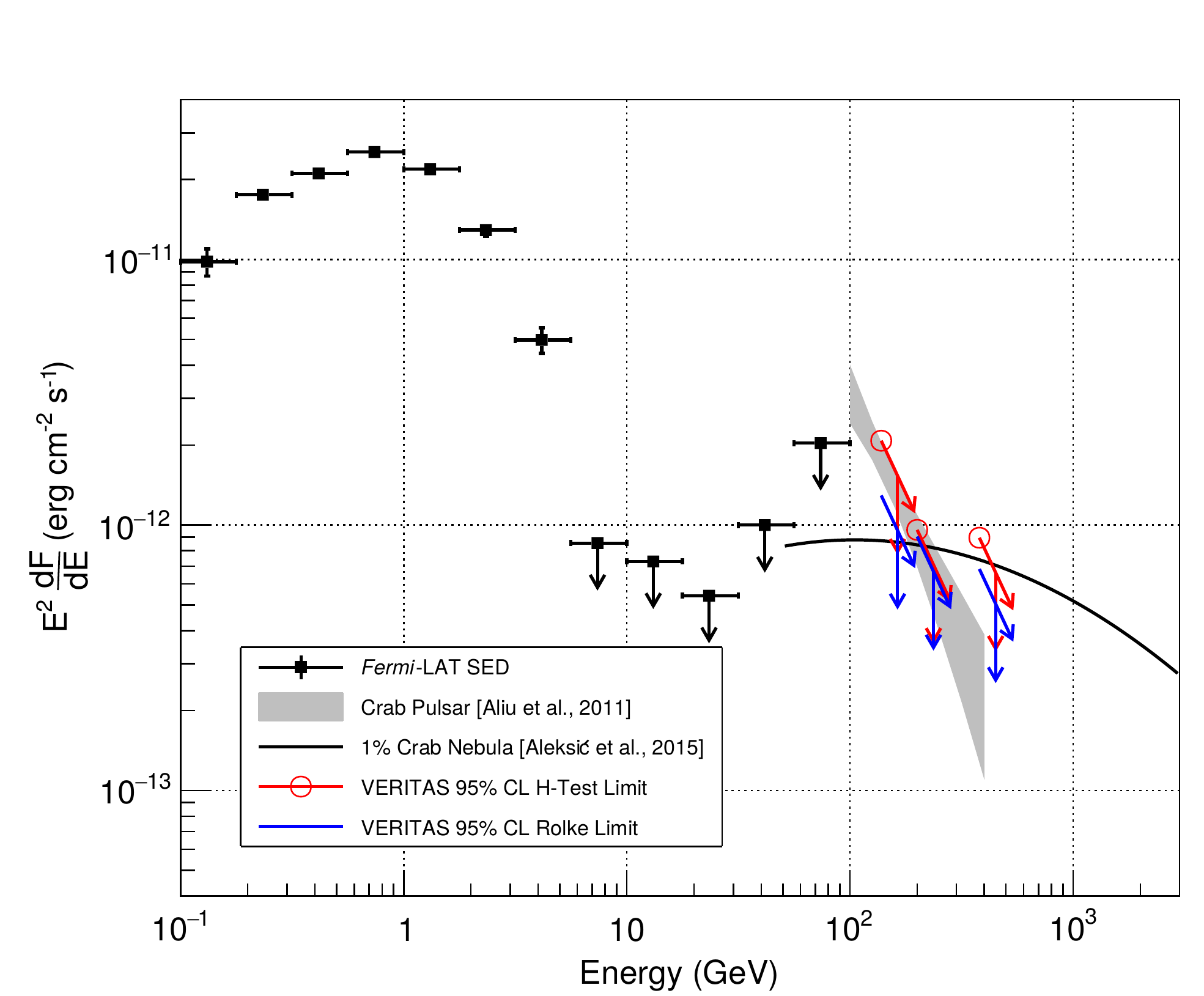}}
      \subfloat[PSR J0631+1036]{\includegraphics[width=0.5\textwidth]{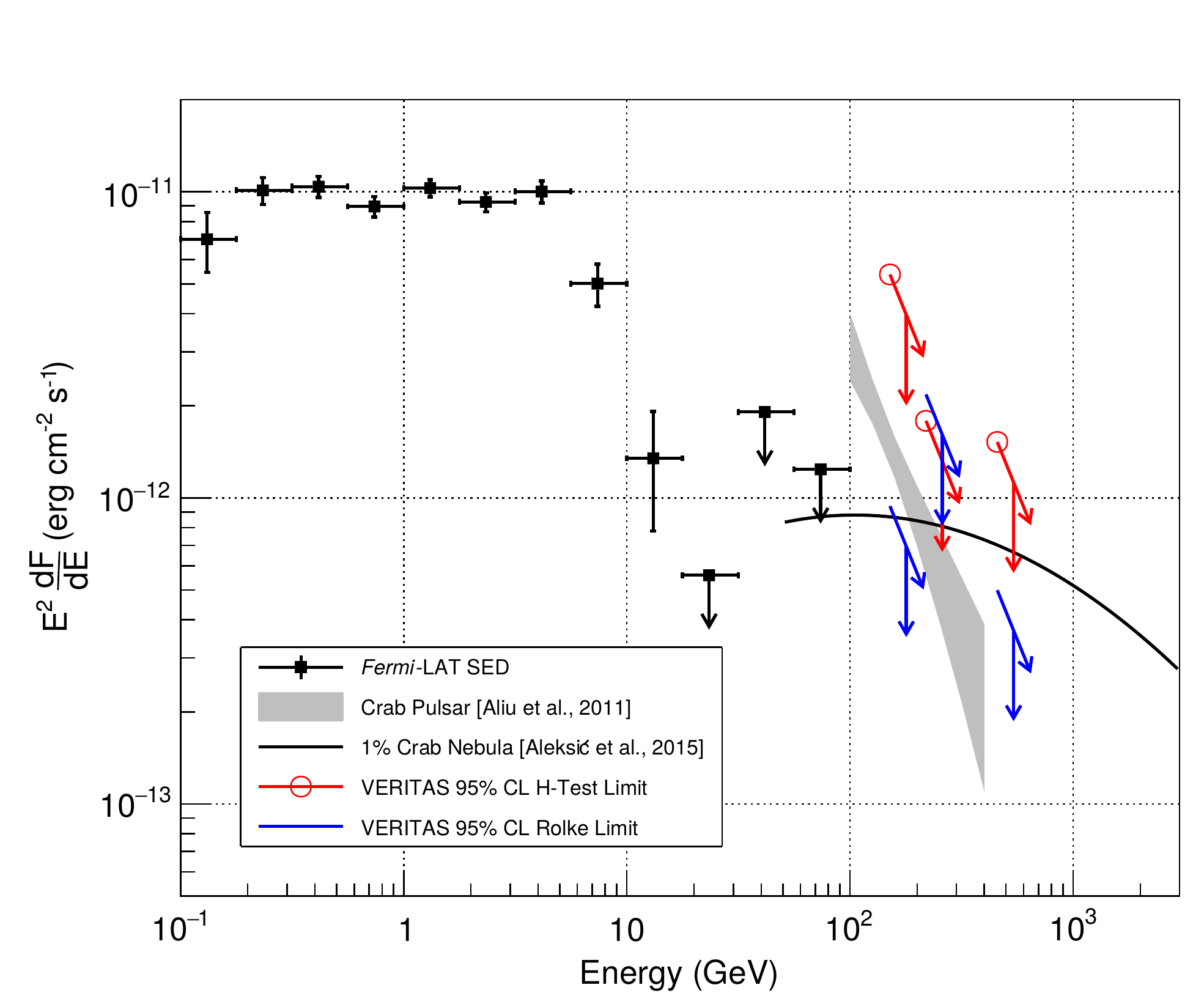}
}

\vspace{10mm}
 \caption{{\it Fermi}-LAT spectra (black squares) with VERITAS differential flux upper limits from the soft, moderate, and hard-cuts analyses (\emph{H}-Test limits: red arrows, Rolke limits: blue arrows) for PSR J0205+6449 (a), PSR J0248+6021 (b), PSR J0357+3205 (c), and PSR J0631+1036 (d).}
 \label{fig:fermiSED_part1}
\end{figure}
\begin{figure}[H]
  \subfloat[PSR J0633+0632]{\includegraphics[width=0.5\textwidth]{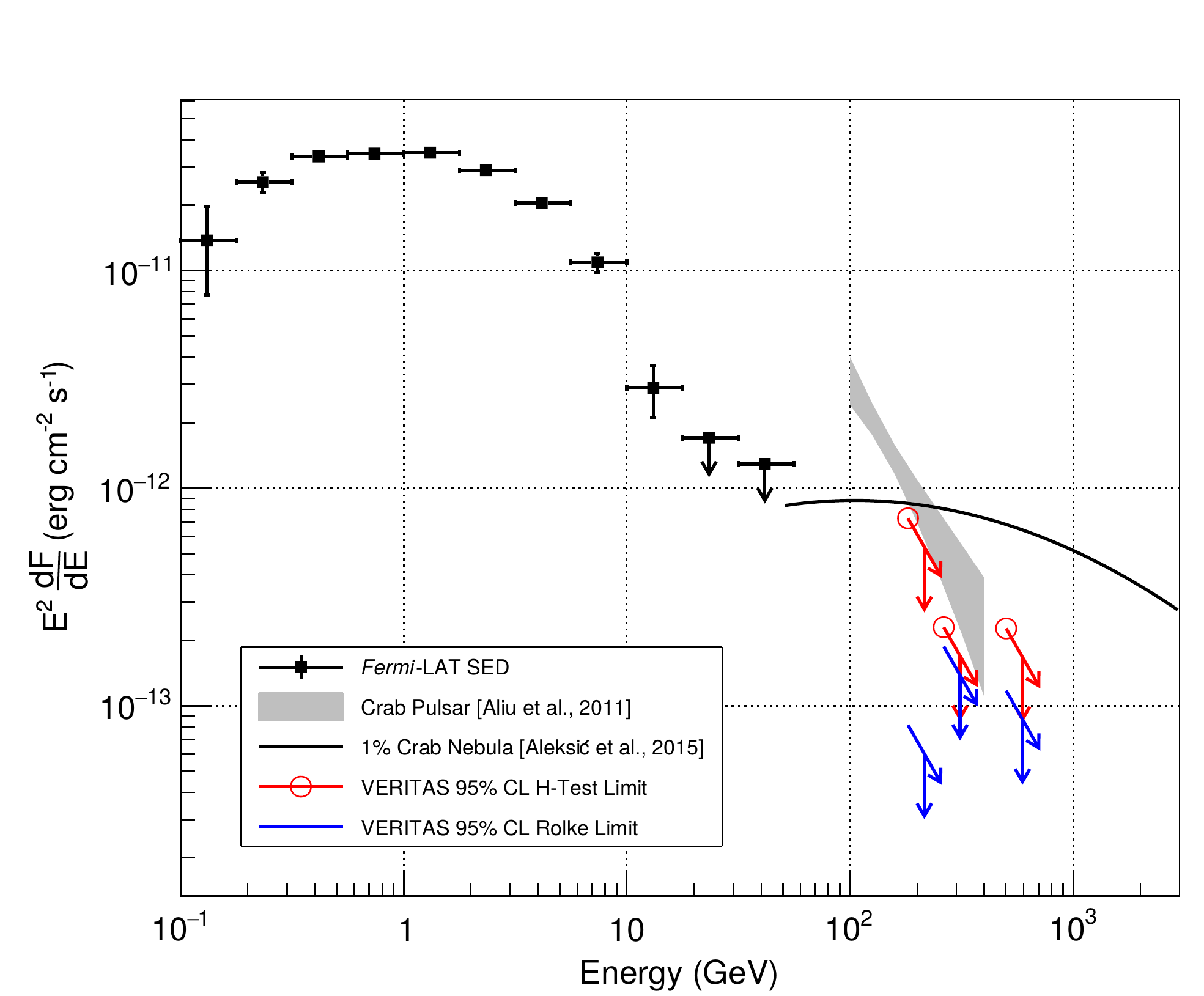}}
    \subfloat[PSR J1907+0602]{\includegraphics[width=0.5\textwidth]{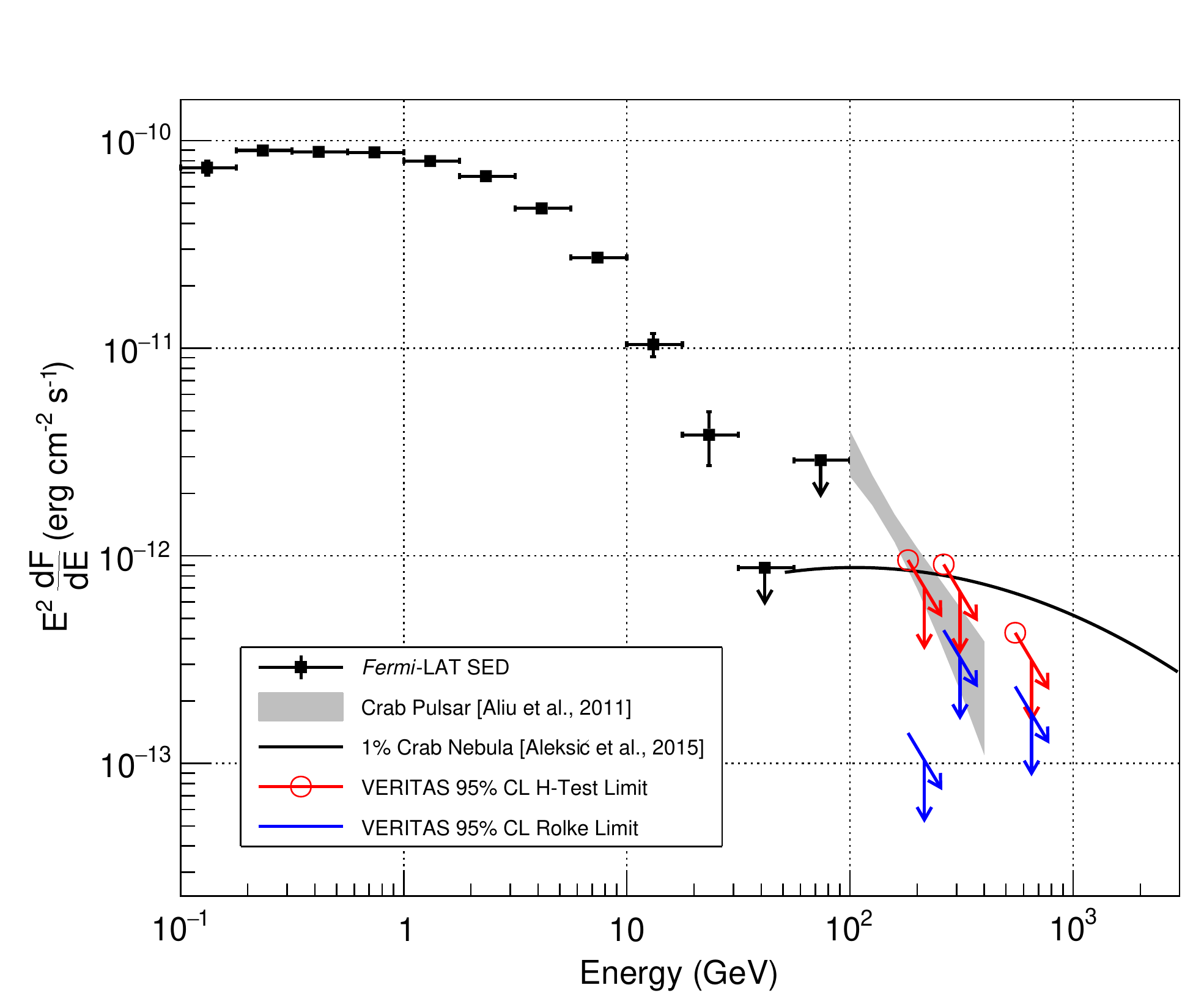}} \\ \\ \\ \\
      \subfloat[PSR J1954+2836]{\includegraphics[width=0.5\textwidth]{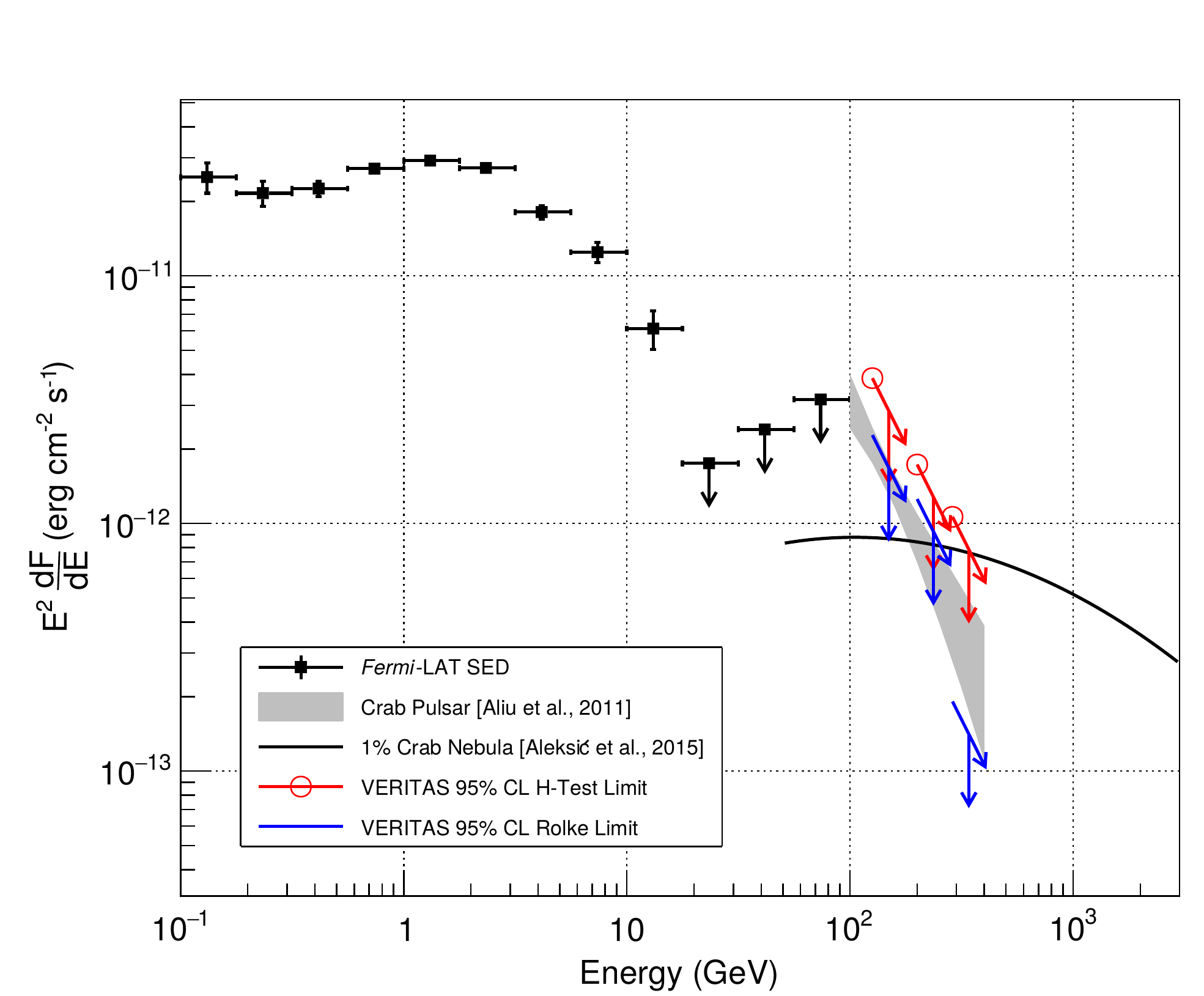}}
      \subfloat[PSR J1958+2846]{\includegraphics[width=0.5\textwidth]{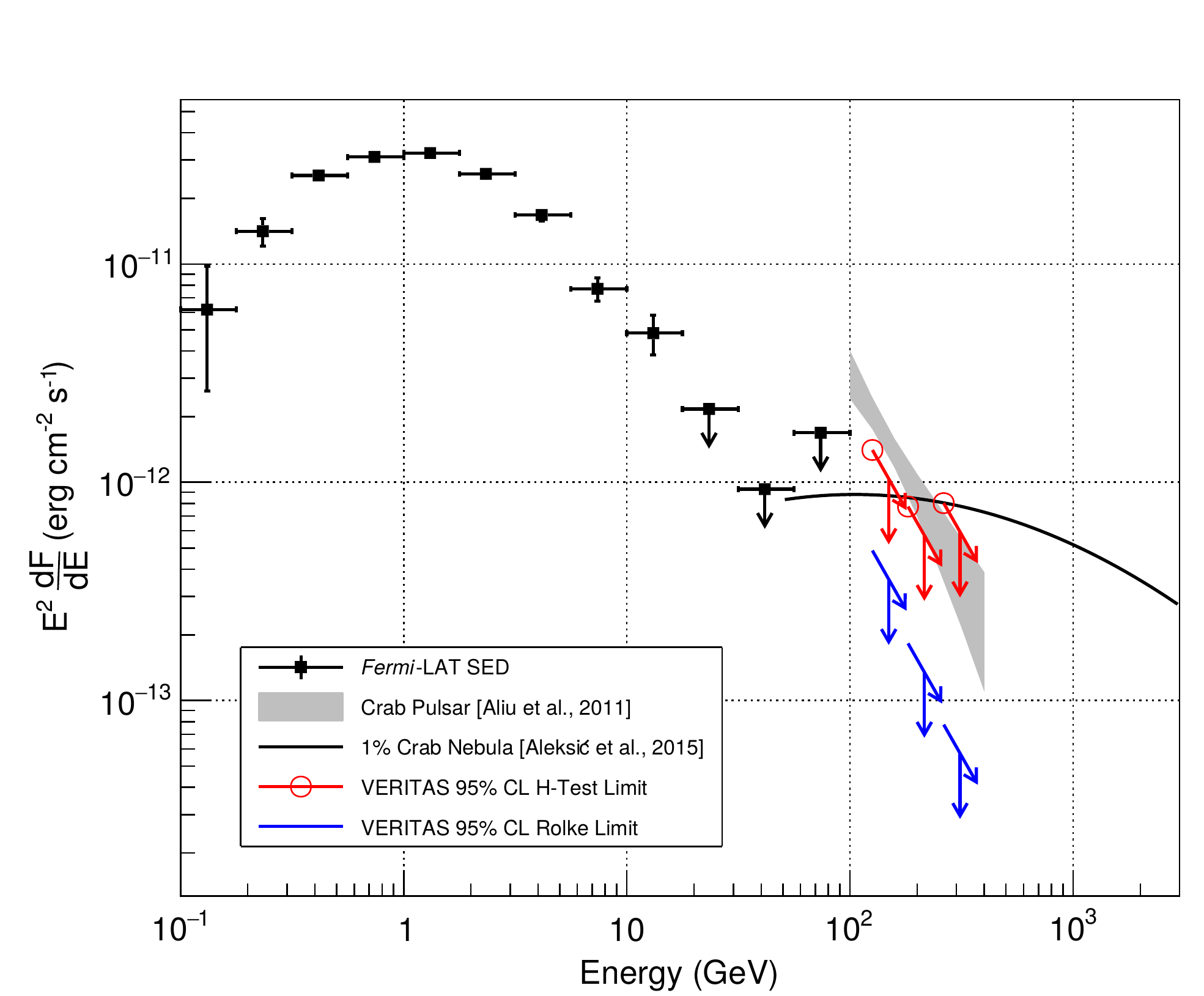}
}

\vspace{10mm}
 \caption{{\it Fermi}-LAT spectra (black squares) with VERITAS differential flux upper limits from the soft, moderate, and hard-cuts analyses (\emph{H}-Test limits: red arrows, Rolke limits: blue arrows) for PSR J0633+0632 (a), PSR J1907+0602 (b), PSR J1954+2836 (c), and PSR J1958+2846 (d).}
 \label{fig:fermiSED_part2}
\end{figure}
\begin{figure}[H]
  \subfloat[PSR J2021+3651]{\includegraphics[width=0.5\textwidth]{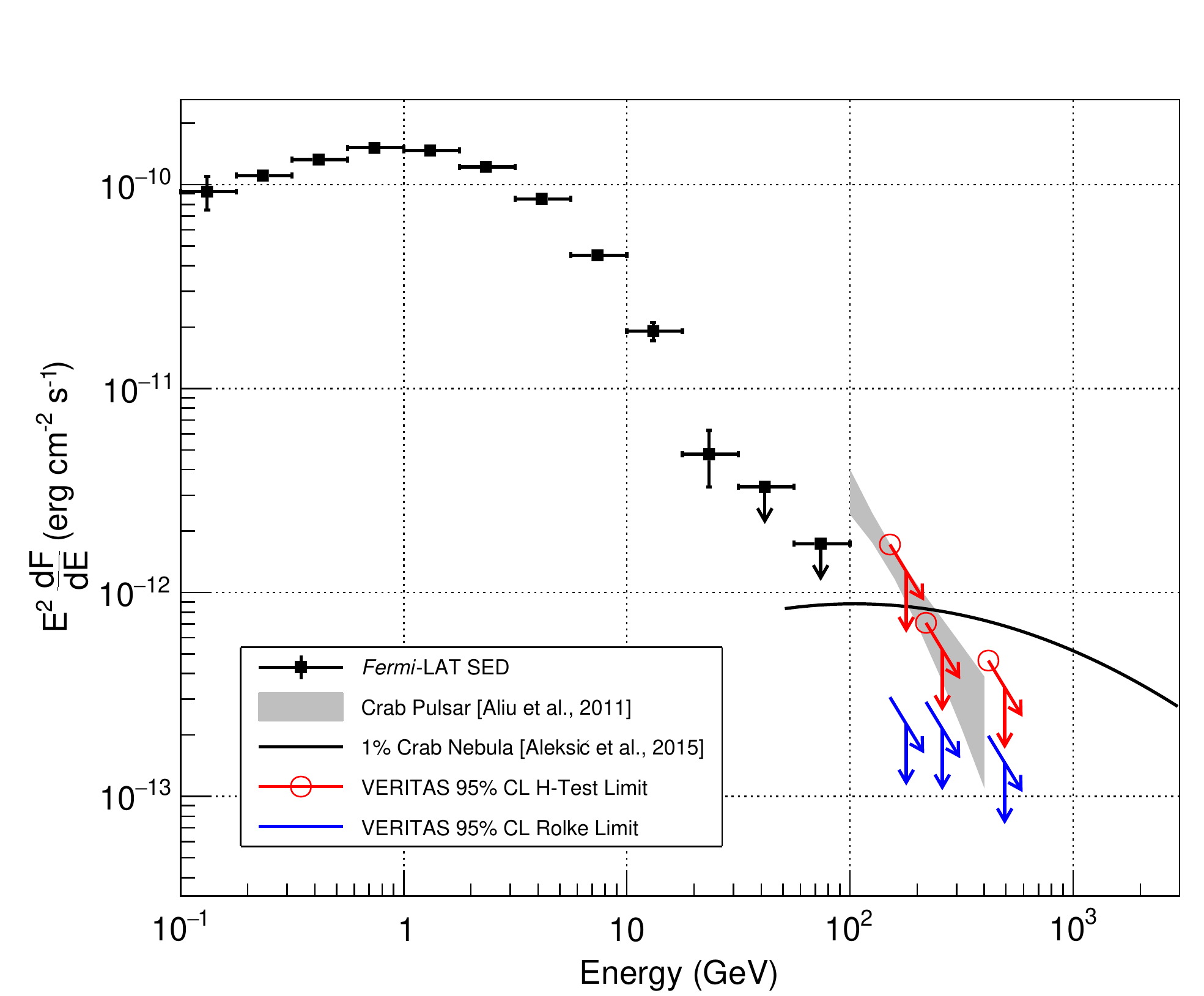}}
    \subfloat[PSR J2021+4026]{\includegraphics[width=0.5\textwidth]{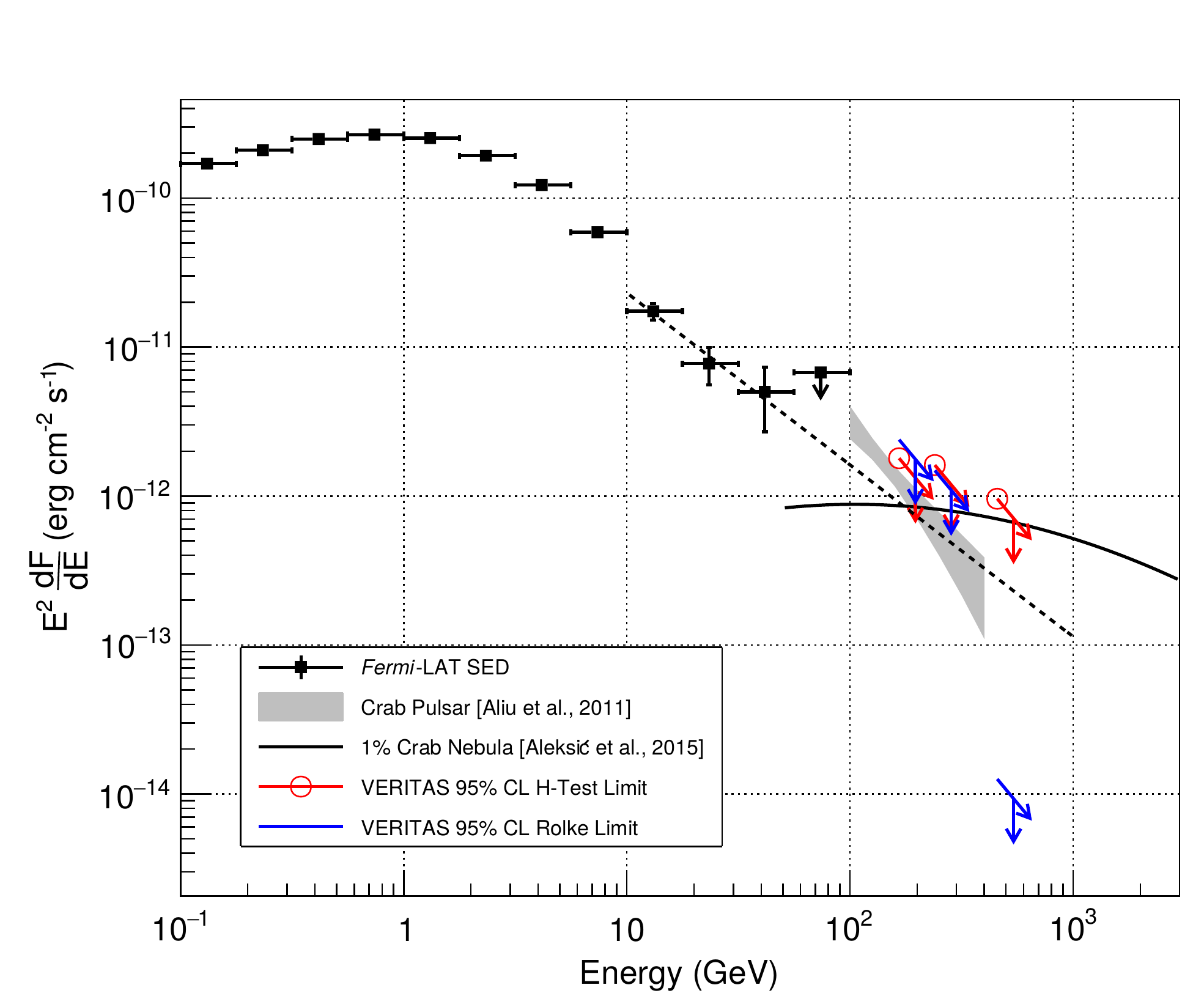}} \\ \\ \\ \\
      \subfloat[PSR J2032+4127]{\includegraphics[width=0.5\textwidth]{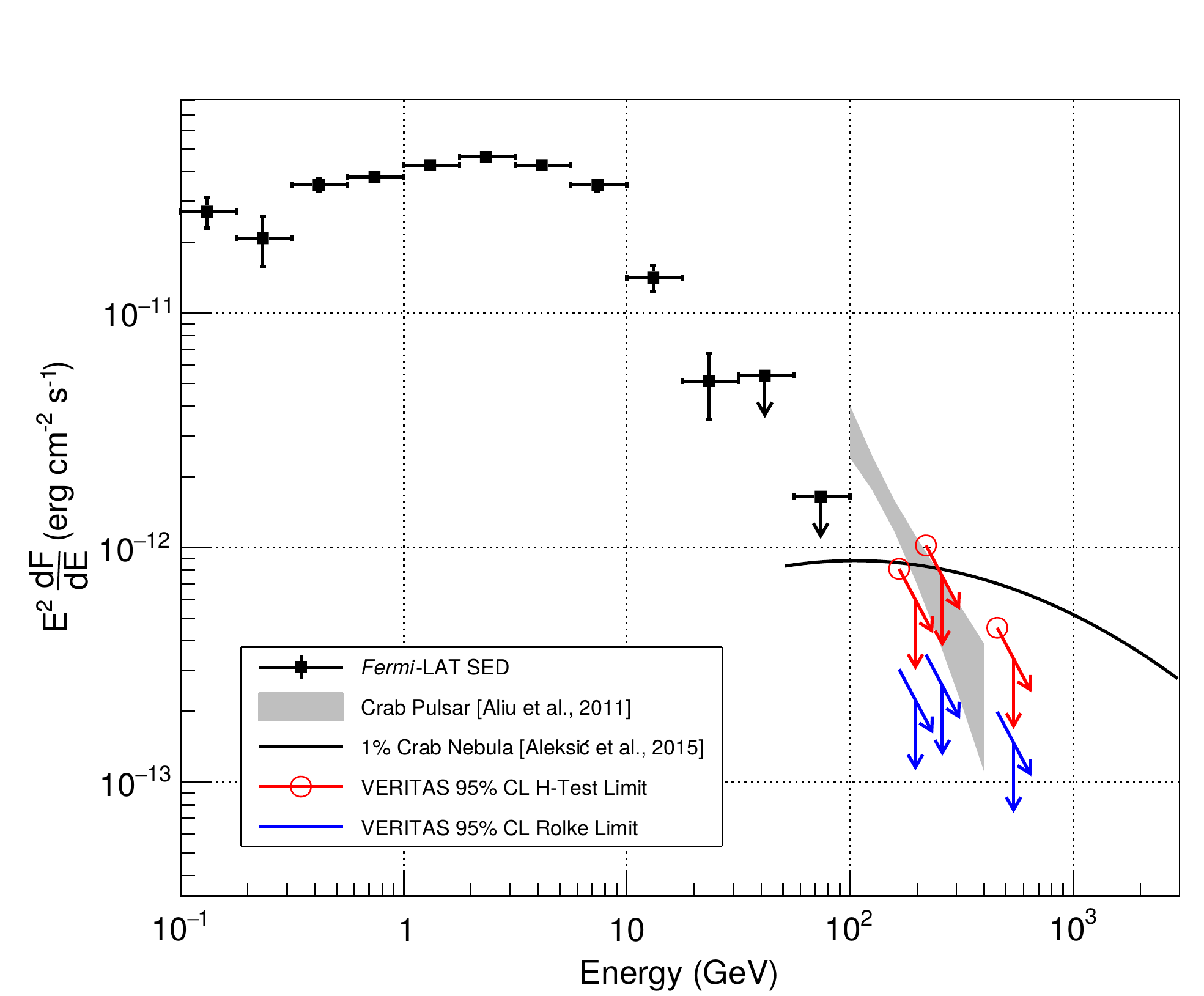}}
      \subfloat[PSR J2229+6114]{\includegraphics[width=0.5\textwidth]{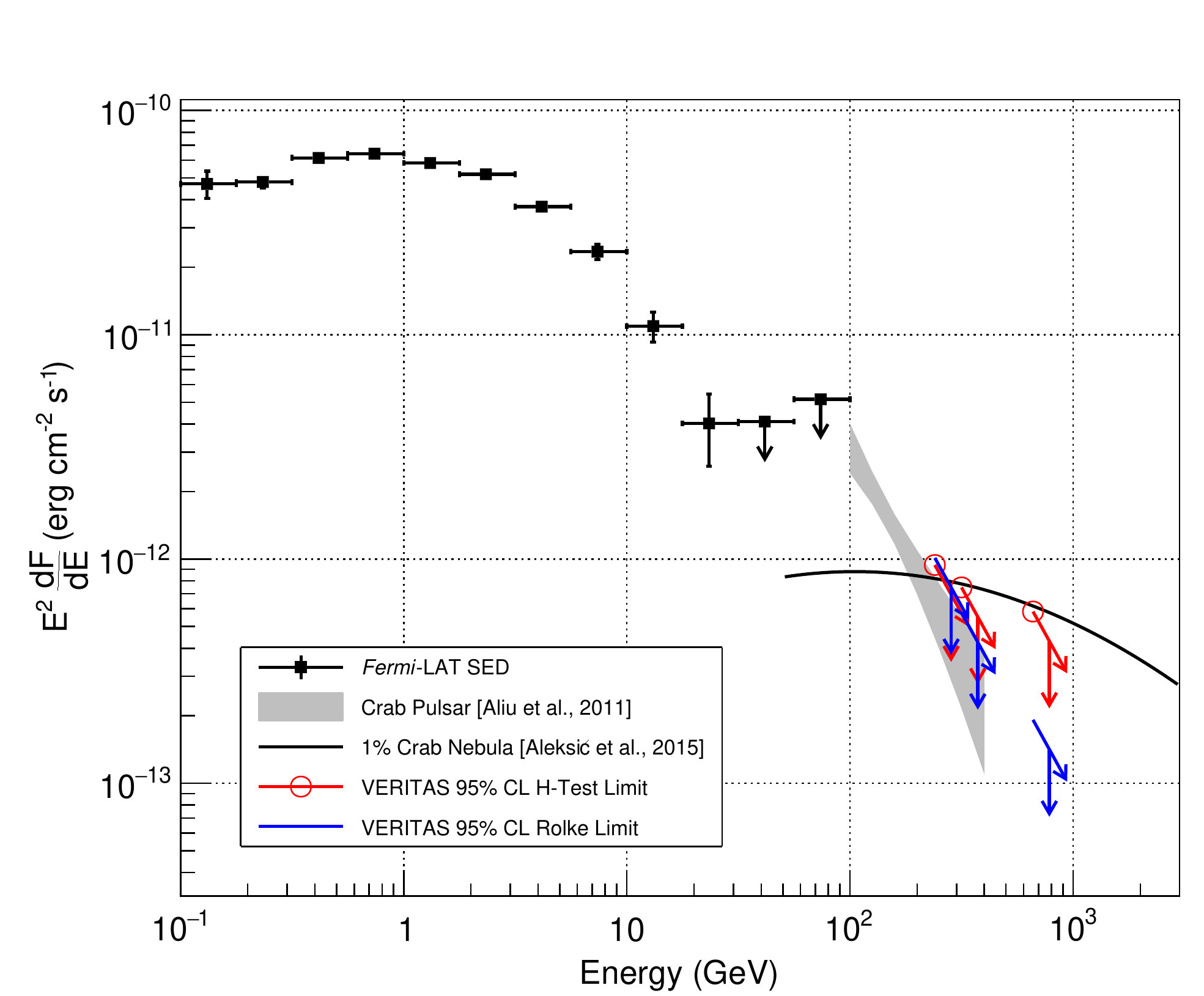}
}

\vspace{10mm}
 \caption{{\it Fermi}-LAT spectra (black squares) with VERITAS differential flux upper limits from the soft, moderate, and hard-cuts analyses (\emph{H}-Test limits: red arrows, Rolke limits: blue arrows) for PSR J2021+3651 (a), PSR J2021+4026 (b), PSR J2032+4127 (c), and PSR J2229+6114 (d).  A power-law fit to the \textit{Fermi} data above 10\,GeV for PSR J2021+4026 is given by the black dashed line.  We note that the highest-energy Rolke flux UL for PSR J2021+4026 appears to constrain a flux level several orders of magnitude below the other ULs; however, this UL corresponds to a large, unphysical negative excess.  See Section~\ref{sec:archival_discussion} for some further discussion.}
 \label{fig:fermiSED_part3}
\end{figure}

The \textit{Fermi}-LAT spectra derived for each pulsar, along with the VERITAS VHE flux upper limits, are presented  in Figures~\ref{fig:PSRJ0007p7303_fermiSED}--\ref{fig:fermiSED_part3}.
The \textit{Fermi}-LAT spectra are all consistent with those reported in the 2PC~\citep{2013ApJS..208...17A}.  For both PSR J0007+7303 and PSR J2021+4026, sufficient spectral points (at least three) are reconstructed to enable reduced $\chi^2$ power-law fits above 10\,GeV of the form given in Equation~\ref{eq:powerlaw}, where $E_0$ is fixed to 20\,GeV, and $\gamma$ and $F_0$ are left free.
The results of the fits are given in Table~\ref{tab:fits}, and the fits are intended to help indicate whether or not a power-law extension of the spectrum from HE to VHE is possible, as has been seen for the Crab pulsar.  The integral flux ULs given in Table~\ref{tab:pulsedfluxtable} have been converted to differential limits assuming a spectral index of $3.8$ and are plotted at the corresponding spectral analysis threshold.  The \textit{Fermi} spectral points are shown in black; the VERITAS 95\% CL flux ULs from the \textit{H} test are indicated by the red arrows; while those from the method of Rolke from the phase-gate test are given by the blue arrows.  The starting point (nock) of the sloped arrow of each pair is set to the energy threshold and flux UL.  These arrows are drawn with a slope to indicate the assumed spectral index of 3.8.  For reference, the Crab pulsar spectral bowtie from~\cite{2011Sci...334...69V} is also shown (gray shaded region), in addition to the Crab Nebula spectral shape from~\cite{2015JHEAp...5...30A} scaled to 1\% (black curved line).  For the two pulsars J0007+7303 and J2021+4026, a reduced-$\chi^2$ power-law fit $>$10\,GeV is given by the black dashed line.

\newcolumntype{C}{>{\centering\arraybackslash}X}
\begin{table*}[h] 
\centering
\caption{Results of the reduced $\chi^2$ power-law fits for the {\it Fermi}-LAT spectra of the two pulsars where a fit $>$10\,GeV was possible.}
\vspace{3mm}
\begin{tabularx}{1.0\textwidth}{CCCCC}
  \hline
   Pulsar & $F_0$ ($\times 10^{-7}$\,$\textrm{GeV}^{-1}$\,$\textrm{cm}^{-2}$\,$\textrm{s}^{-1}$)  & $\gamma$ & $\chi^2$ / n.d.f. & Probability \\ \hline
  
   J0007+7303 &    $2.96 \pm 0.29$  &     $3.98 \pm 0.24$      &    0.47 / 1  &  0.49 \\
   J2021+4026 &    $1.60 \pm 0.24$  &     $3.23 \pm 0.38$      &    0.25 / 1  &  0.62 \\
\hline
\end{tabularx}

\label{tab:fits}
\end{table*}

\pagebreak

\pagebreak

\pagebreak

\pagebreak

\section{Discussion and Conclusion} \label{sec:archival_discussion}

Six searches for pulsed VHE emission from each of 13 pulsars appearing in archival VERITAS data have been performed.  No evidence of pulsed VHE emission is found from any pulsar.  This search for VHE pulsed emission from the present set of 13 archival pulsars is the first ever done in the VHE band and represents the first comprehensive northern-sky survey of its kind.  We note that the upper limits constrain a flux that is in many cases below the level of the Crab pulsar, so the broad statement can be made that potential pulsed VHE emission from the majority of the pulsars must be more faint than the VHE flux from the Crab pulsar ($\sim$1\% Crab Nebula level).   Further, the \textit{Fermi}-LAT spectral reconstruction did not result in sufficient photon statistics to allow any firm statement about the shapes of the spectra above 10\,GeV. 

The upper limits presented here constrain potential spectral hardening or a new spectral component to be at or below the level of the limits.  Although pulsar models generally predict a component of VHE emission that is several orders of magnitude below the flux levels probed in this search (e.g.,~\cite{2015ApJ...811...63H}), another VHE component from the highest energy particles scattering infrared to optical emission may be present at higher energies~\citep{2018ApJ...869L..18H}.  The flux ULs for each pulsar are consistent with synchro-curvature radiation emission scenarios, where the HE gamma-ray spectra are expected to display a power law with an exponential cutoff at a few GeV.  One flux UL  calculated for PSR J2021+4026 (hard cuts; Rolke method) appears to constrain a possible power-law continuation $>$10\,GeV into the VHE band.  We note that this upper limit corresponds to   the largest down-fluctuation ($-1.9\sigma$) in our results, though this low significance is not unexpected given the total number of tests for signal we perform.\footnote{That this UL constrains a much lower flux level than the other limits for PSR J2021+4026 is a result of the steep down-fluctuation into a regime where~\cite{2005NIMPA.551..493R} caution against overinterpretation of obtained ULs.  The method of~\cite{1998PhRvD..57.3873F} experiences similar difficulty here and gives an upper limit of zero on the excess counts.} 
We caution that the possibility of power-law extensions from the HE band $>$10\,GeV to the VHE band is a rather tenuous assumption, given that it is based on what is observed for the only known VHE pulsar, the Crab.  The spectral characteristics of other hitherto undetected VHE pulsars may involve significantly different characteristics $>$10\,GeV, such as the presence of the expected exponential cutoff in the HE band in addition to the emergence of a new component at VHEs.

\begin{figure}[h]
  \centering
  \includegraphics[scale=0.65]{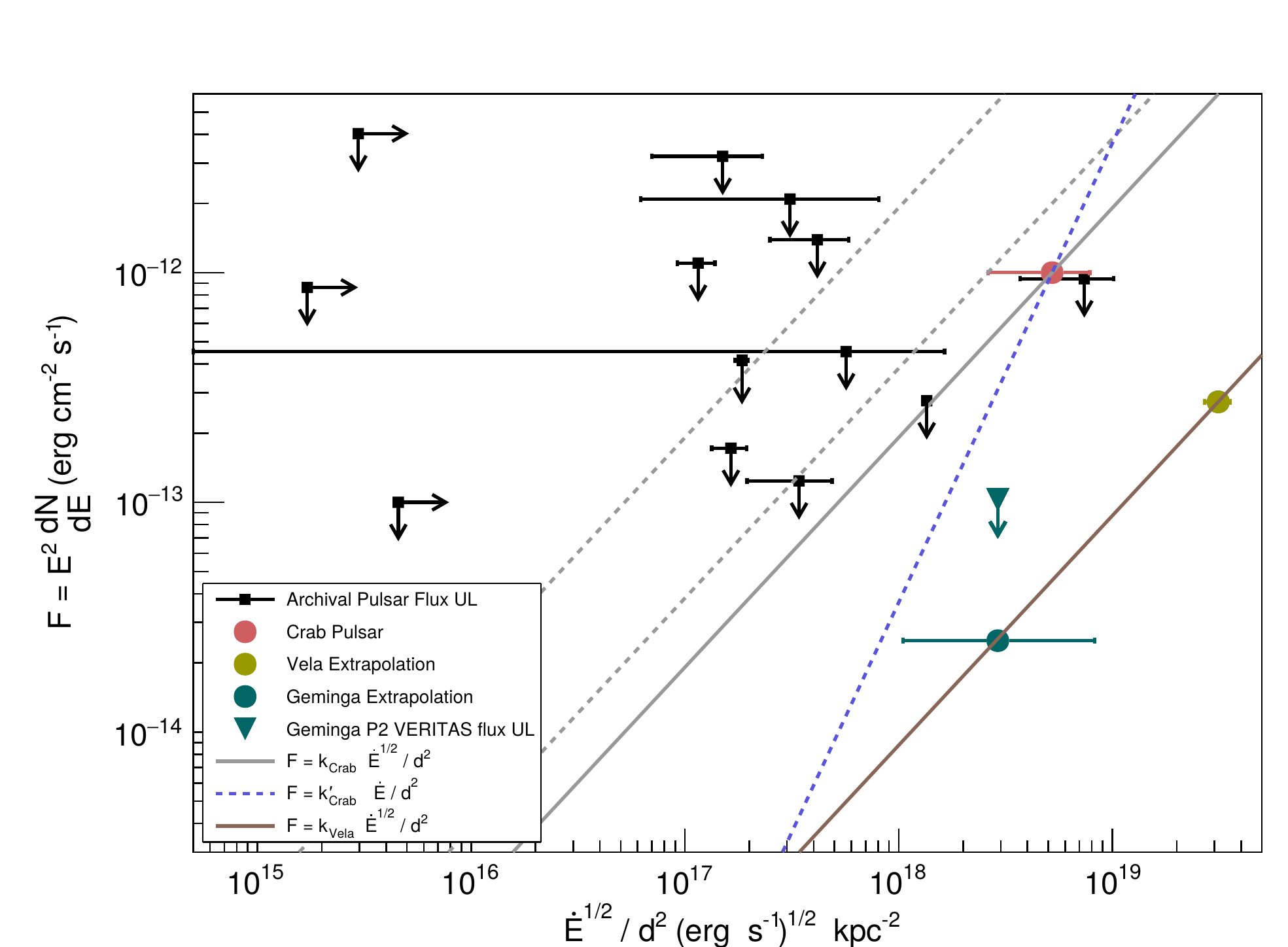}
 \caption{Flux upper limits from the phase-gate test (Rolke method) versus $\sqrt{{\dot E}} / d^2$  for soft cuts.  The VHE flux limits are shown by the black squares, and the right-pointing arrows indicate a lower limit on $\sqrt{{\dot E}} / d^2$ for pulsars where only a distance limit is available.  Error bars come from propagation of the uncertainty on the distance as given in Table~\ref{tab:pulsar_properties}.  The flux and $\sqrt{{\dot E}} / d^2$ for the Crab pulsar are represented by the red dot.  Extrapolated fluxes (see text) for the Geminga and Vela pulsars are shown as a teal and olive circles, respectively.  The teal triangle and arrow is a Geminga VHE flux upper limit from VERITAS at 135\,GeV from~\cite{2015ApJ...800...61A}.    The Crab pulsar flux shown here is calculated according to the method given in the text.  The solid gray line corresponds to $F = k_{\textrm{Crab}} \sqrt{{\dot E}}/d^{2}$ (see text).  The gray dashed lines have the same proportionality but indicate a flux level of two and ten times the Crab pulsar flux.  The blue dashed line corresponds to $F = k^{\prime}_{\textrm{Crab}} {\dot E}/d^2$, which is equivalent to the prediction $L_{\gamma} \propto {\dot E}$ that has been made for the gamma-ray luminosity in some models (e.g.,~\cite{2012ApJ...754...33L}).  The solid brown line corresponds to $F = k_{\textrm{Vela}} \sqrt{{\dot E}}/d^{2}$ (see text).}
  \label{fig:flux_vs_sqrtEdot_soft}
\end{figure}

\begin{figure}[h]
  \centering
  \subfloat{\includegraphics[scale=0.65]{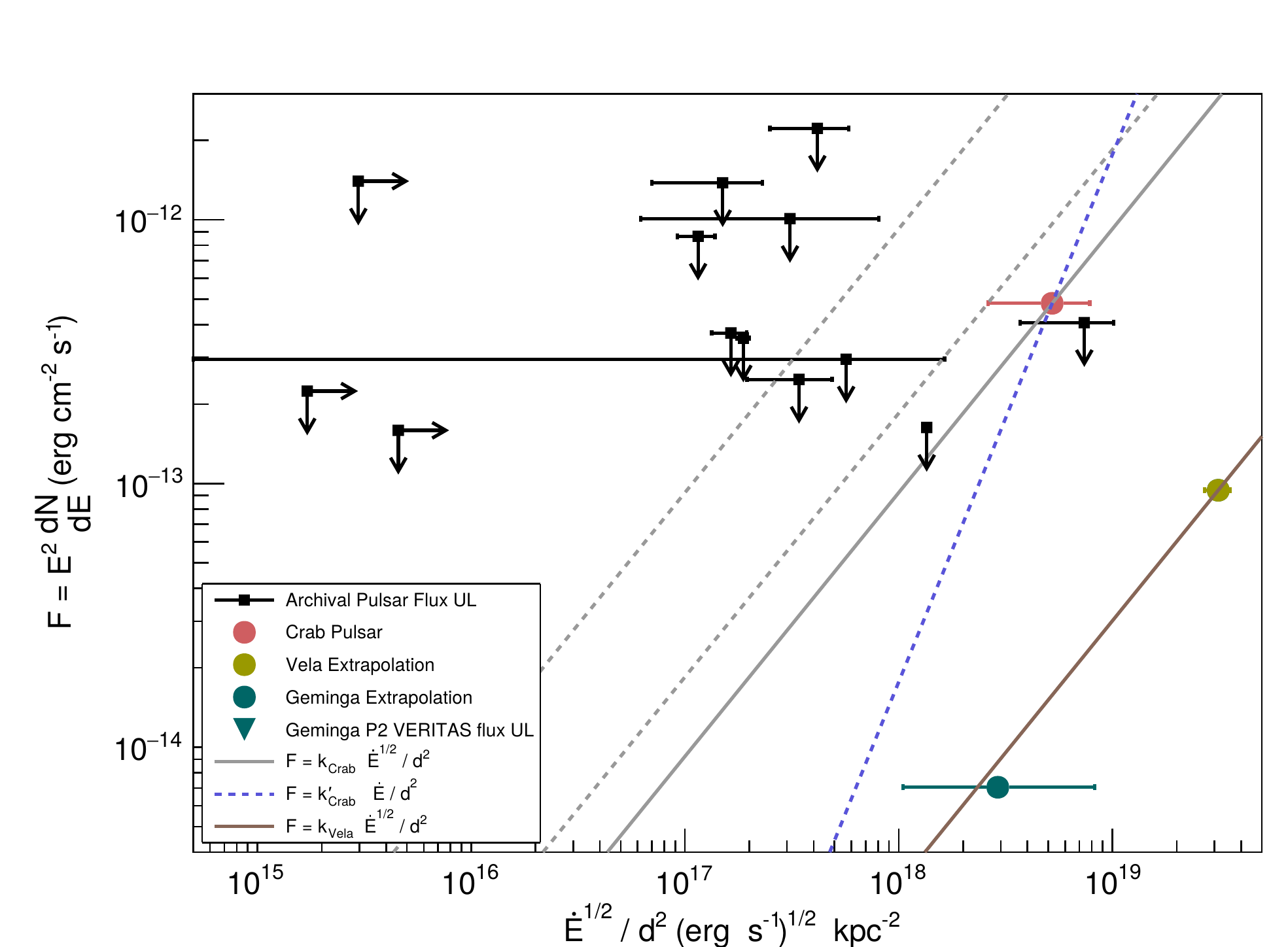}} \\
  \subfloat{\includegraphics[scale=0.65]{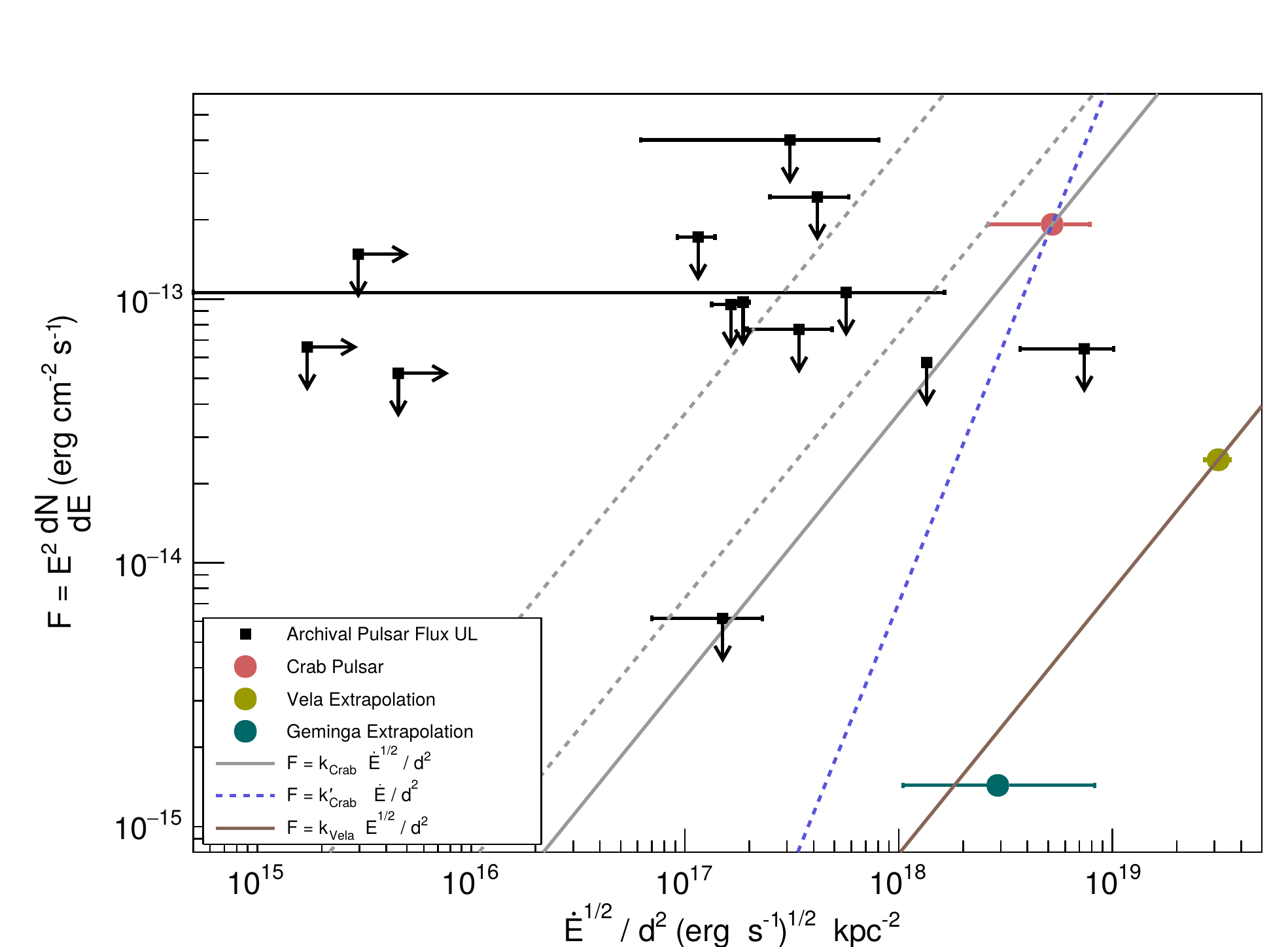}}   \\
 \caption{Same as Figure~\ref{fig:flux_vs_sqrtEdot_soft} for moderate cuts ({\it top}) and hard cuts ({\it bottom}).  See Figure~\ref{fig:flux_vs_sqrtEdot_soft} caption for details.}
 \label{fig:flux_vs_sqrtEdot_medium_hard}
\end{figure}

A population study is conducted using the flux upper limits derived from the VERITAS data.  The VHE flux limits from the phase-gate tests for the three sets of cuts are shown as a function of $\sqrt{{\dot E}} / d^2$ in Figures~\ref{fig:flux_vs_sqrtEdot_soft} and~\ref{fig:flux_vs_sqrtEdot_medium_hard}.  The Crab pulsar flux that is shown for soft cuts is chosen to be $1 \times 10^{-12}$\,erg\,cm$^{-2}$\,s$^{-1}$, which is the approximate flux measured at 200\,GeV by VERITAS~\citep{2011Sci...334...69V}.  For moderate and hard cuts, the Crab flux is extrapolated to 300 and 500\,GeV, respectively, according to a power law with $\Gamma = 3.8$ (the same spectral index for the Crab pulsar measured in~\cite{2011Sci...334...69V}).  These energies are chosen to approximately match the average energy thresholds given in Table~\ref{tab:pulsedfluxtable}.

An assumption that the flux $F$ is proportional to $\sqrt{{\dot E}} / d^2$ (gray and brown lines in Figures~\ref{fig:flux_vs_sqrtEdot_soft} and~\ref{fig:flux_vs_sqrtEdot_medium_hard}) is a restatement of the equivalent assumption $L_{\gamma} \propto \sqrt{\dot E}$, where $L_{\gamma}$ is the gamma-ray luminosity, since $F \propto L_{\gamma} / d^2$.  
The prediction that $L_{\gamma} \propto \sqrt{\dot E}$ is expected in models that assume a linear dependence of $L_{\gamma}$ on the open-field-line voltage in the gap~\citep{1996A&AS..120C..49A}. 
The 2PC \textit{Fermi}-LAT gamma-ray pulsar population plotted as $L_{\gamma}$ against ${\dot E}$ roughly follow a power-law trend~\citep{2013ApJS..208...17A}, though there is likely too much scatter in the data to make a firm empirical claim about the proportionality of $L_{\gamma}$ on $\sqrt{{\dot E}}$ in HE gamma rays. 

Adopting the assumption that young gamma-ray pulsars should all have a flux $F = k \sqrt{\dot E}/d^2,$ where $k$ is a constant of proportionality, we can set $k$ based on what is known for the Crab pulsar:
\begin{equation}
k_{\textrm{Crab}} = F_{\textrm{Crab}} \hspace{1mm} d_{\textrm{Crab}}^2 / \sqrt{\dot E_{\textrm{Crab}}}.
\end{equation}
If true, this assumption  would result in VHE pulsar fluxes trending around the solid gray line corresponding to $F = k_{\textrm{Crab}} \sqrt{\dot E}/d^2$ in Figures~\ref{fig:flux_vs_sqrtEdot_soft} and~\ref{fig:flux_vs_sqrtEdot_medium_hard}.  Almost all of the VERITAS flux ULs lie well above this line, so this prediction for the fluxes of VHE pulsars remains unconstrained in most cases.  However, the three flux limits for one pulsar (PSR J2229+6114) all fall below the gray line.  The error on $\sqrt{\dot E}/d^2$ due to the uncertainty on the distance measurement places the PSR J2229+6114 limits within $1\sigma$ (soft and moderate cuts) or $2\sigma$ (hard cuts) of the gray line, so no firm claim regarding the validity of the trend can be made with the UL for this pulsar.  Furthermore, systematic errors on distance measurements are typically quite large.  VHE flux estimates for the Vela and Geminga pulsars are derived by power-law extrapolation of their {\it Fermi}-LAT spectra above 10\,GeV to 200, 300, and 500\,GeV for soft, moderate, and hard cuts, respectively.  These flux estimates are shown in Figures~\ref{fig:flux_vs_sqrtEdot_soft} and~\ref{fig:flux_vs_sqrtEdot_medium_hard} and lie well below the $F = k \sqrt{\dot E}/d^2$ expectation for $k = k_{\textrm{Crab}}$.  
Measures of ${\dot E}$ and the distances for Geminga and Vela are taken from the 2PC~\citep{2013ApJS..208...17A} (references therein for the distances are~\cite{2012ApJ...755...39V} and~\cite{2003ApJ...596.1137D}, respectively).  That these flux extrapolations lie well below the gray lines in Figures~\ref{fig:flux_vs_sqrtEdot_soft} and~\ref{fig:flux_vs_sqrtEdot_medium_hard} challenges the assumption that the fluxes of VHE pulsars may follow the assumed trend.  
If instead the assumption $k = k_{\textrm{Vela}}$ is made such that $F = k_{\textrm{Vela}} \sqrt{\dot E}/d^2$, the VHE fluxes should trend around the brown lines in Figures~\ref{fig:flux_vs_sqrtEdot_soft} and~\ref{fig:flux_vs_sqrtEdot_medium_hard}.  We note that for all three sets of cuts, the brown line intersects the extrapolated Geminga flux point error.  
It could be the case that other VHE pulsars follow this assumed $\sqrt{\dot E}$ trend, though with a much lower value for $k$ than $k_{\textrm{Crab}}$. 

The upcoming Cherenkov Telescope Array (CTA) will boast a significantly reduced low-energy threshold and significantly improved sensitivity over current-generation instruments.  Observations with CTA will therefore be able to obtain far more constraining flux ULs for the same exposure time~\citep{2017arXiv170907997C}, and the results presented here can help guide future observations with CTA.  The firm detection of more pulsars above 100\,GeV remains an important scientific endeavor, since for now the nature of their VHE emission mostly remains unresolved. \\ \\


This research is supported by grants from the U.S. Department of Energy Office of Science, the U.S. National Science Foundation and the Smithsonian Institution, and by NSERC in Canada. We acknowledge the excellent work of the technical support staff at the Fred Lawrence Whipple Observatory and at the collaborating institutions in the construction and operation of the instrument.  

\software{Tempo2~\citep{2006MNRAS.369..655H}, {\it Fermi}-LAT Science Tools (v10r0p5)}, ROOT (5.34.36)"

\bibliography{./references}

\pagebreak
\section{Appendix}

\subsection{Pulse Profiles from the VERITAS Data}

The pulsar light curves obtained by phase folding the VERITAS data for each of the 13 archival pulsars are shown in this section beginning on the next page.  Each figure shows the P1 and P2 (where applicable) phase gates from Table~\ref{tab:gates} in green, with the background gate shown in gray.  The red dashed line indicates the average background counts, with the two red dotted lines giving the $\pm1\sigma$ deviation.  The inset text box gives $N_{\textrm{on}}$, $\alpha N_{\textrm{off}}$, $N_{\textrm{excess}}$, and the significance from equation 17 in~\cite{1983ApJ...272..317L} in that order from top to bottom.

\begin{figure}[t]
  \centering
  \subfloat{\includegraphics[scale=0.7]{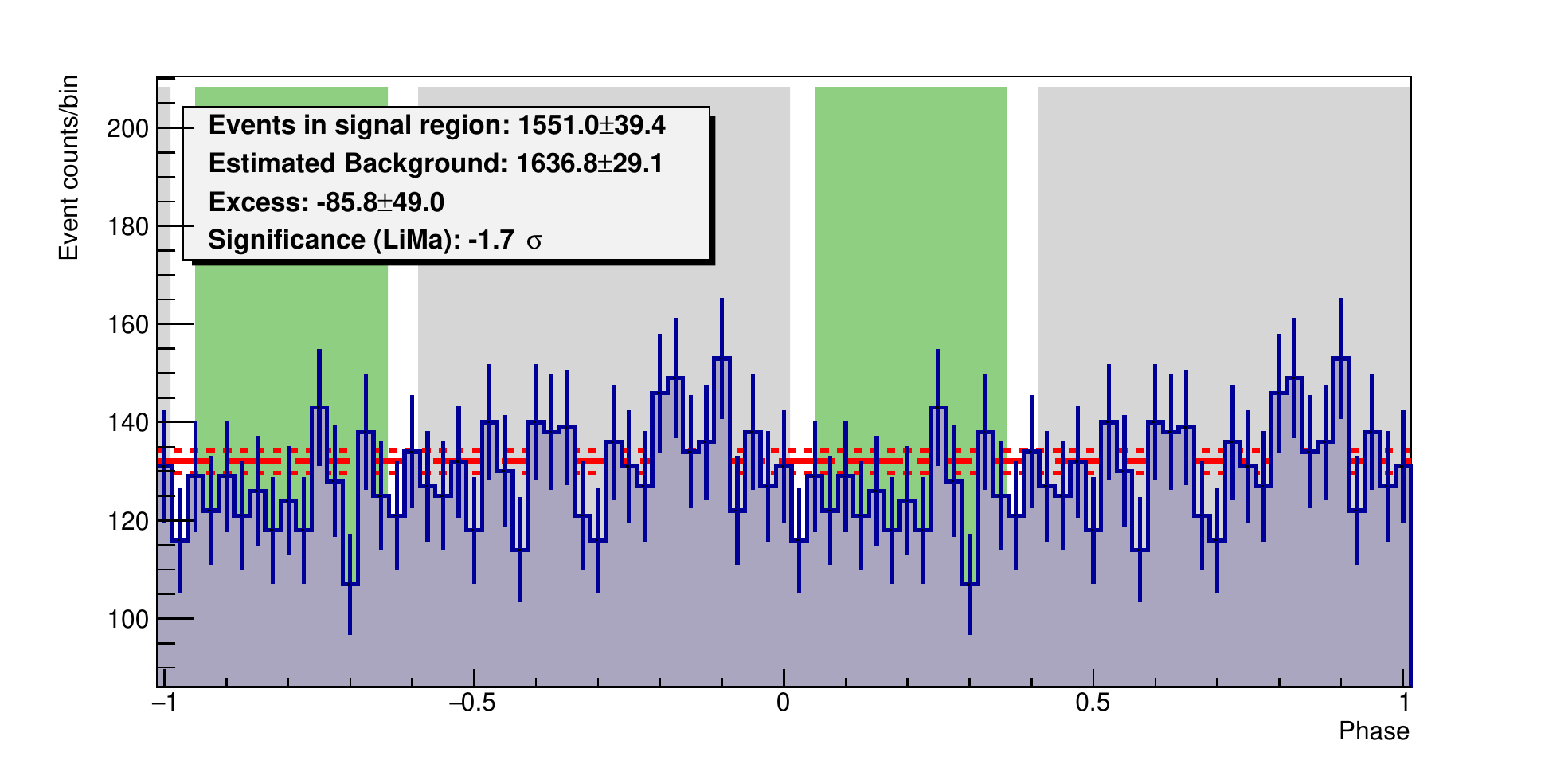}}   \\
  \subfloat{\includegraphics[scale=0.7]{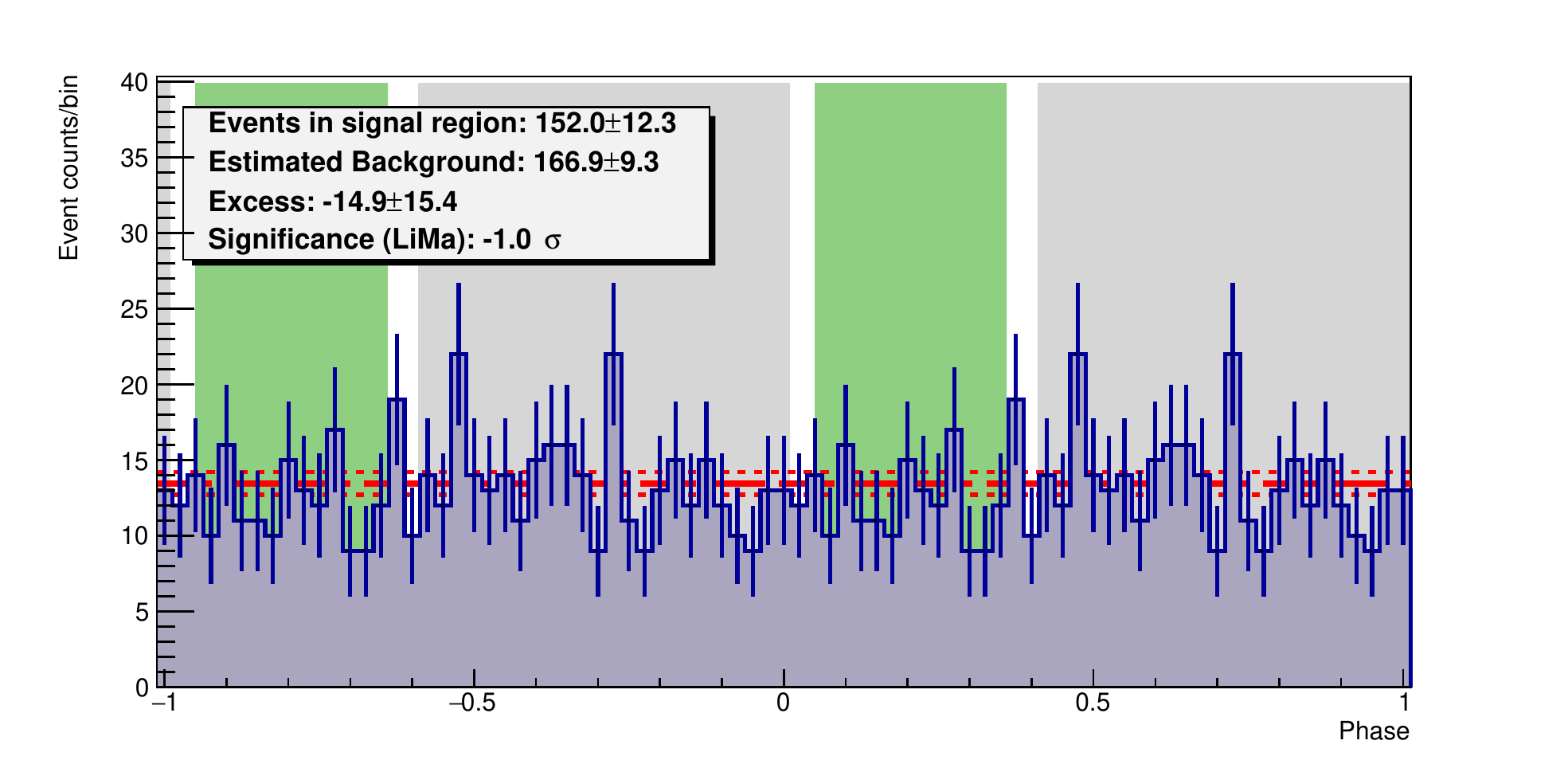}} \\
  \subfloat{\includegraphics[scale=0.7]{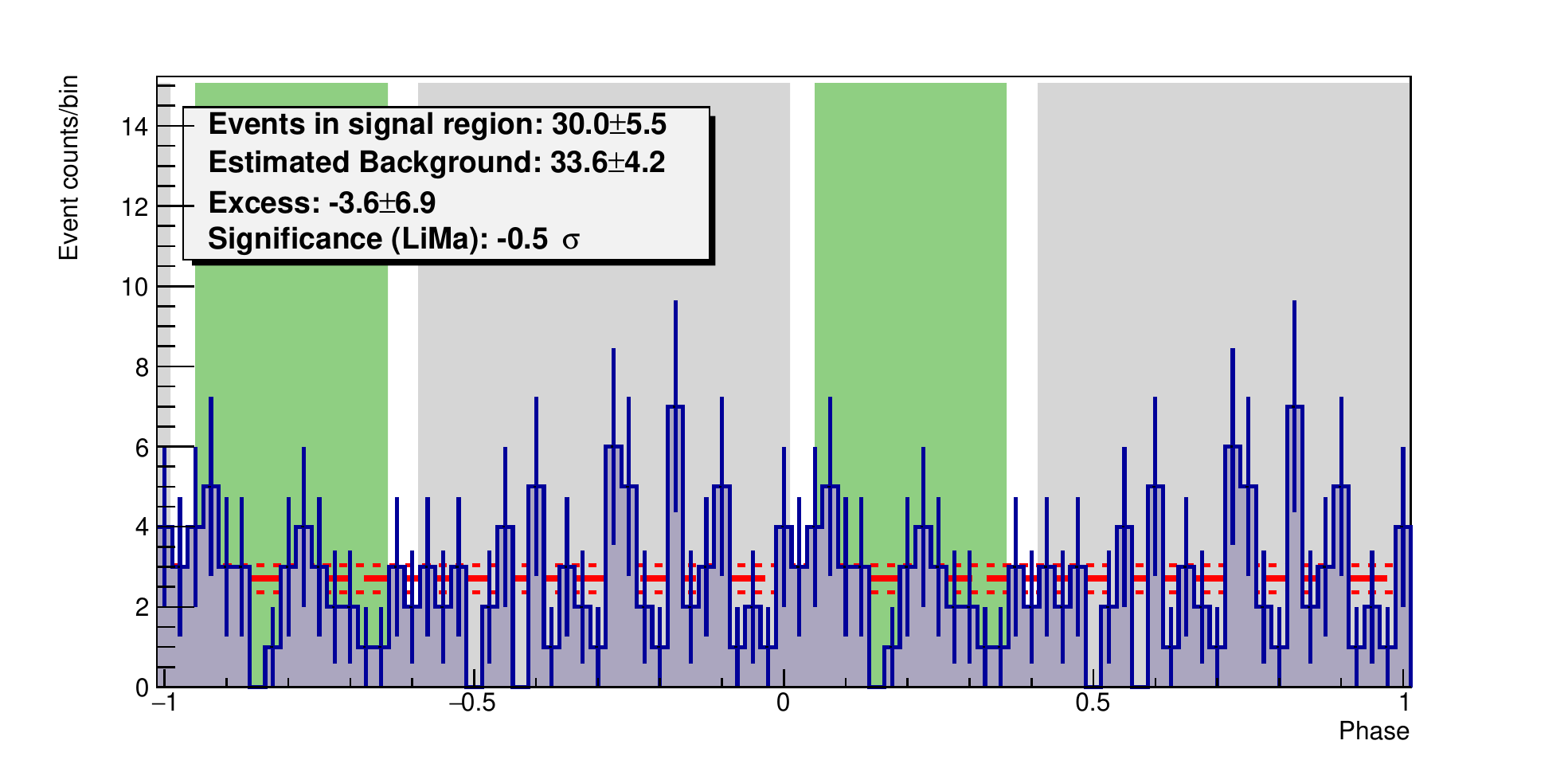}}   \\
 \caption{Pulse profiles of \textbf{PSR J0007+7303} from VERITAS data for soft cuts (top panel), moderate cuts (middle panel), and hard cuts (bottom panel).}
 \label{fig:PSRJ0007p7303_lcs}
\end{figure}

\pagebreak

\begin{figure}[t]
  \centering
  \subfloat{\includegraphics[scale=0.7]{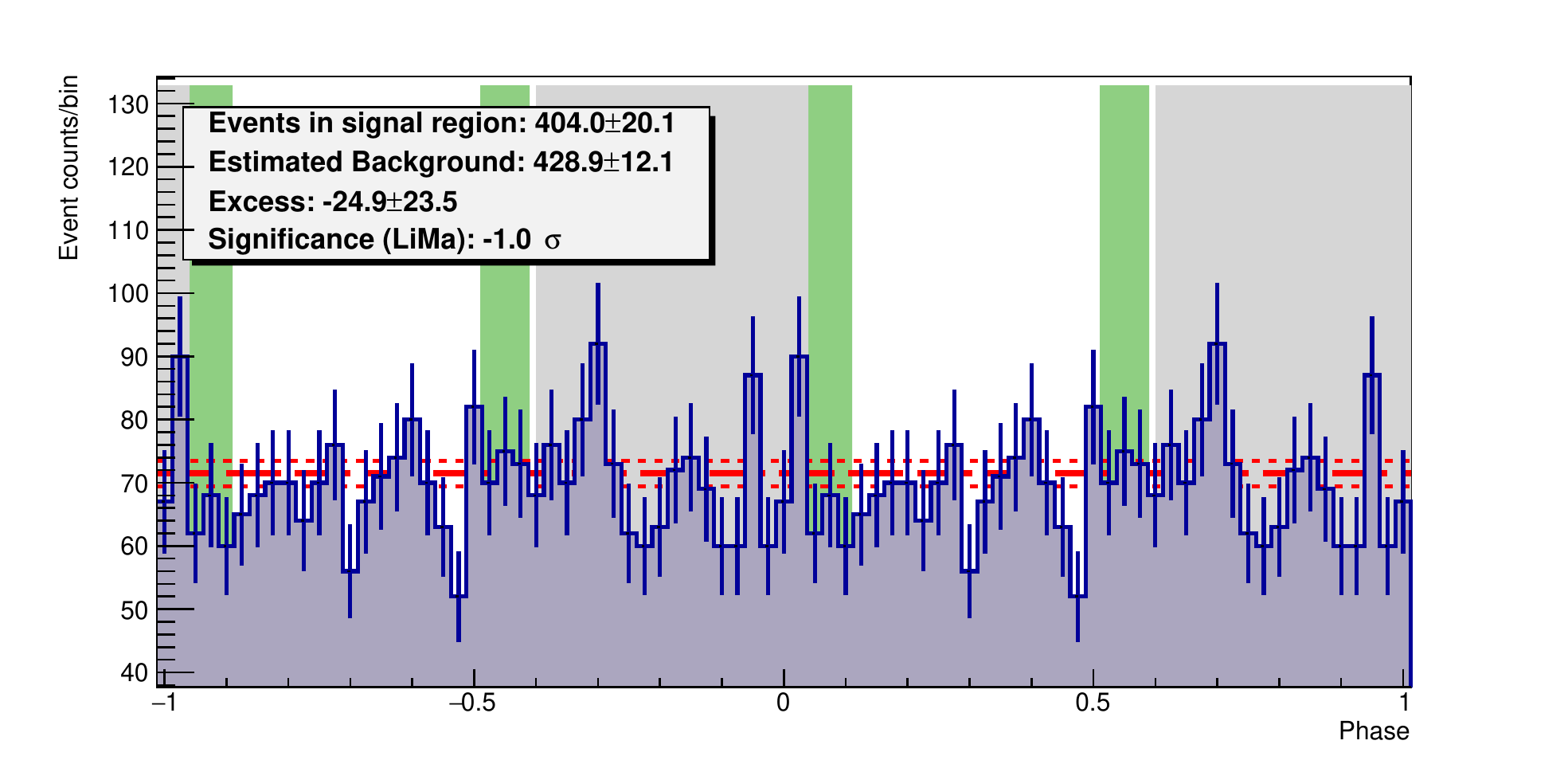}}   \\
  \subfloat{\includegraphics[scale=0.7]{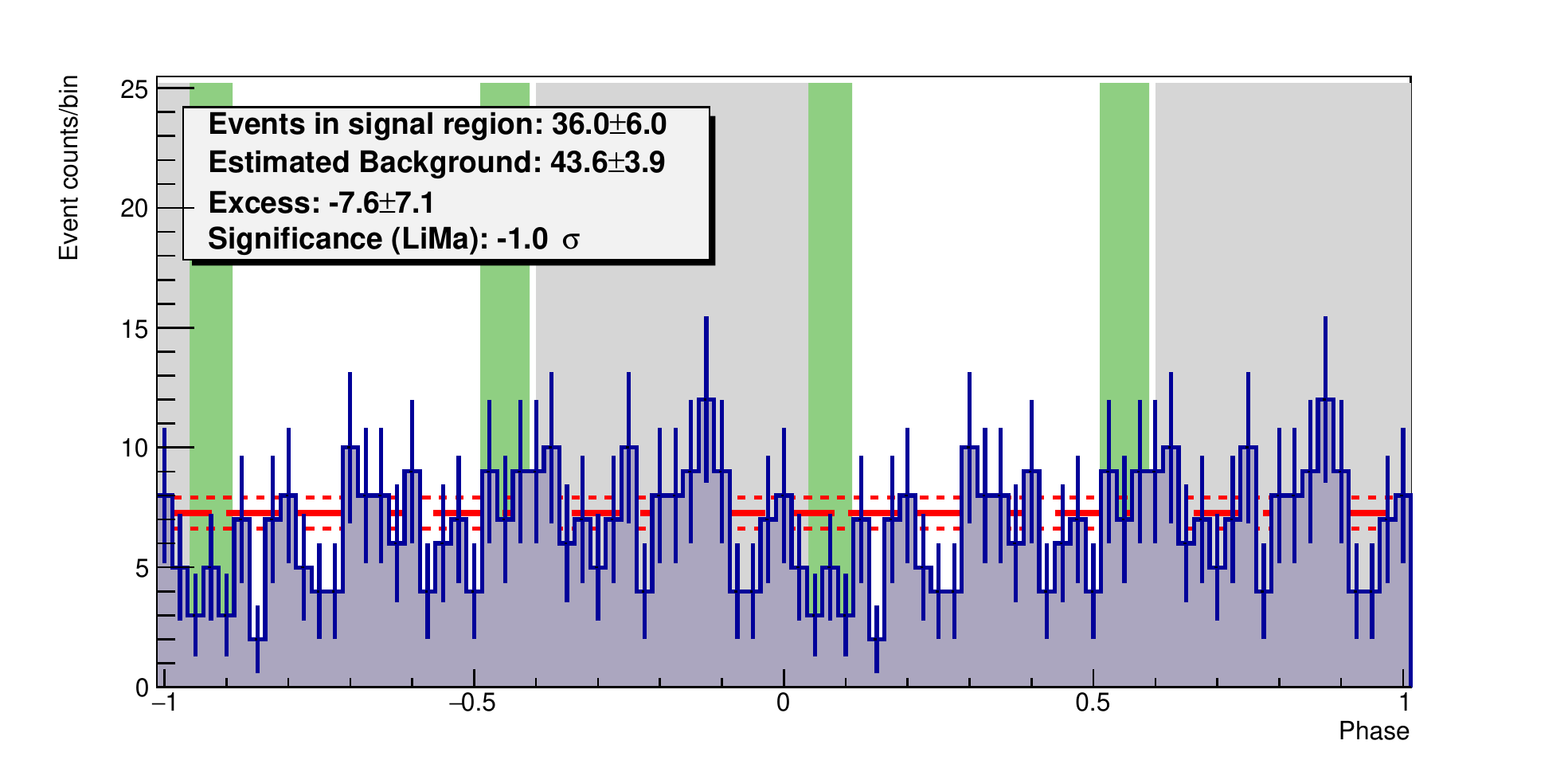}} \\
  \subfloat{\includegraphics[scale=0.7]{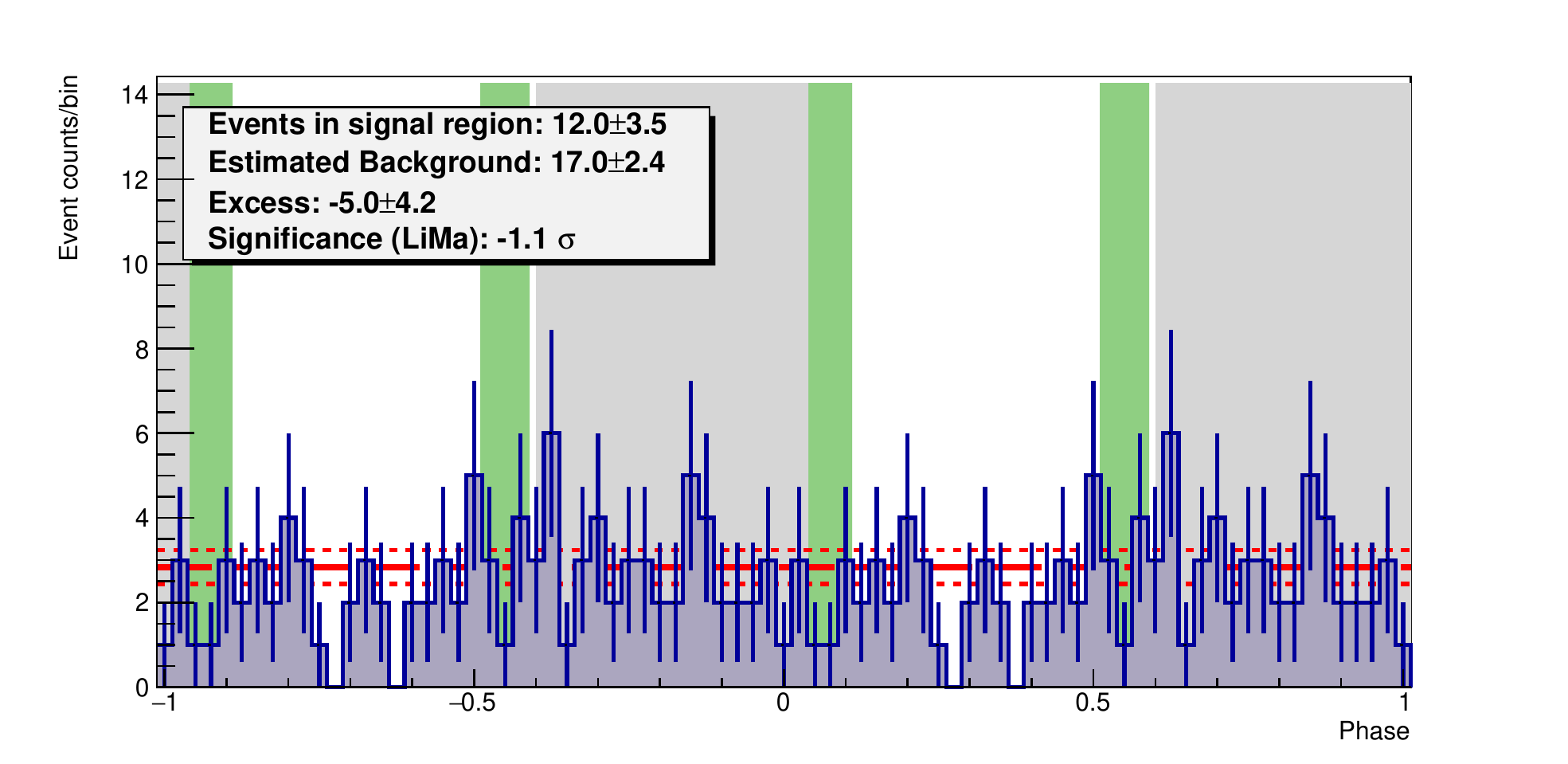}}   \\
 \caption{Pulse profiles of \textbf{PSR J0205+6449} from VERITAS data for soft cuts (top panel), moderate cuts (middle panel), and hard cuts (bottom panel).}
 \label{fig:PSRJ0205p6449_lcs}
\end{figure}

\pagebreak

\begin{figure}[t]
  \centering
  \subfloat{\includegraphics[scale=0.7]{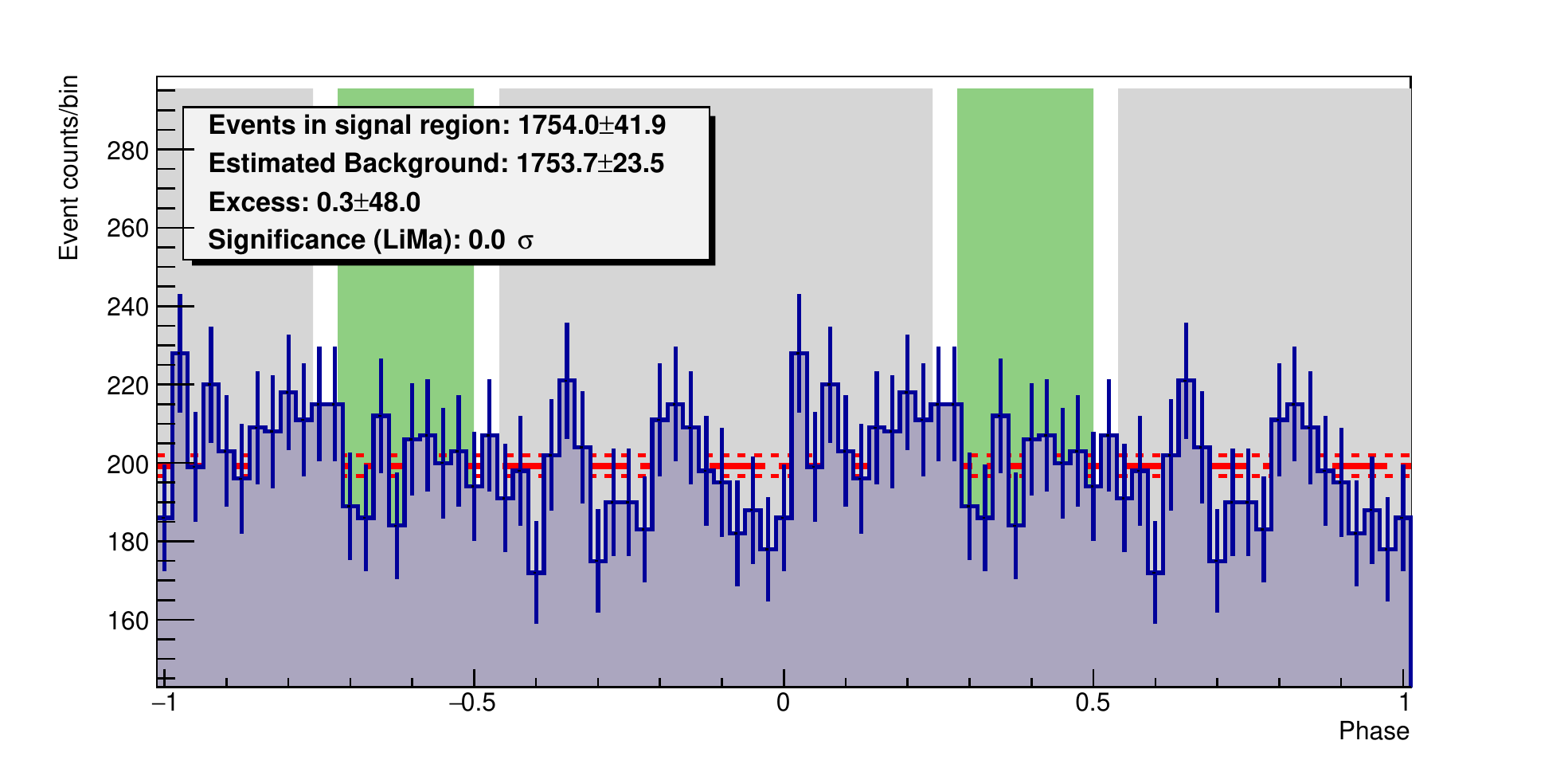}}   \\
  \subfloat{\includegraphics[scale=0.7]{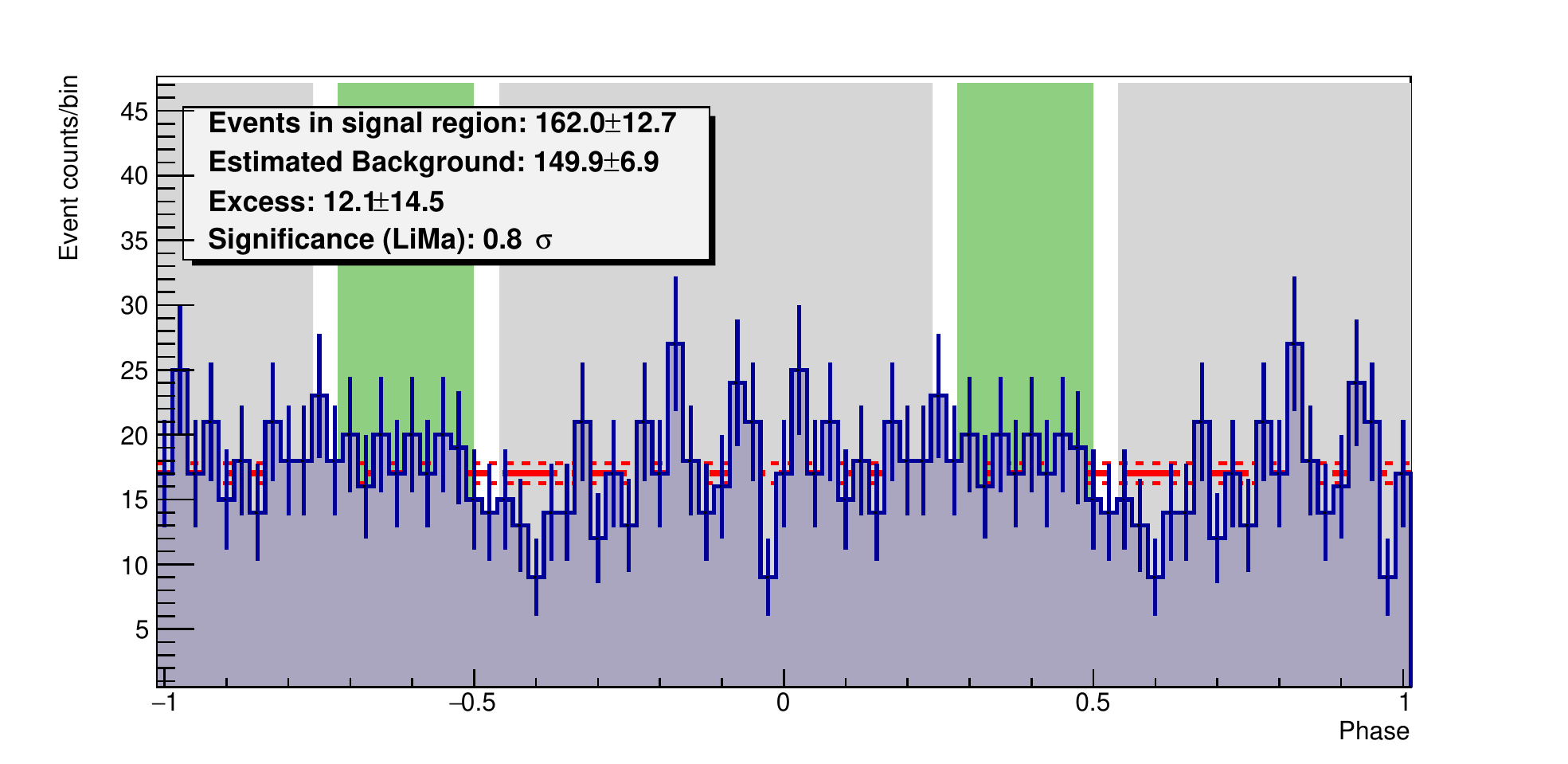}} \\
  \subfloat{\includegraphics[scale=0.7]{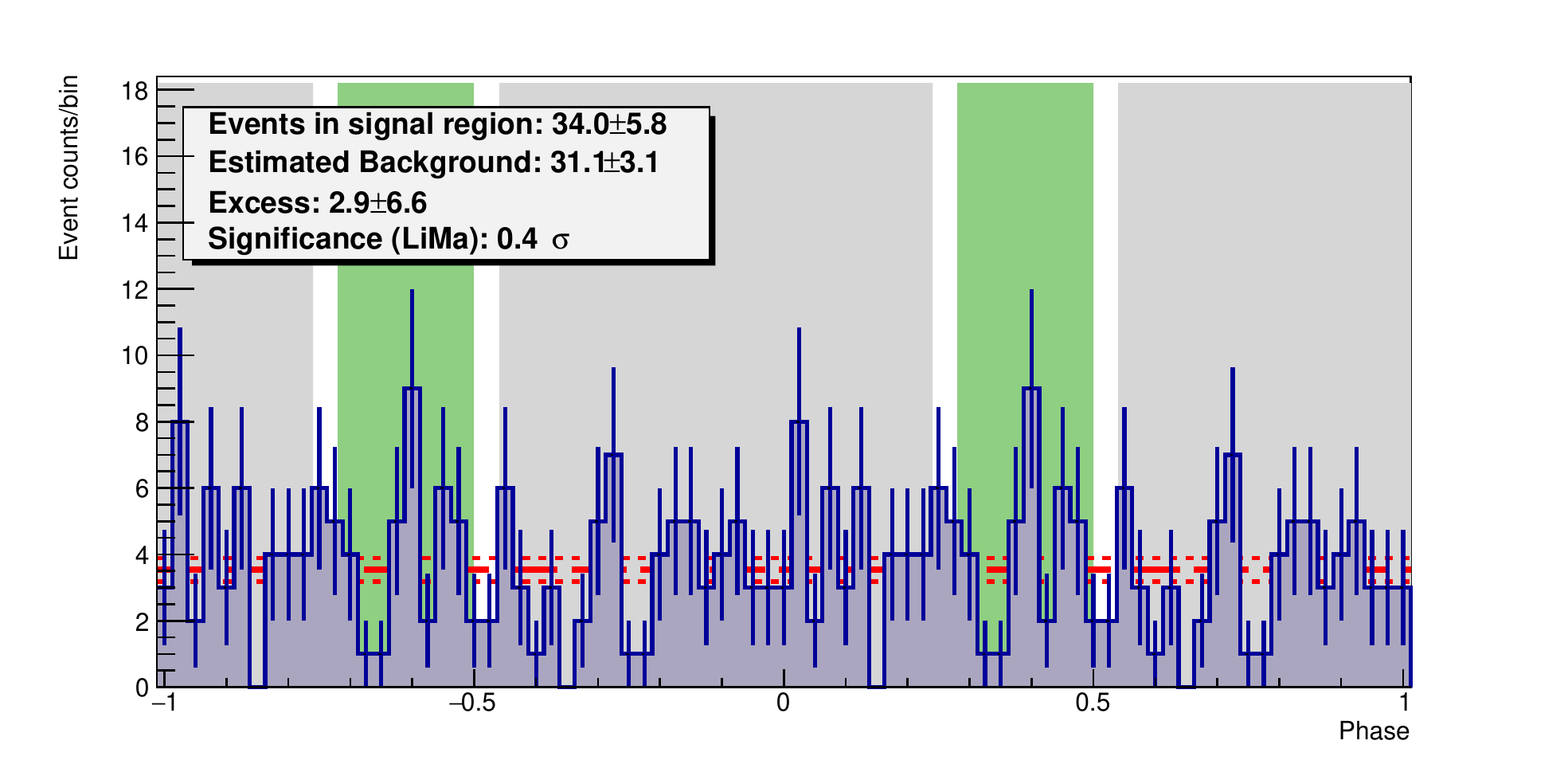}}   \\
 \caption{Pulse profiles of \textbf{PSR J0248+6021} from VERITAS data for soft cuts (top panel), moderate cuts (middle panel), and hard cuts (bottom panel).}
 \label{fig:PSRJ0248p6021_lcs}
\end{figure}

\pagebreak

\begin{figure}[t]
  \centering
  \subfloat{\includegraphics[scale=0.7]{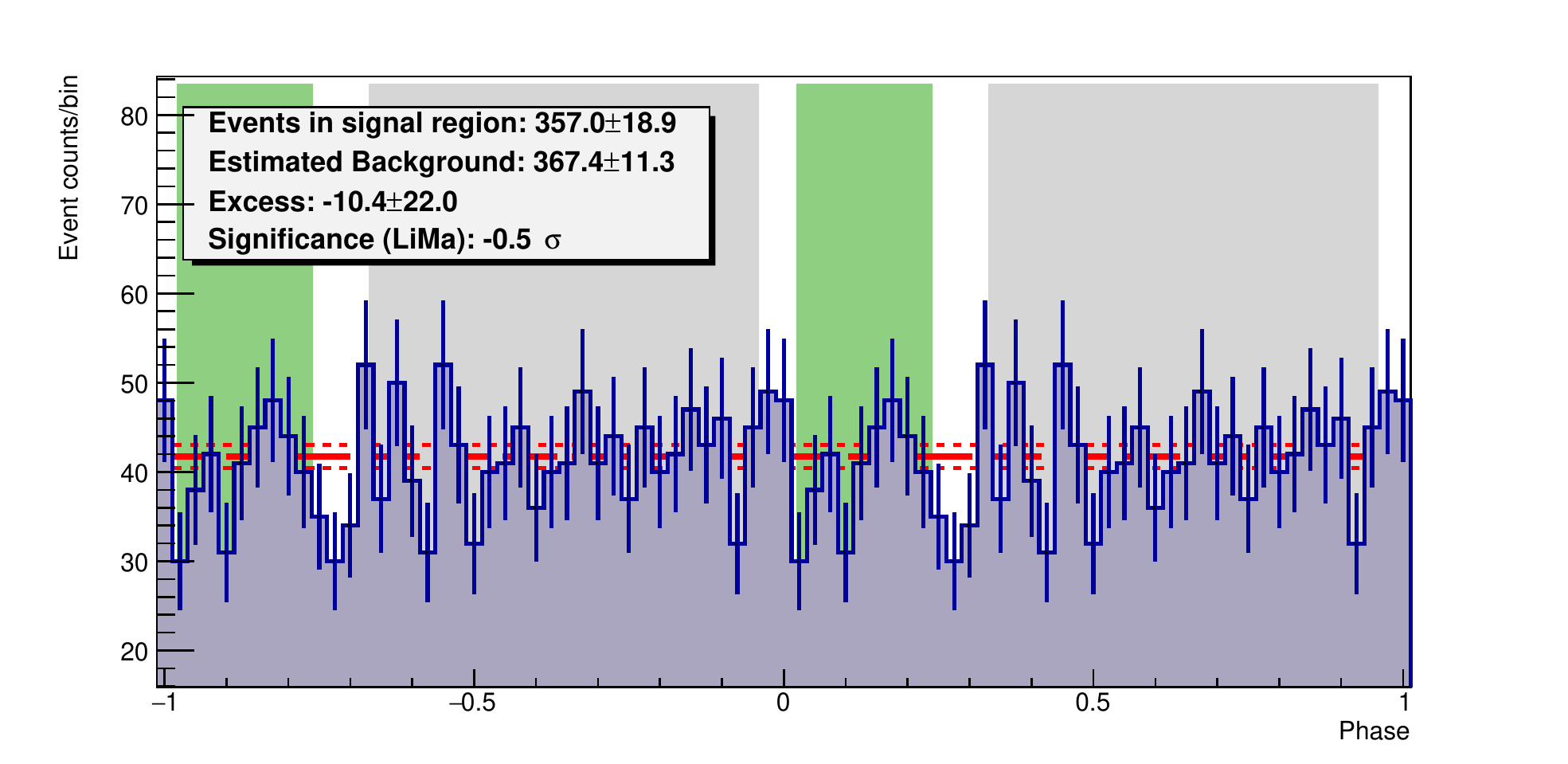}}   \\
  \subfloat{\includegraphics[scale=0.7]{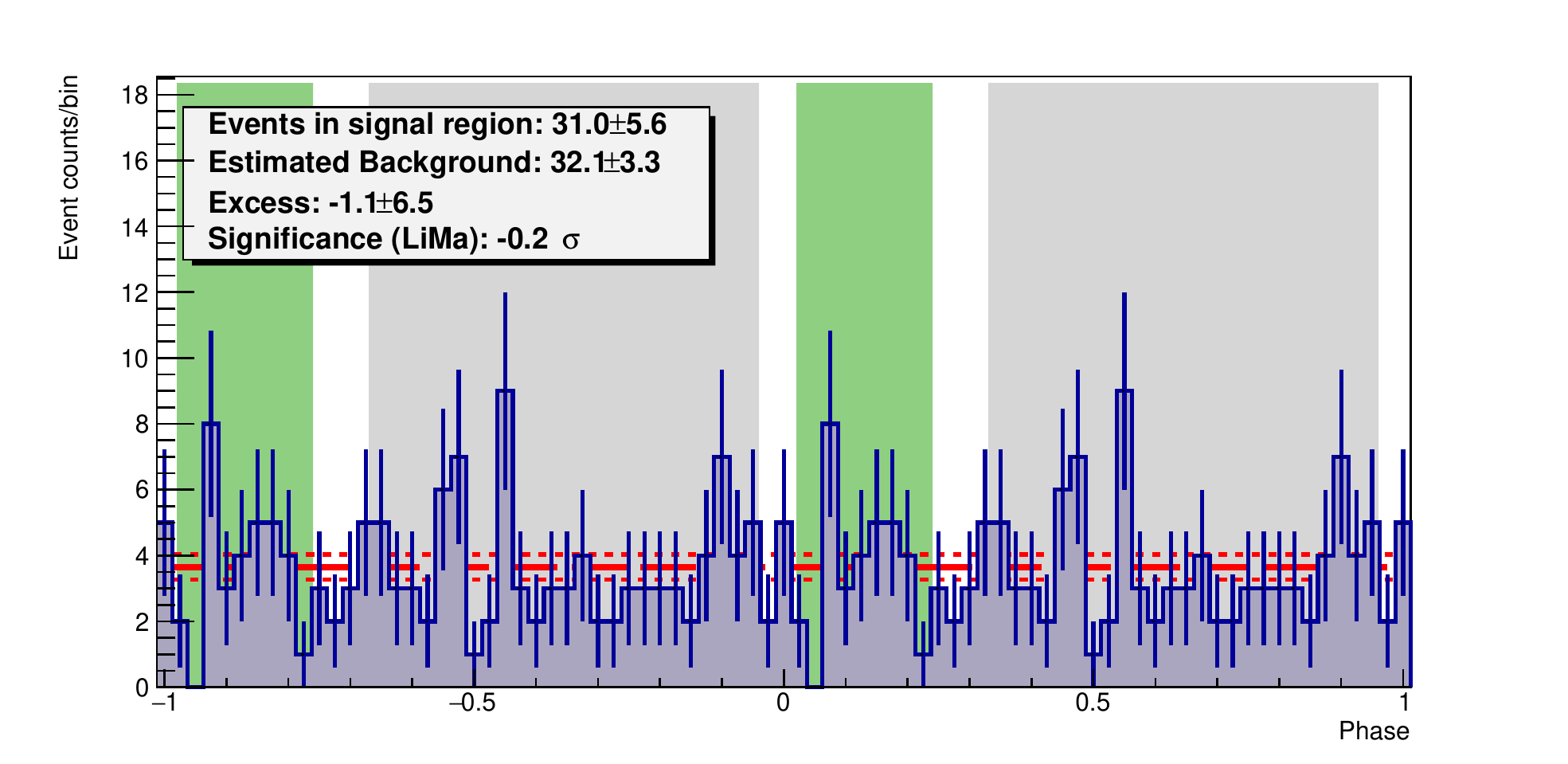}} \\
  \subfloat{\includegraphics[scale=0.7]{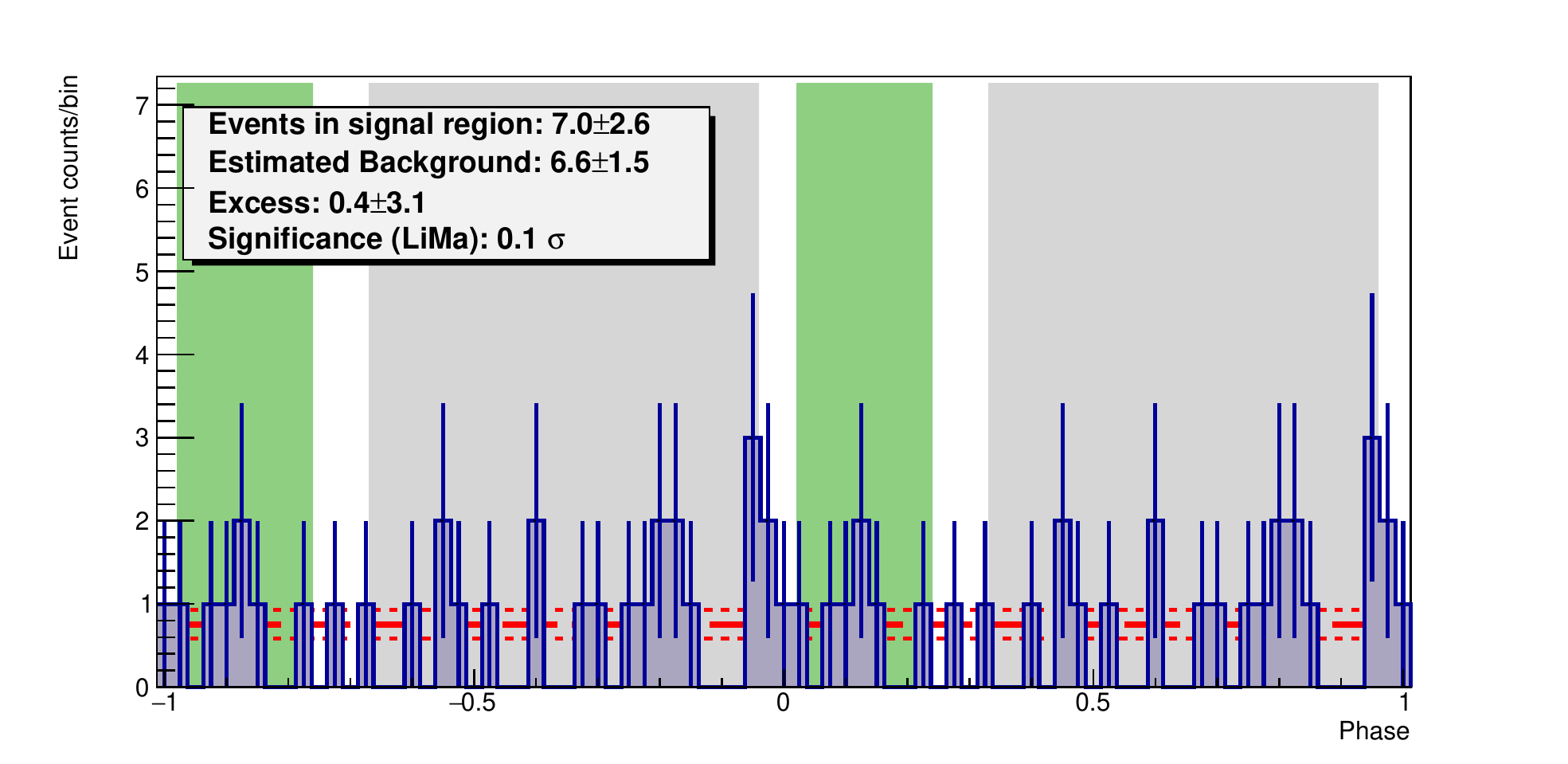}}   \\
 \caption{Pulse profiles of \textbf{PSR J0357+3205} from VERITAS data for soft cuts (top panel), moderate cuts (middle panel), and hard cuts (bottom panel).}
 \label{fig:PSRJ0357p3205_lcs}
\end{figure}

\pagebreak

\begin{figure}[t]
  \centering
  \subfloat{\includegraphics[scale=0.7]{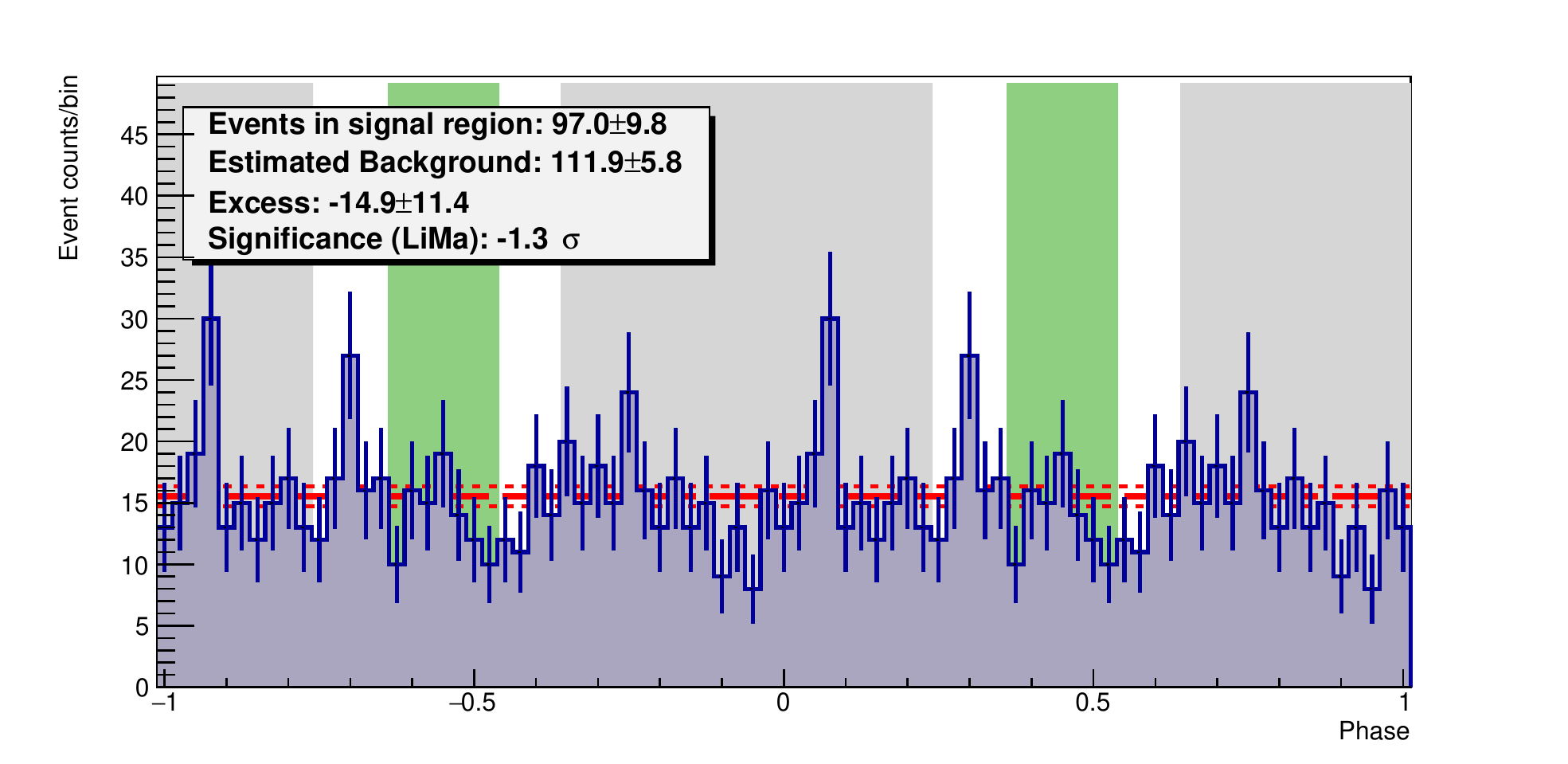}}   \\
  \subfloat{\includegraphics[scale=0.7]{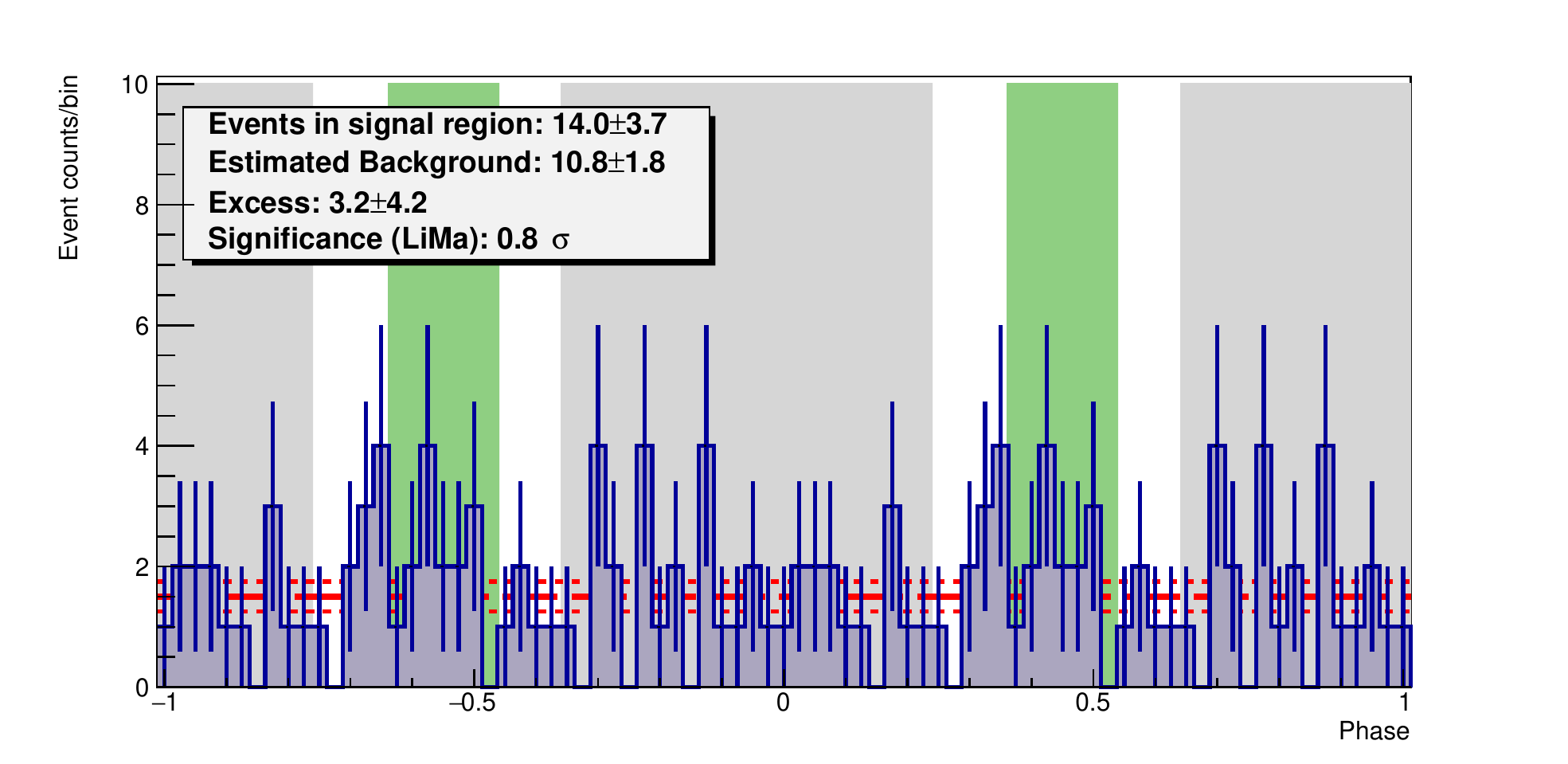}} \\
  \subfloat{\includegraphics[scale=0.7]{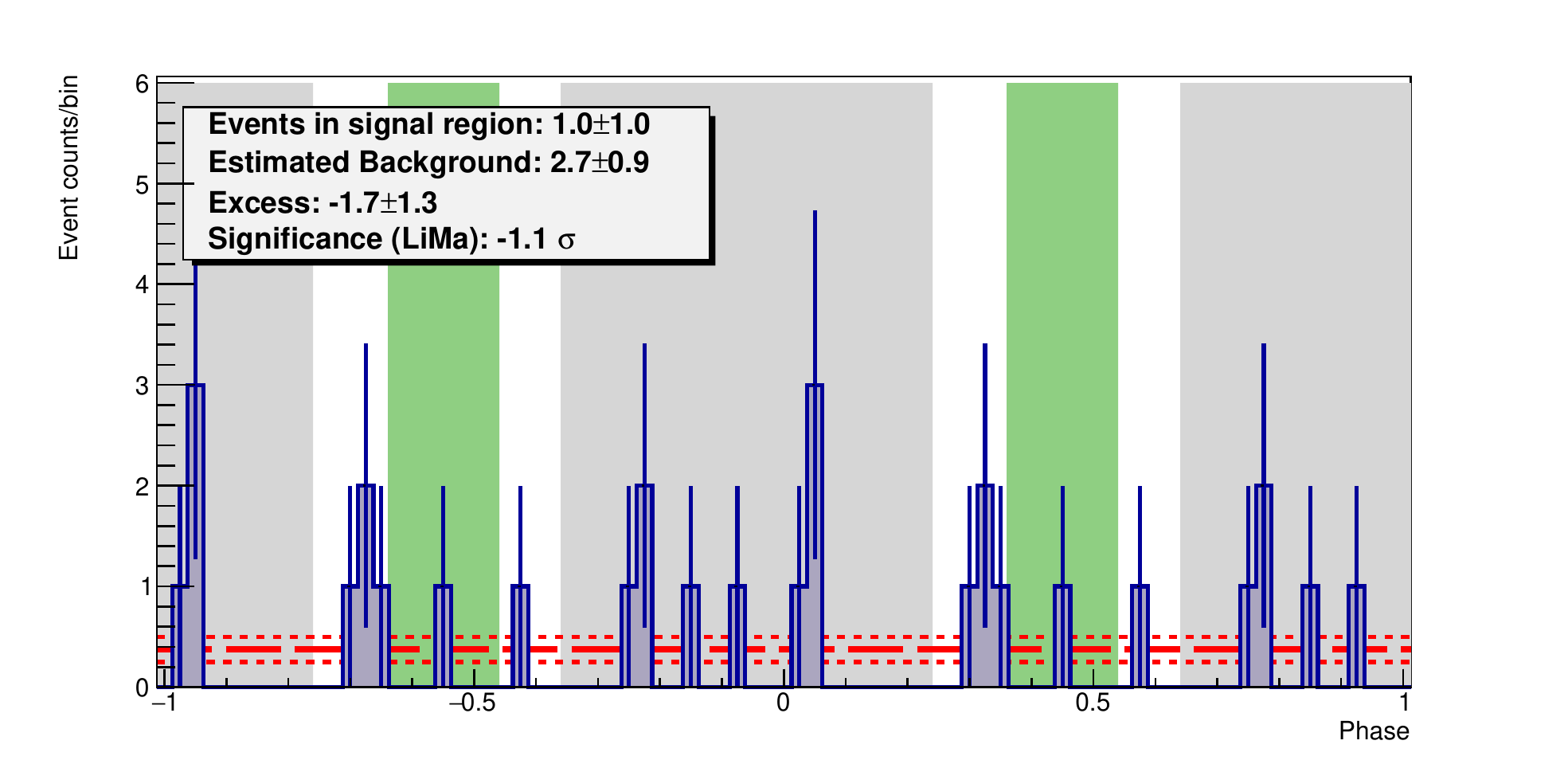}}   \\
 \caption{Pulse profiles of \textbf{PSR J0631+1036} from VERITAS data for soft cuts (top panel), moderate cuts (middle panel), and hard cuts (bottom panel).}
 \label{fig:PSRJ0631p1036_lcs}
\end{figure}

\pagebreak

\begin{figure}[t]
  \centering
  \subfloat{\includegraphics[scale=0.7]{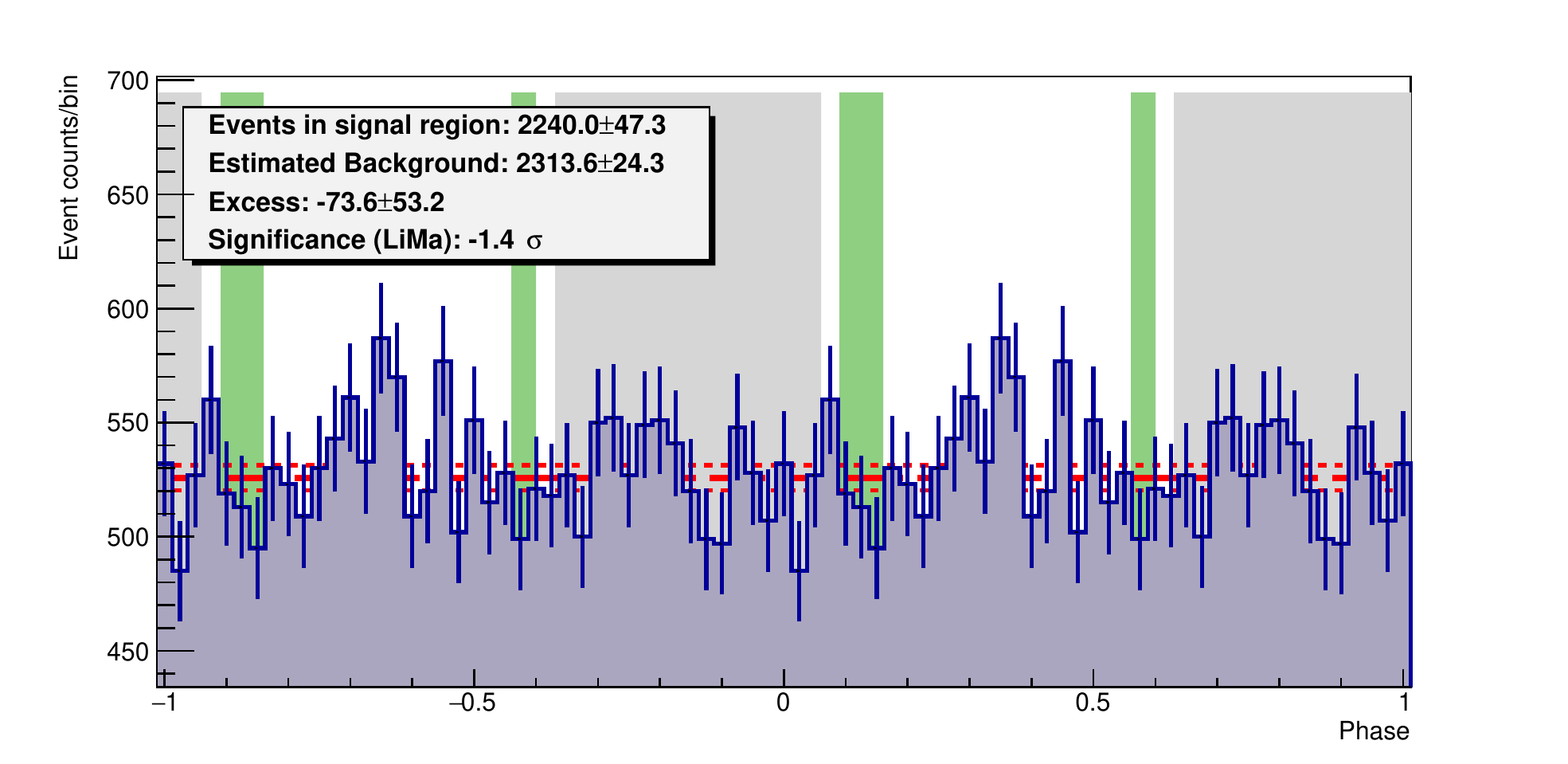}}   \\
  \subfloat{\includegraphics[scale=0.7]{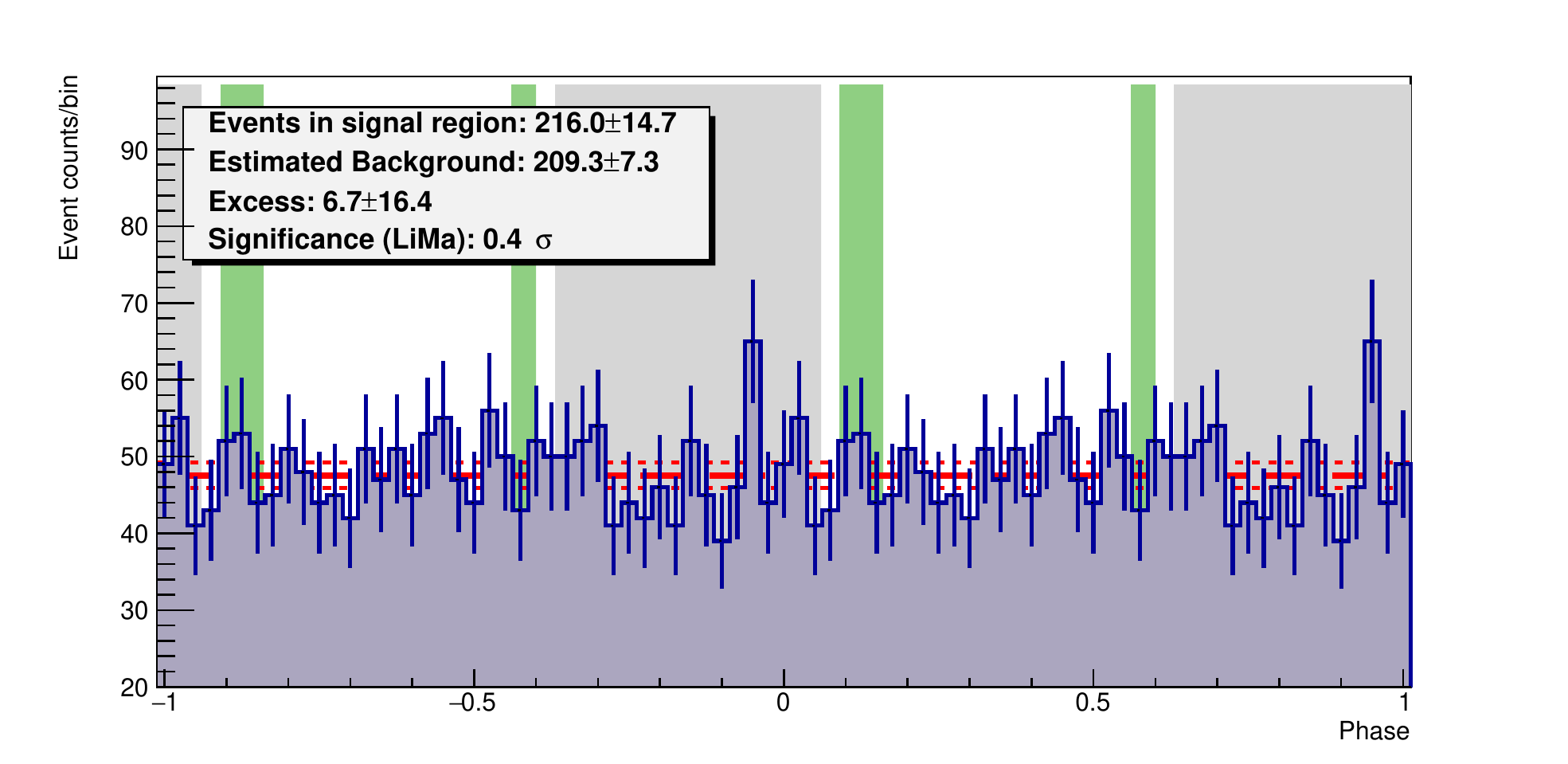}} \\
  \subfloat{\includegraphics[scale=0.7]{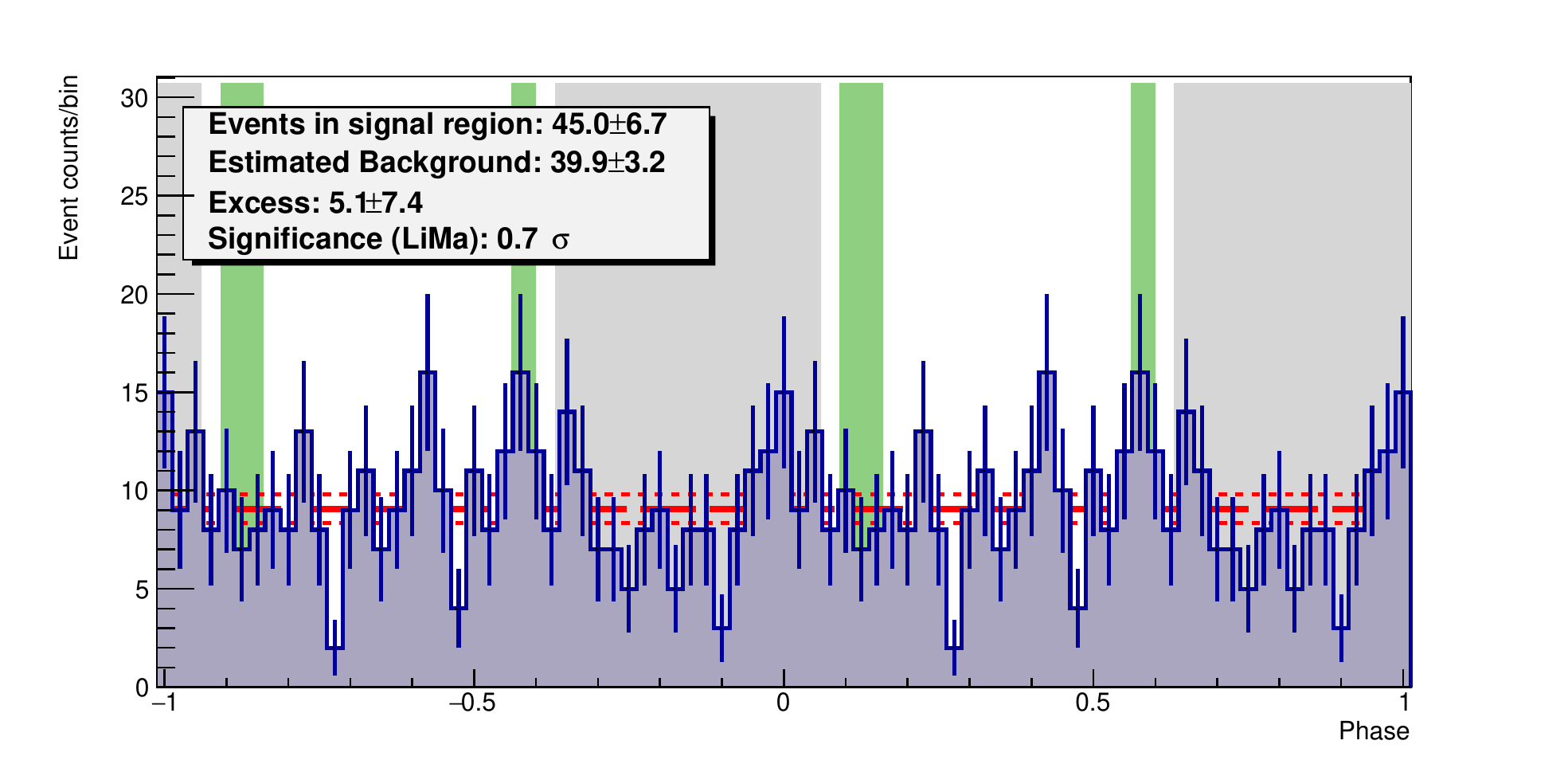}}   \\
 \caption{Pulse profiles of \textbf{PSR J0633+0632} from VERITAS data for soft cuts (top panel), moderate cuts (middle panel), and hard cuts (bottom panel).}
 \label{fig:PSRJ0633p0632_lcs}
\end{figure}

\pagebreak

\begin{figure}[t]
  \centering
  \subfloat{\includegraphics[scale=0.7]{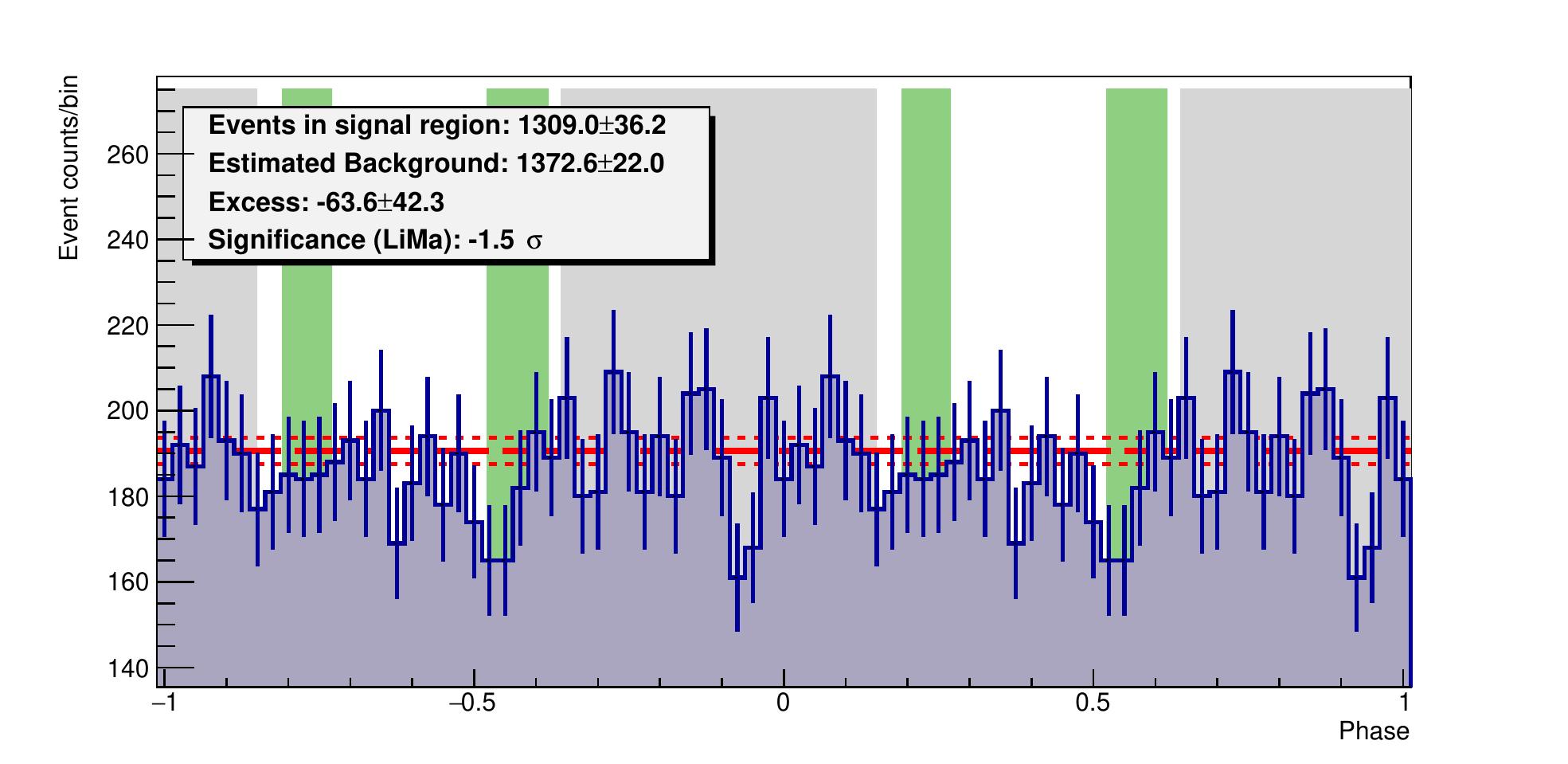}}   \\
  \subfloat{\includegraphics[scale=0.7]{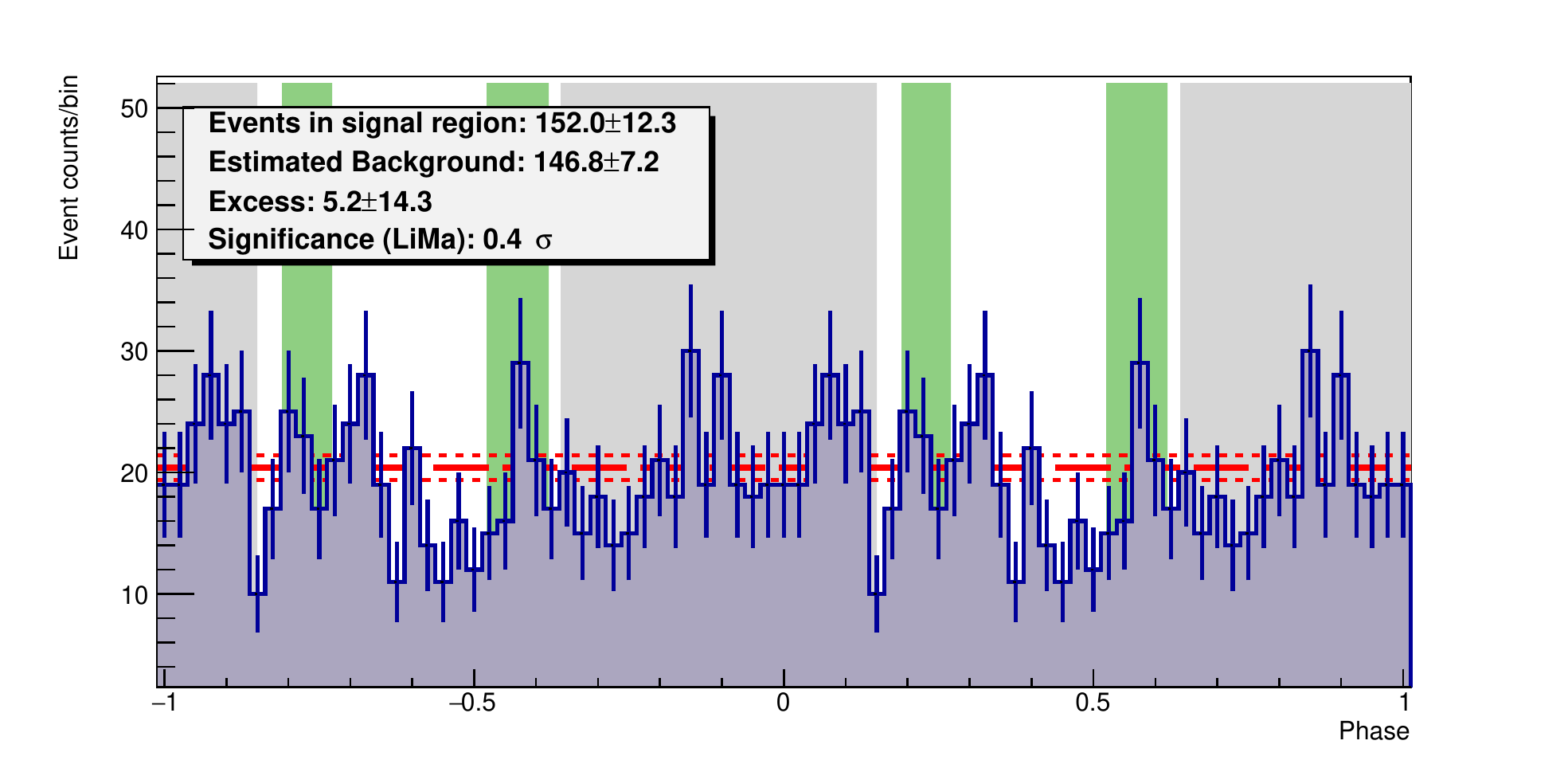}} \\
  \subfloat{\includegraphics[scale=0.7]{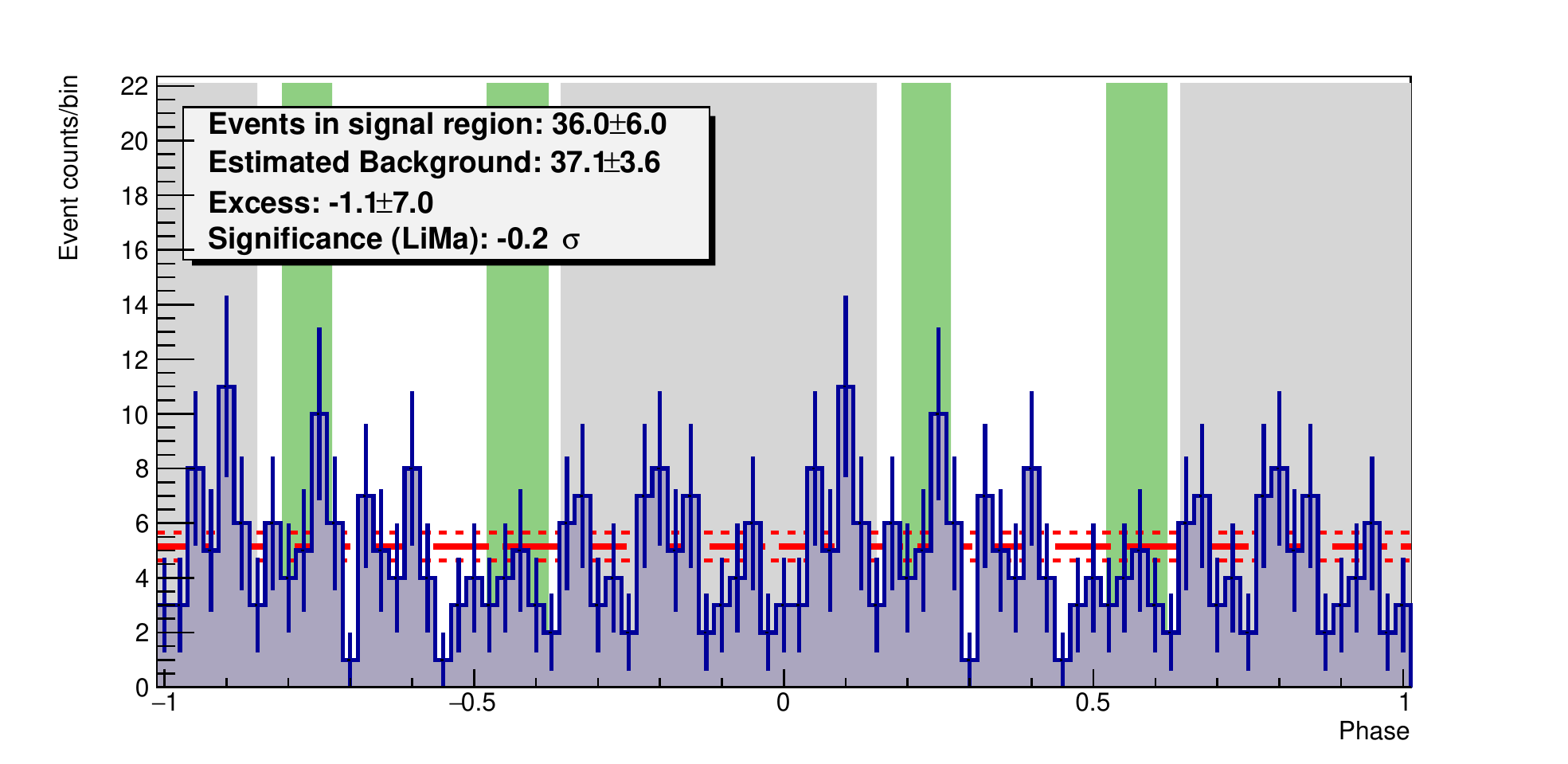}}   \\
 \caption{Pulse profiles of \textbf{PSR J1907+0602} from VERITAS data for soft cuts (top panel), moderate cuts (middle panel), and hard cuts (bottom panel).}
 \label{fig:PSRJ1907p0602_lcs}
\end{figure}

\pagebreak

\begin{figure}[t]
  \centering
  \subfloat{\includegraphics[scale=0.7]{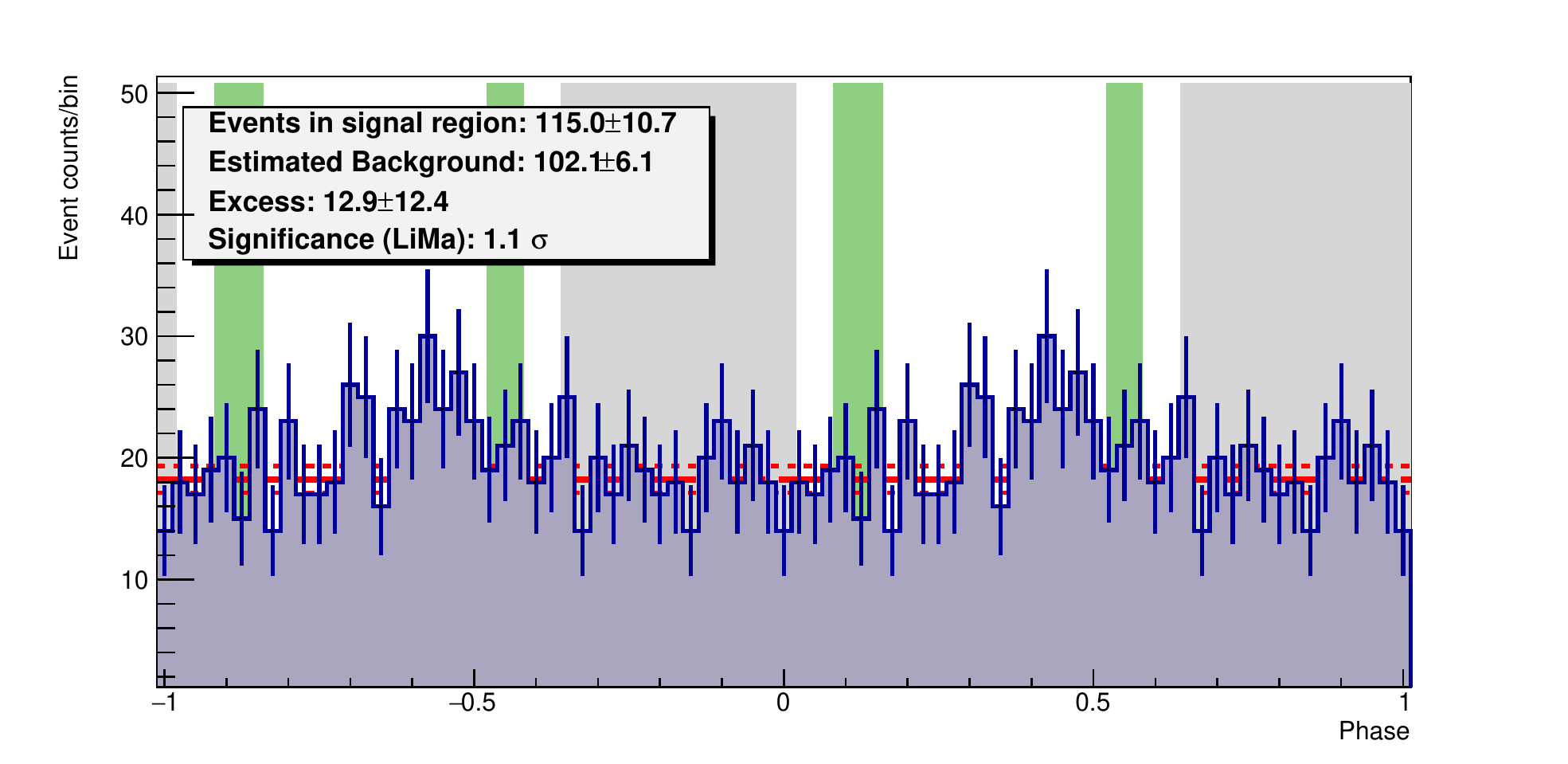}}   \\
  \subfloat{\includegraphics[scale=0.7]{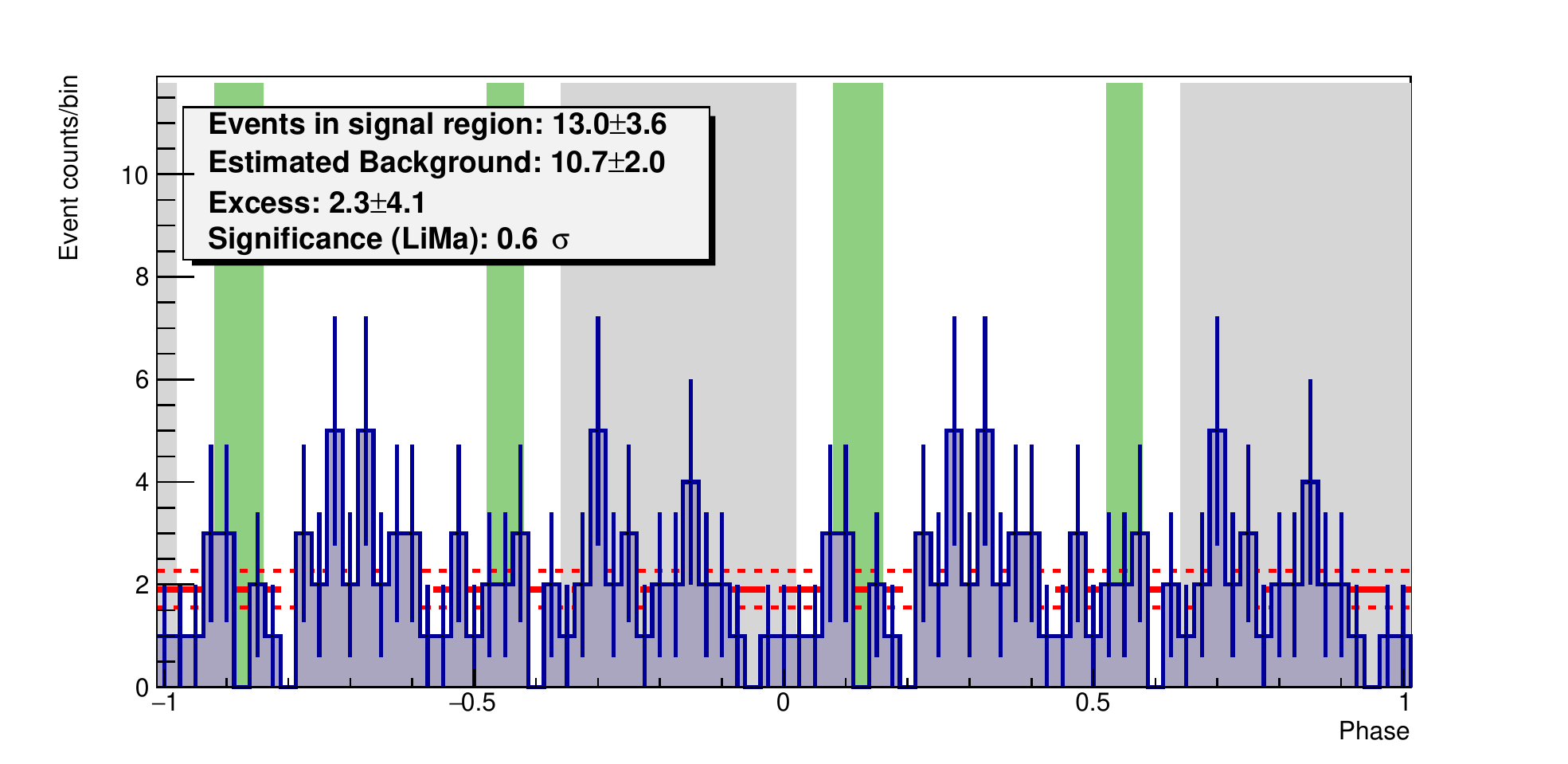}} \\
  \subfloat{\includegraphics[scale=0.7]{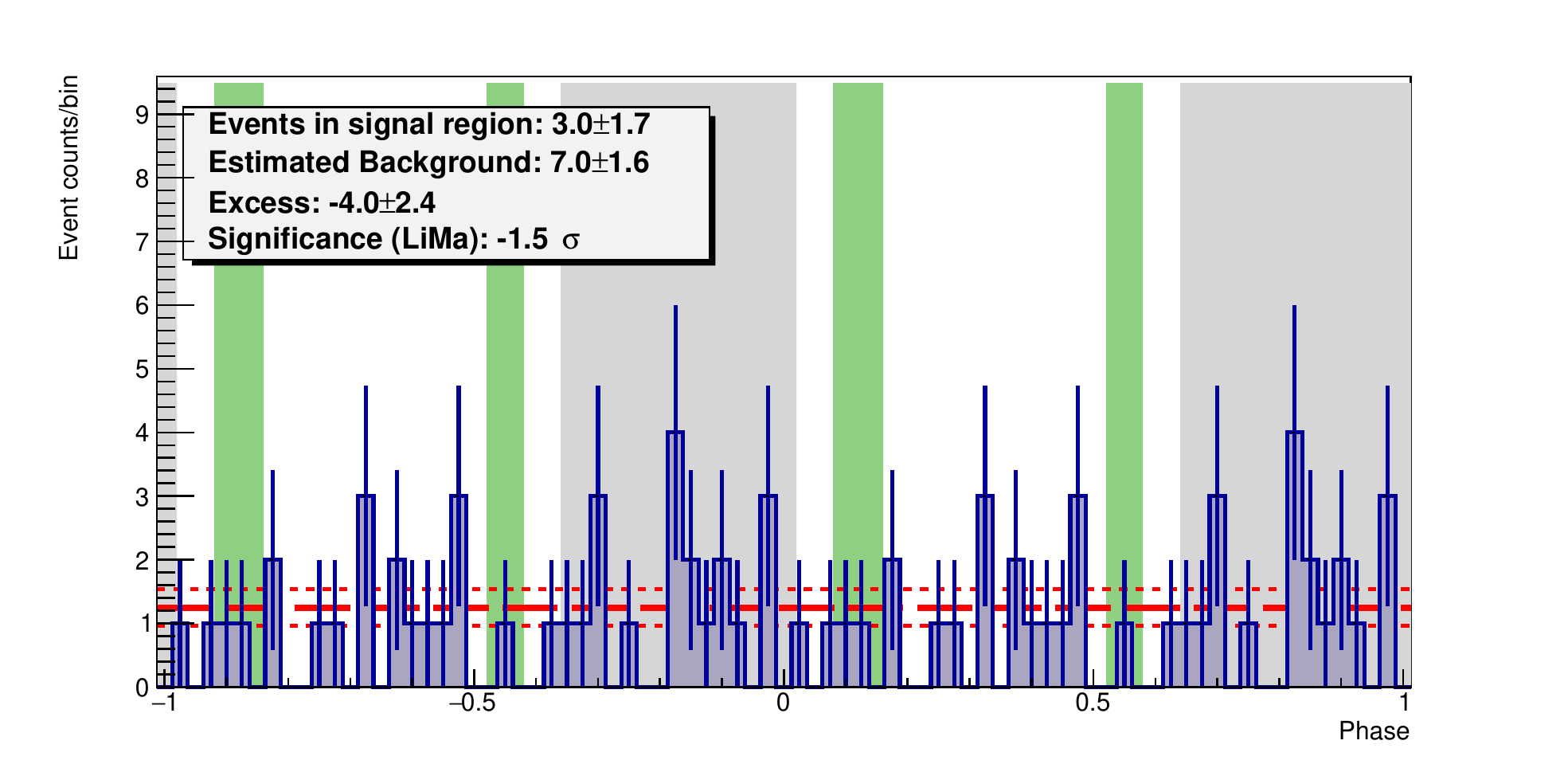}}   \\
 \caption{Pulse profiles of \textbf{PSR J1954+2836} from VERITAS data for soft cuts (top panel), moderate cuts (middle panel), and hard cuts (bottom panel).}
 \label{fig:PSRJ1954p2836_lcs}
\end{figure}

\pagebreak

\begin{figure}[t]
  \centering
  \subfloat{\includegraphics[scale=0.7]{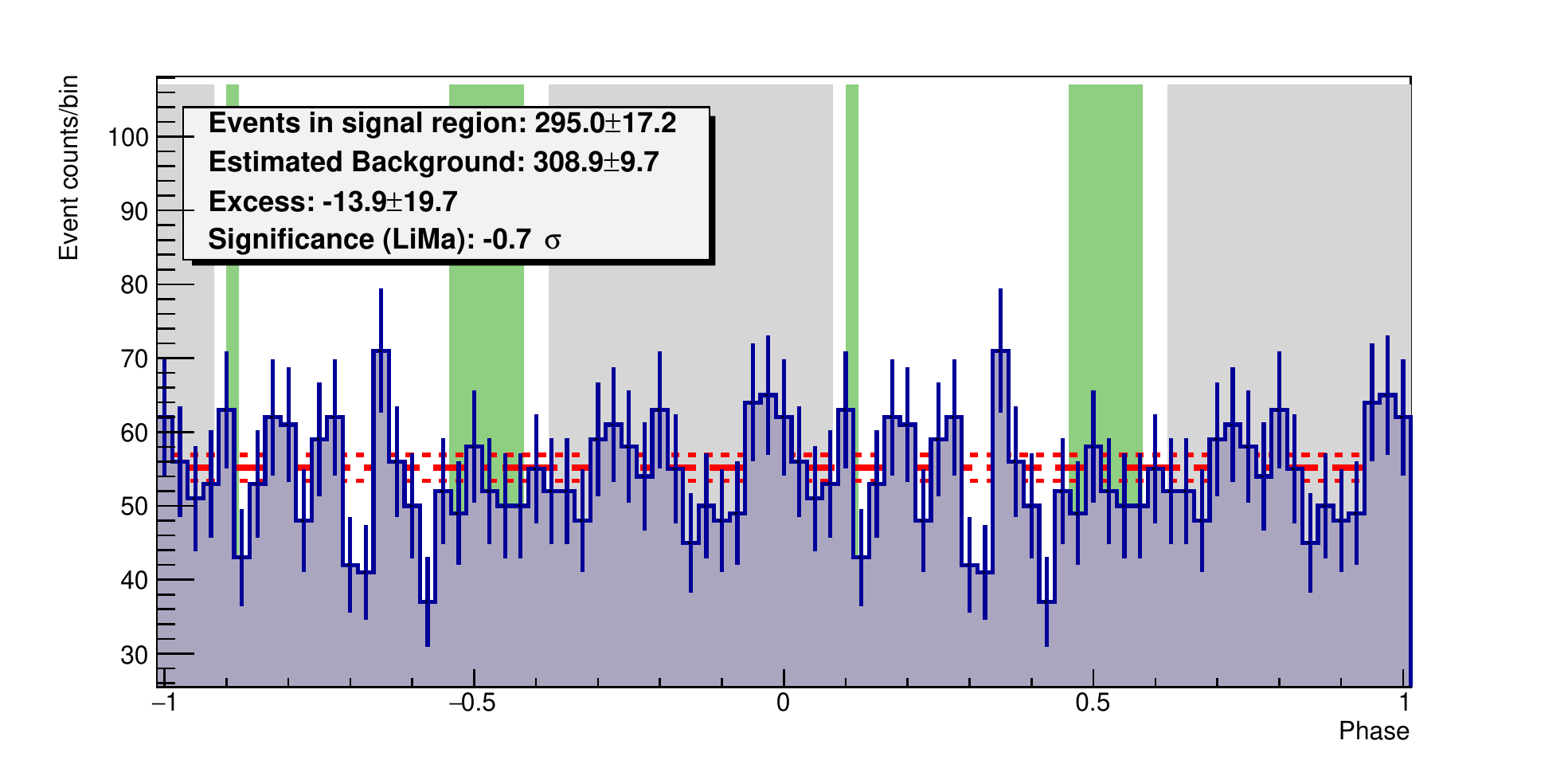}}   \\
  \subfloat{\includegraphics[scale=0.7]{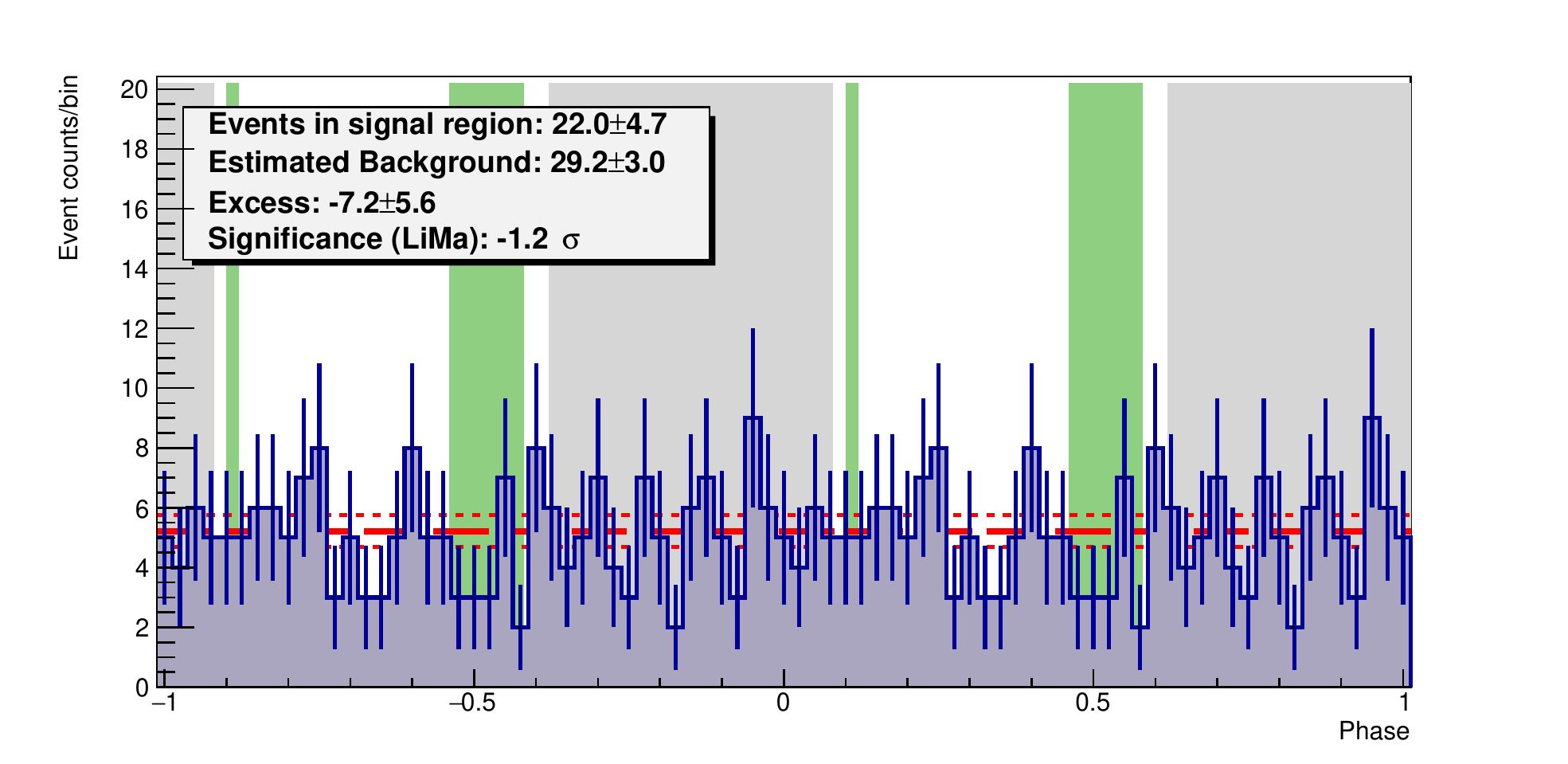}} \\
  \subfloat{\includegraphics[scale=0.7]{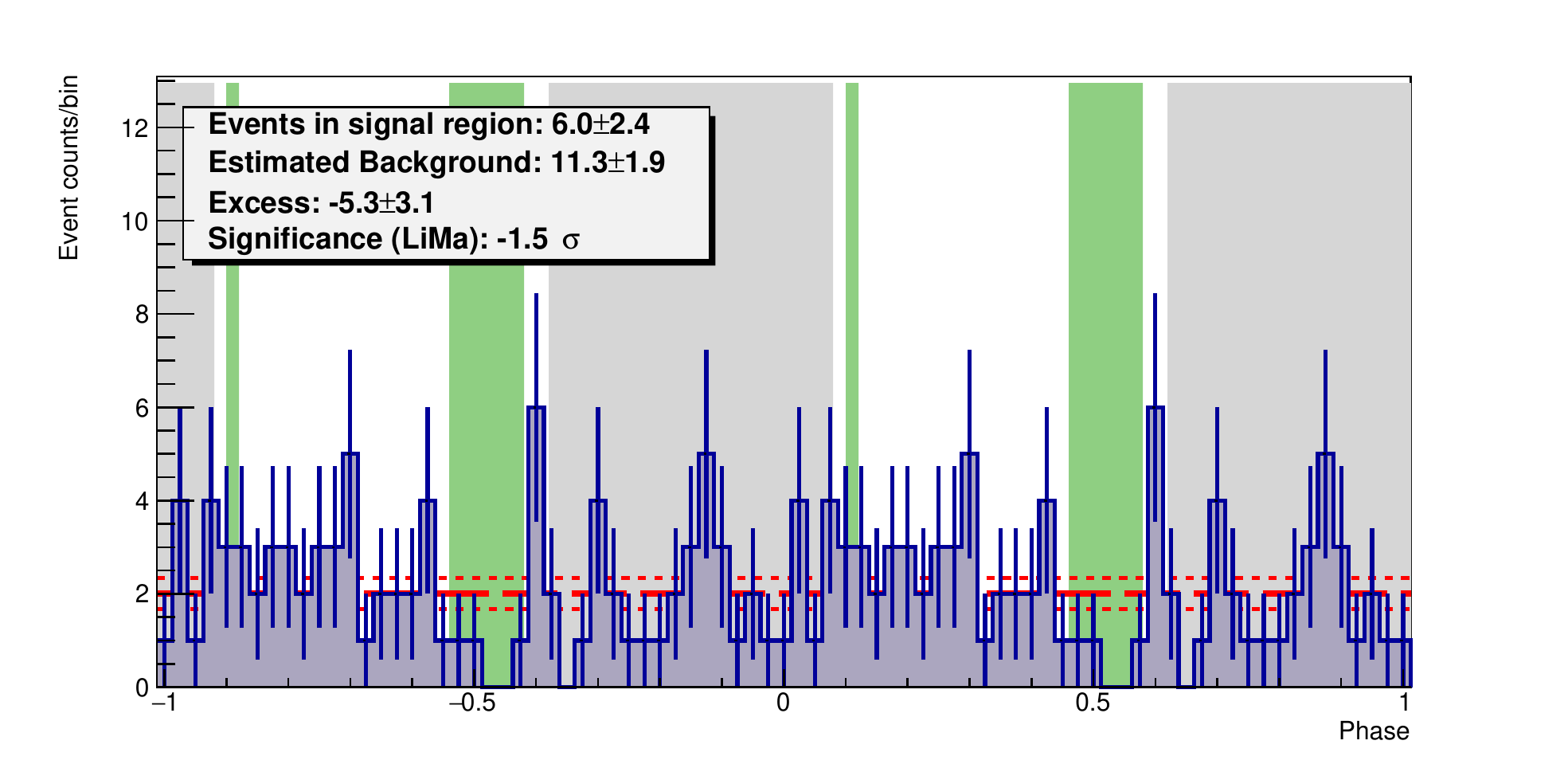}}   \\
 \caption{Pulse profiles of \textbf{PSR J1958+2846} from VERITAS data for soft cuts (top panel), moderate cuts (middle panel), and hard cuts (bottom panel).}
 \label{fig:PSRJ1958p2846_lcs}
\end{figure}

\pagebreak

\begin{figure}[t]
  \centering
  \subfloat{\includegraphics[scale=0.7]{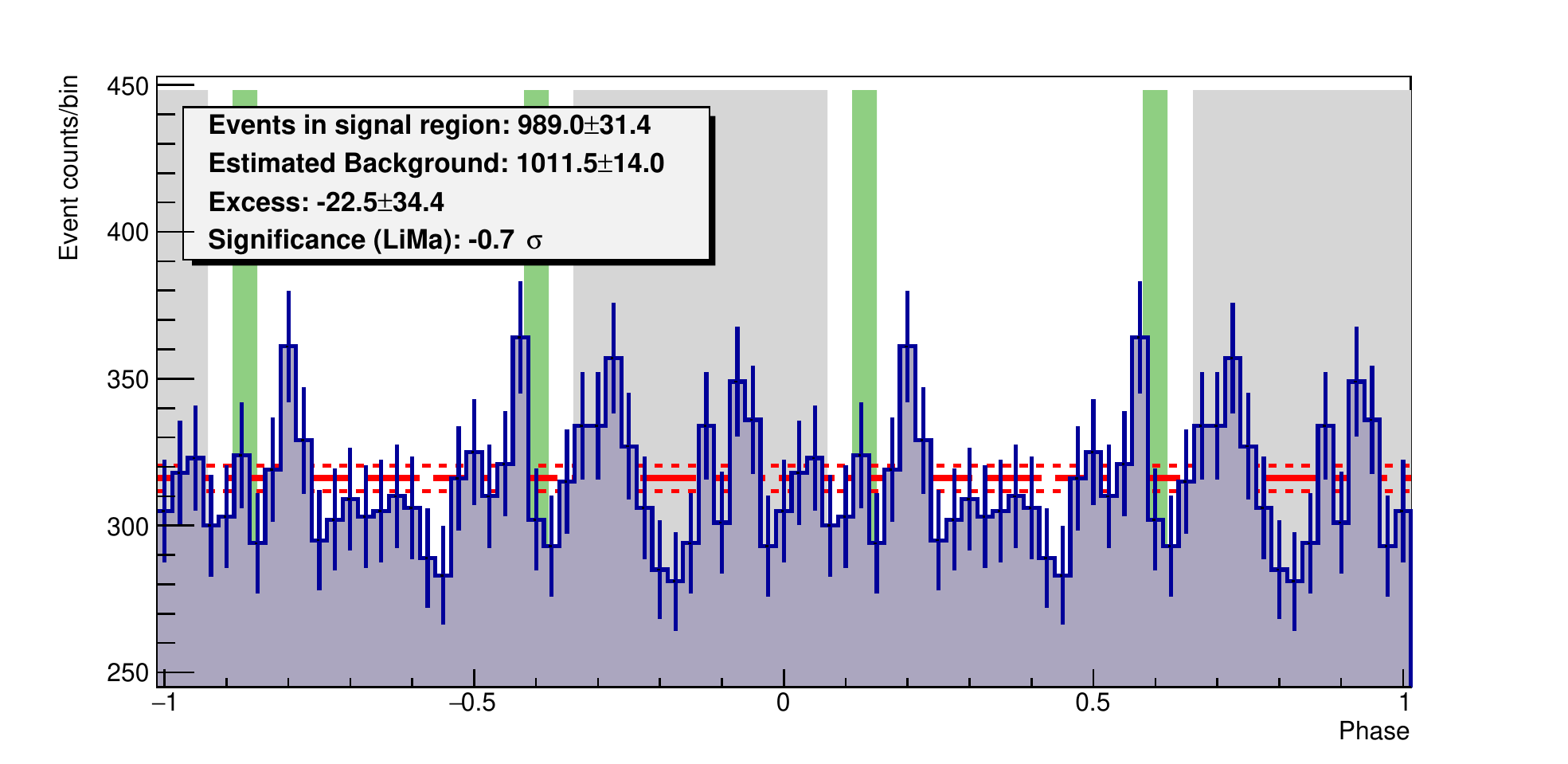}}   \\
  \subfloat{\includegraphics[scale=0.7]{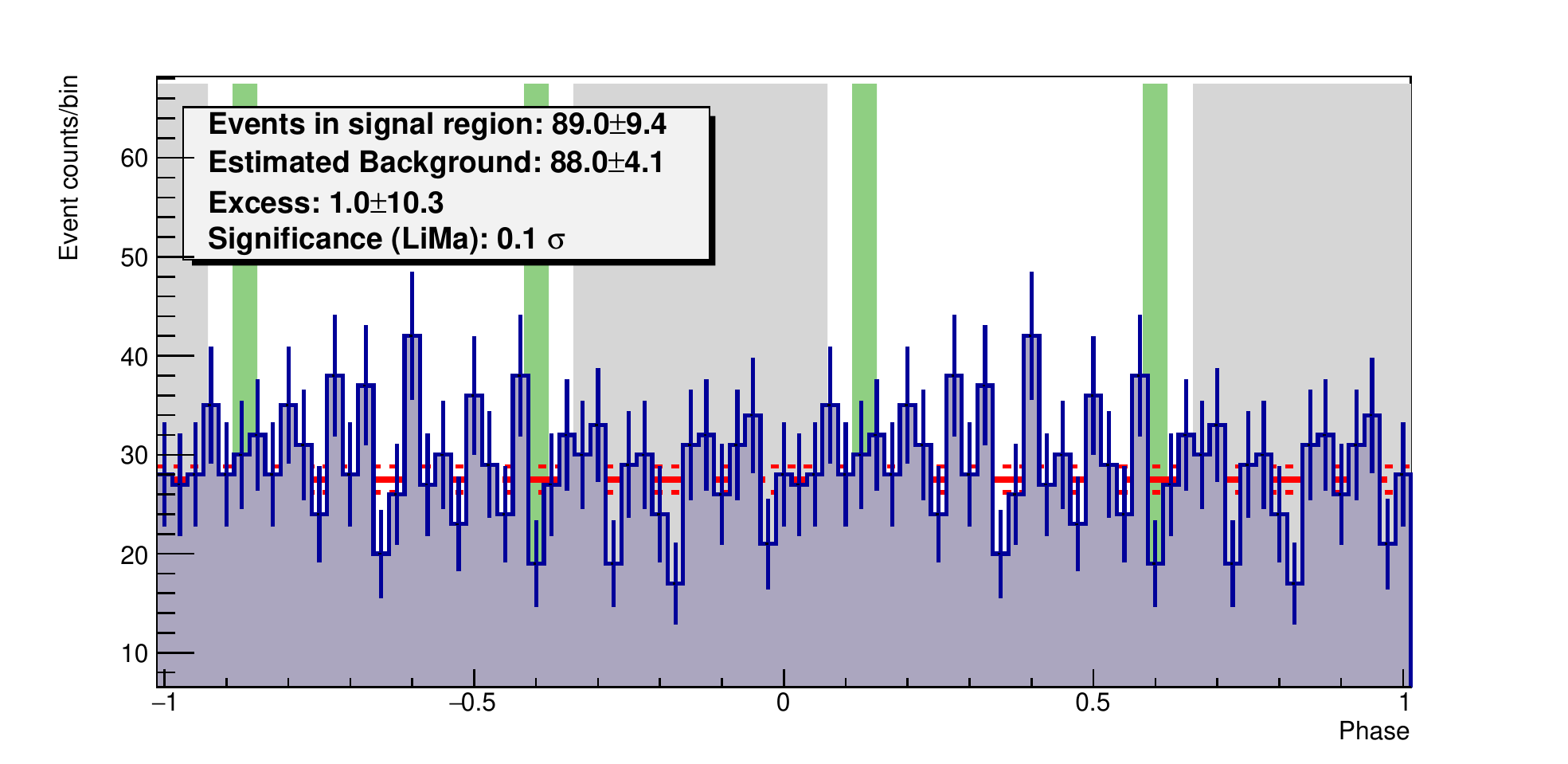}} \\
  \subfloat{\includegraphics[scale=0.7]{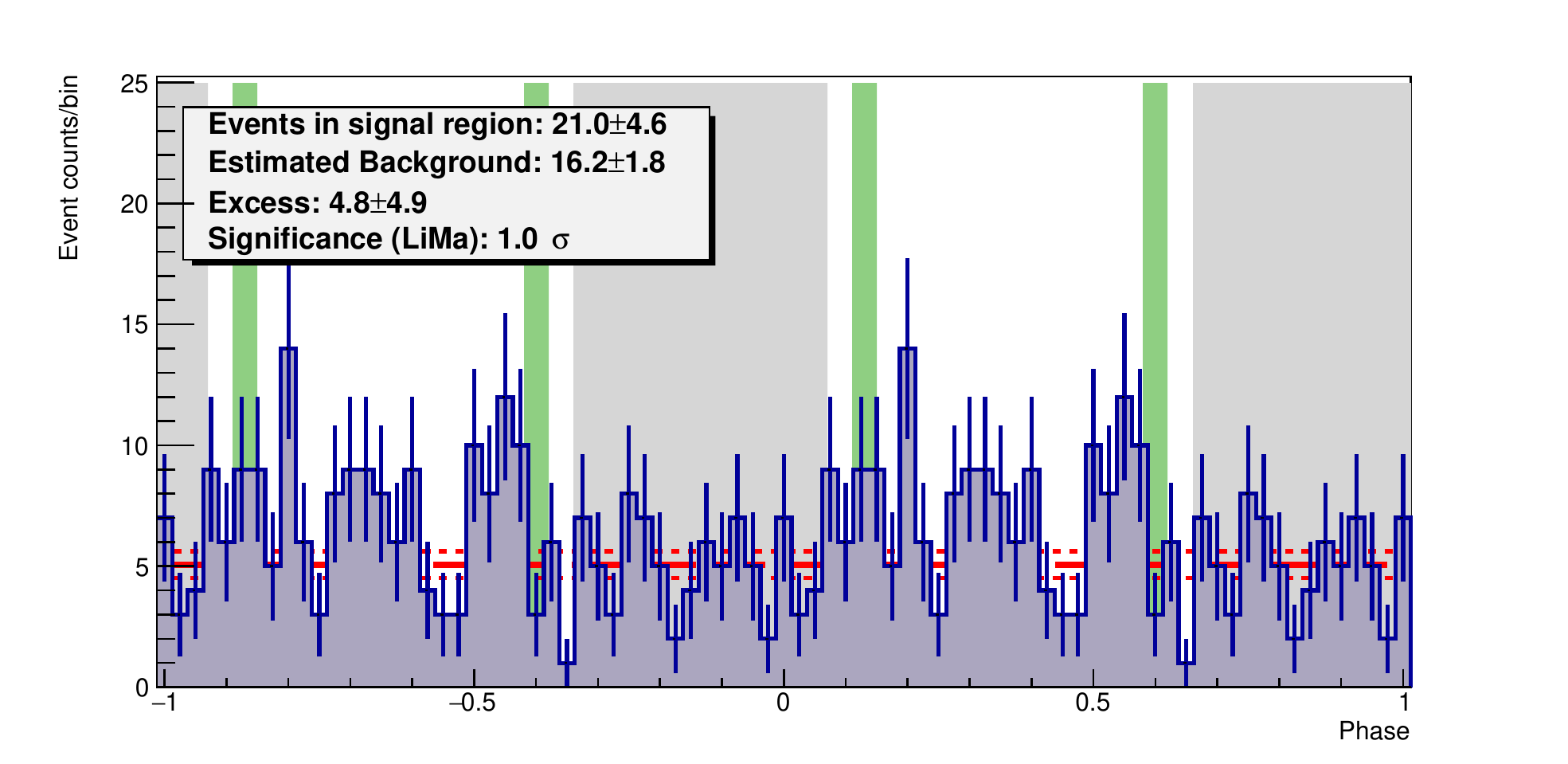}}   \\
 \caption{Pulse profiles of \textbf{PSR J2021+3651} from VERITAS data for soft cuts (top panel), moderate cuts (middle panel), and hard cuts (bottom panel).}
 \label{fig:PSRJ2021p3651_lcs}
\end{figure}

\pagebreak

\begin{figure}[t]
  \centering
  \subfloat{\includegraphics[scale=0.7]{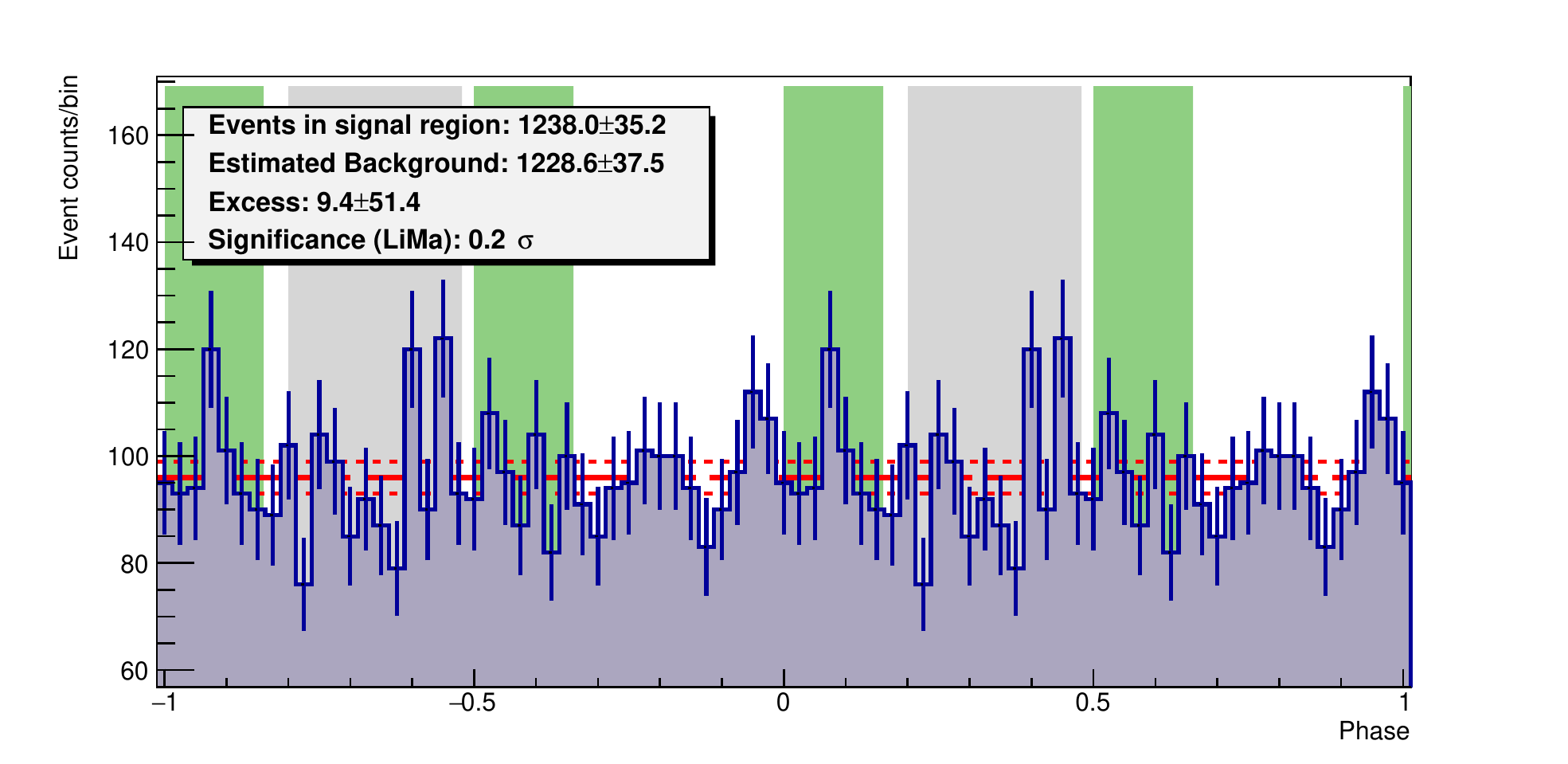}}   \\
  \subfloat{\includegraphics[scale=0.7]{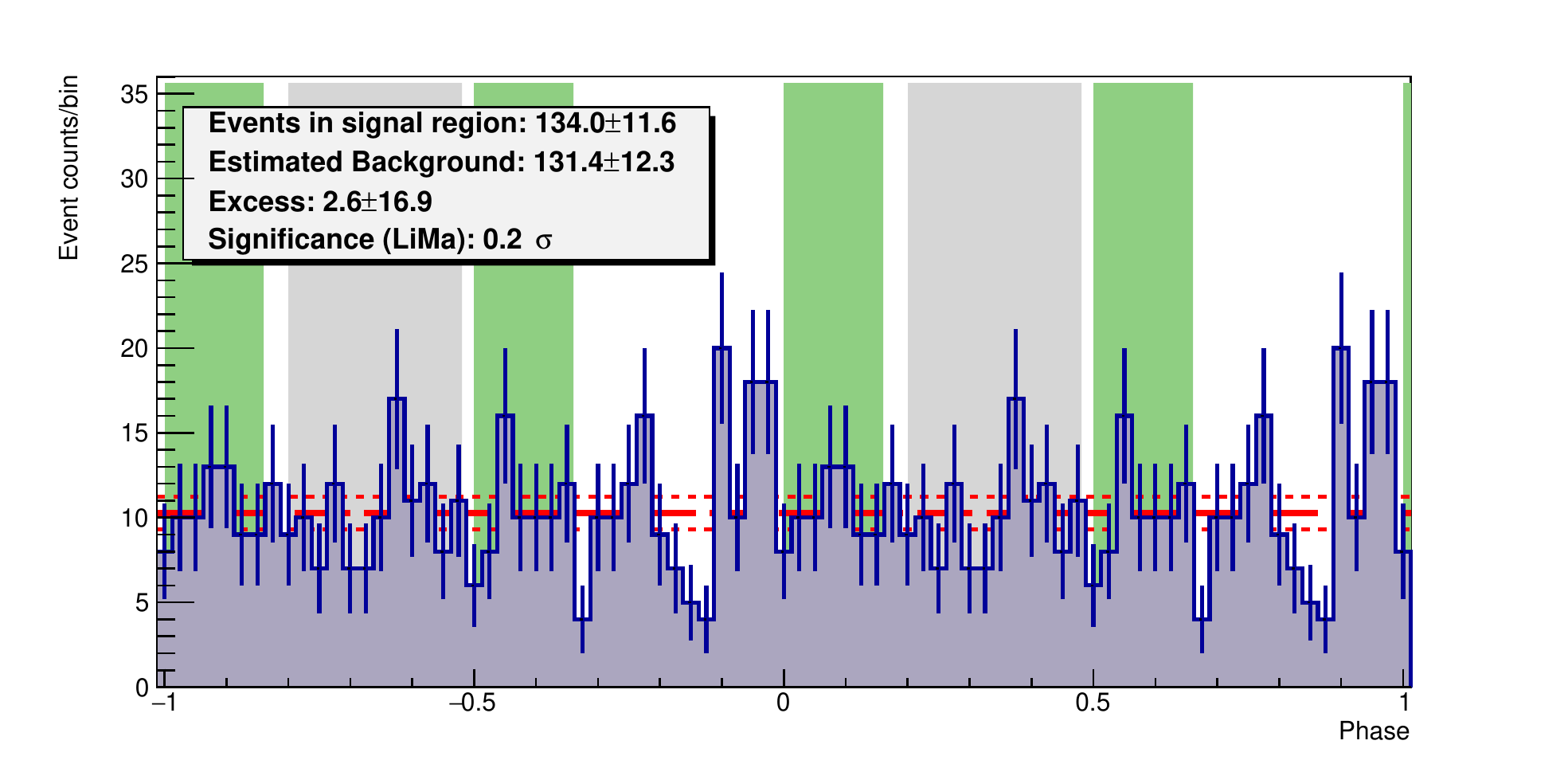}} \\
  \subfloat{\includegraphics[scale=0.7]{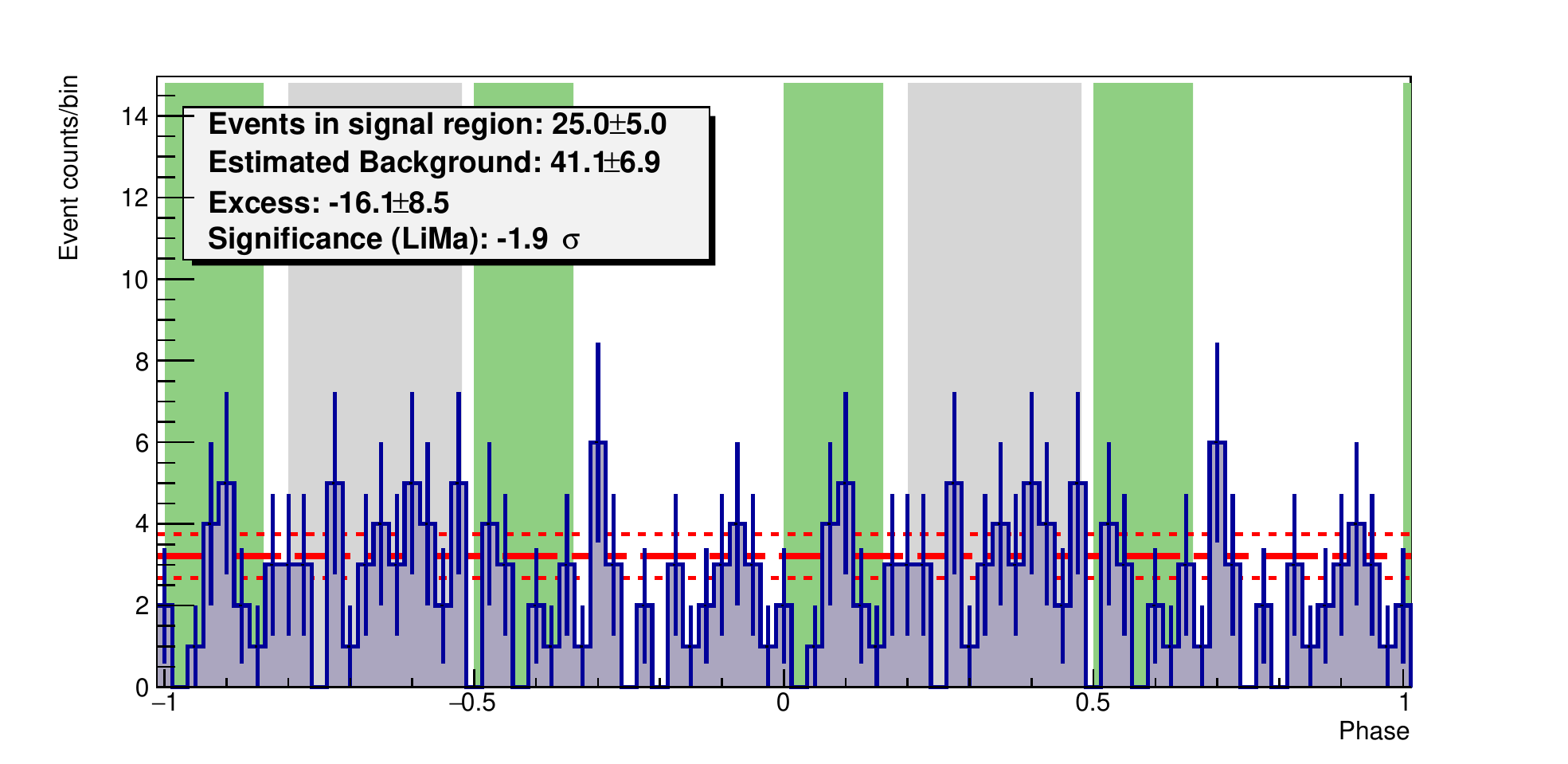}}   \\
 \caption{Pulse profiles of \textbf{PSR J2021+4026} from VERITAS data for soft cuts (top panel), moderate cuts (middle panel), and hard cuts (bottom panel).}
 \label{fig:PSRJ2021p4026_lcs}
\end{figure}

\pagebreak

\begin{figure}[t]
  \centering
  \subfloat{\includegraphics[scale=0.7]{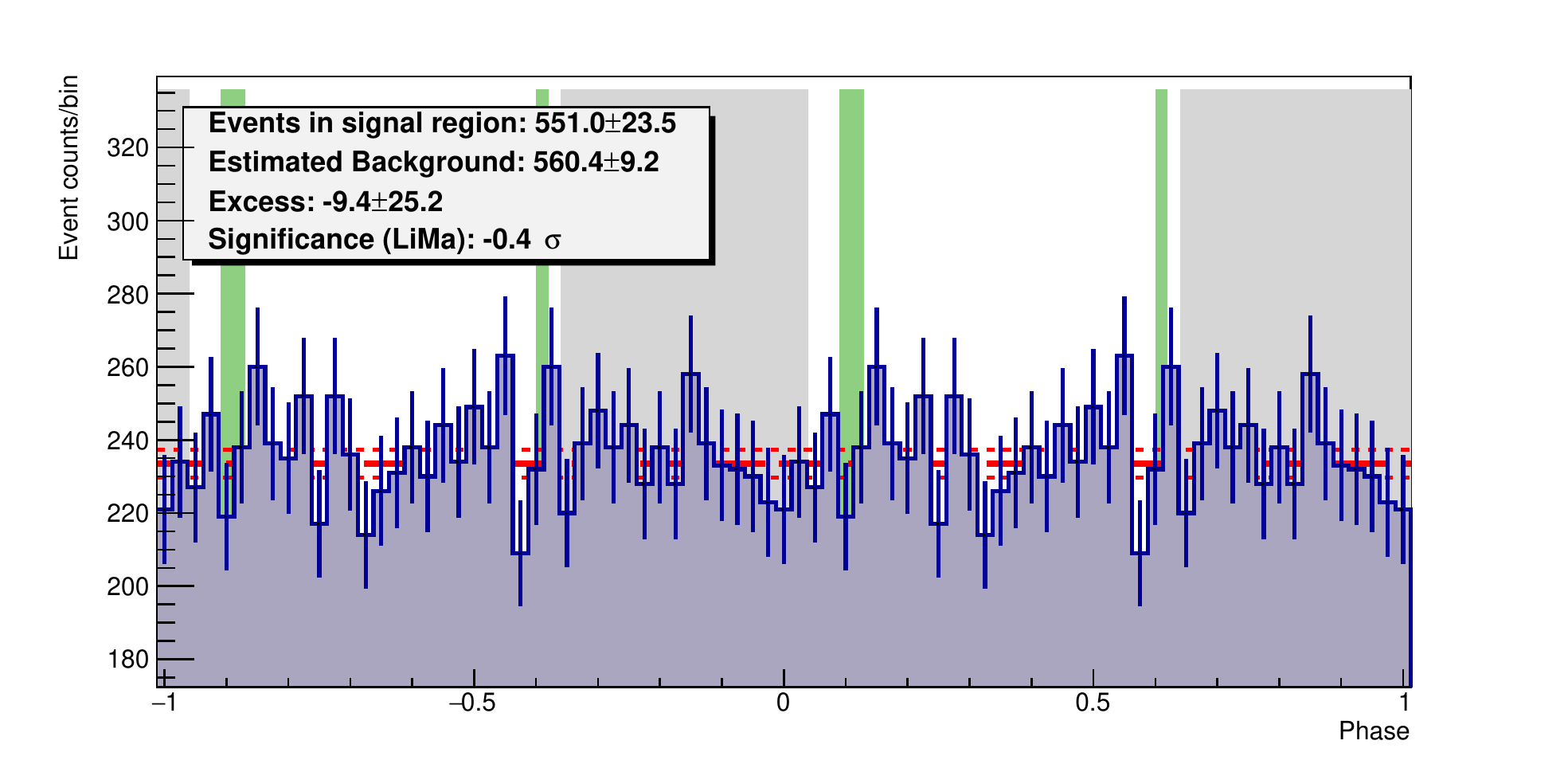}}   \\
  \subfloat{\includegraphics[scale=0.7]{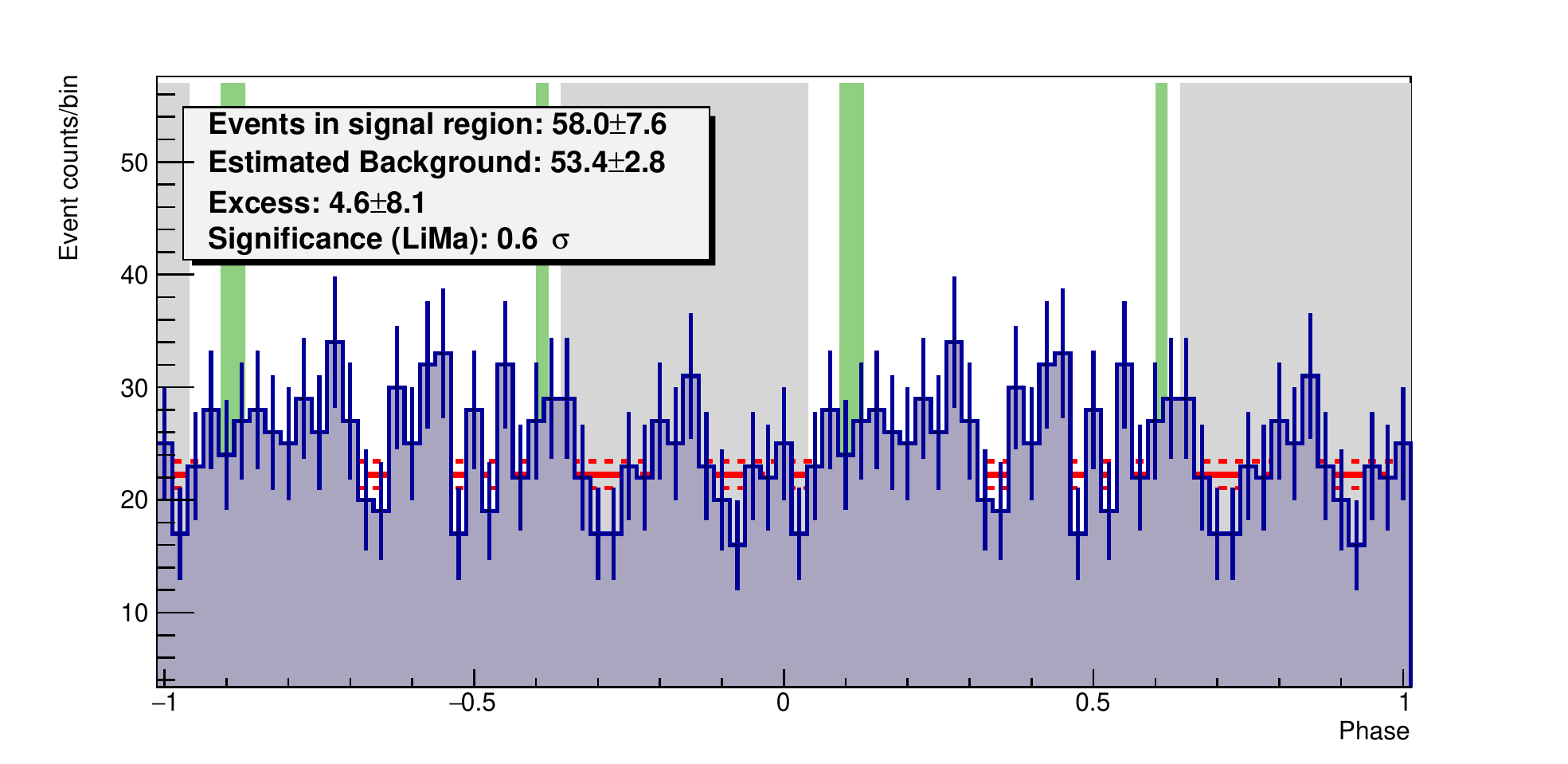}} \\
  \subfloat{\includegraphics[scale=0.7]{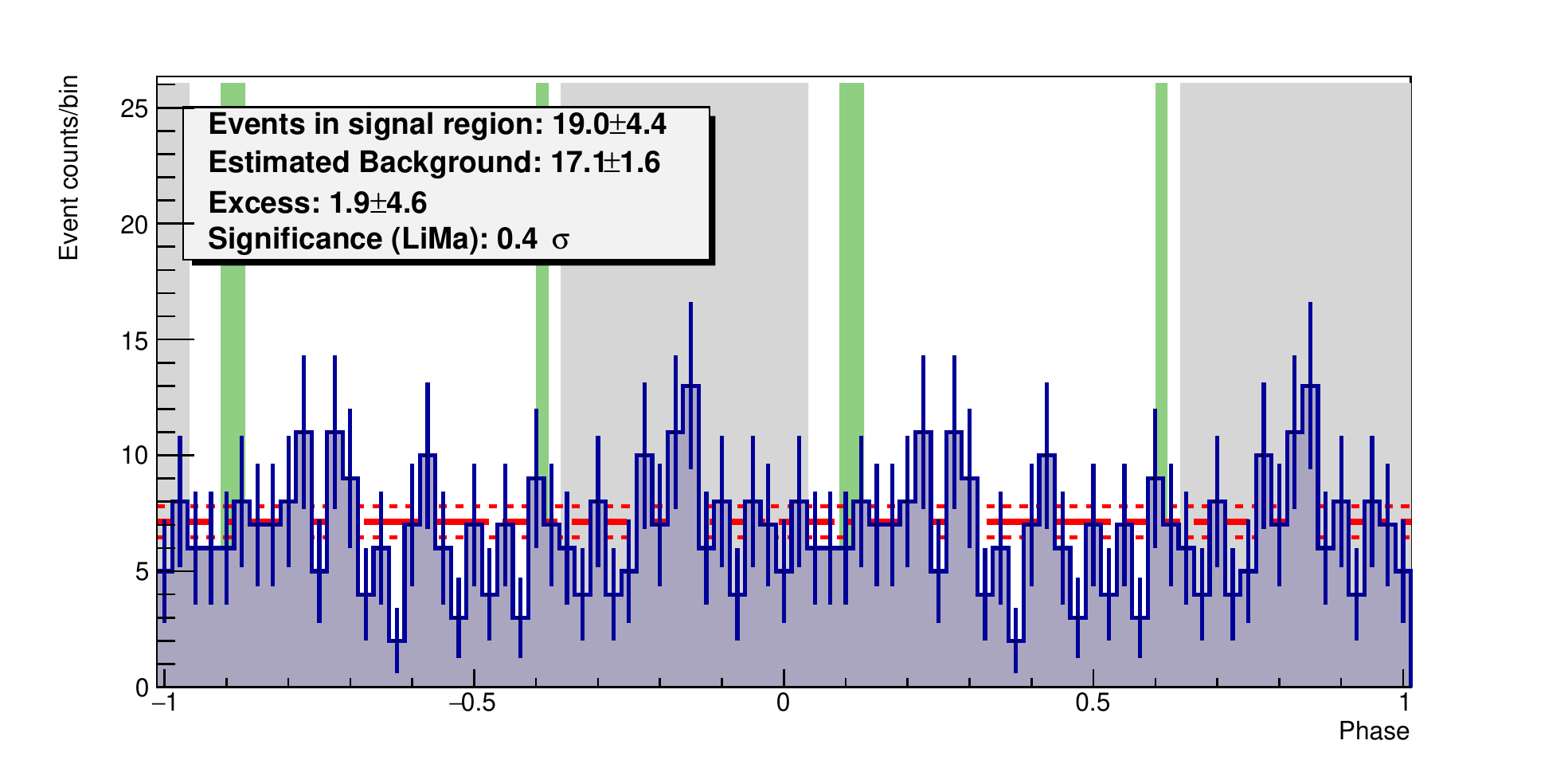}}   \\
 \caption{Pulse profiles of \textbf{PSR J2032+4127} from VERITAS data for soft cuts (top panel), moderate cuts (middle panel), and hard cuts (bottom panel).}
 \label{fig:PSRJ2032p4127_lcs}
\end{figure}

\pagebreak

\begin{figure}[t]
  \centering
  \subfloat{\includegraphics[scale=0.7]{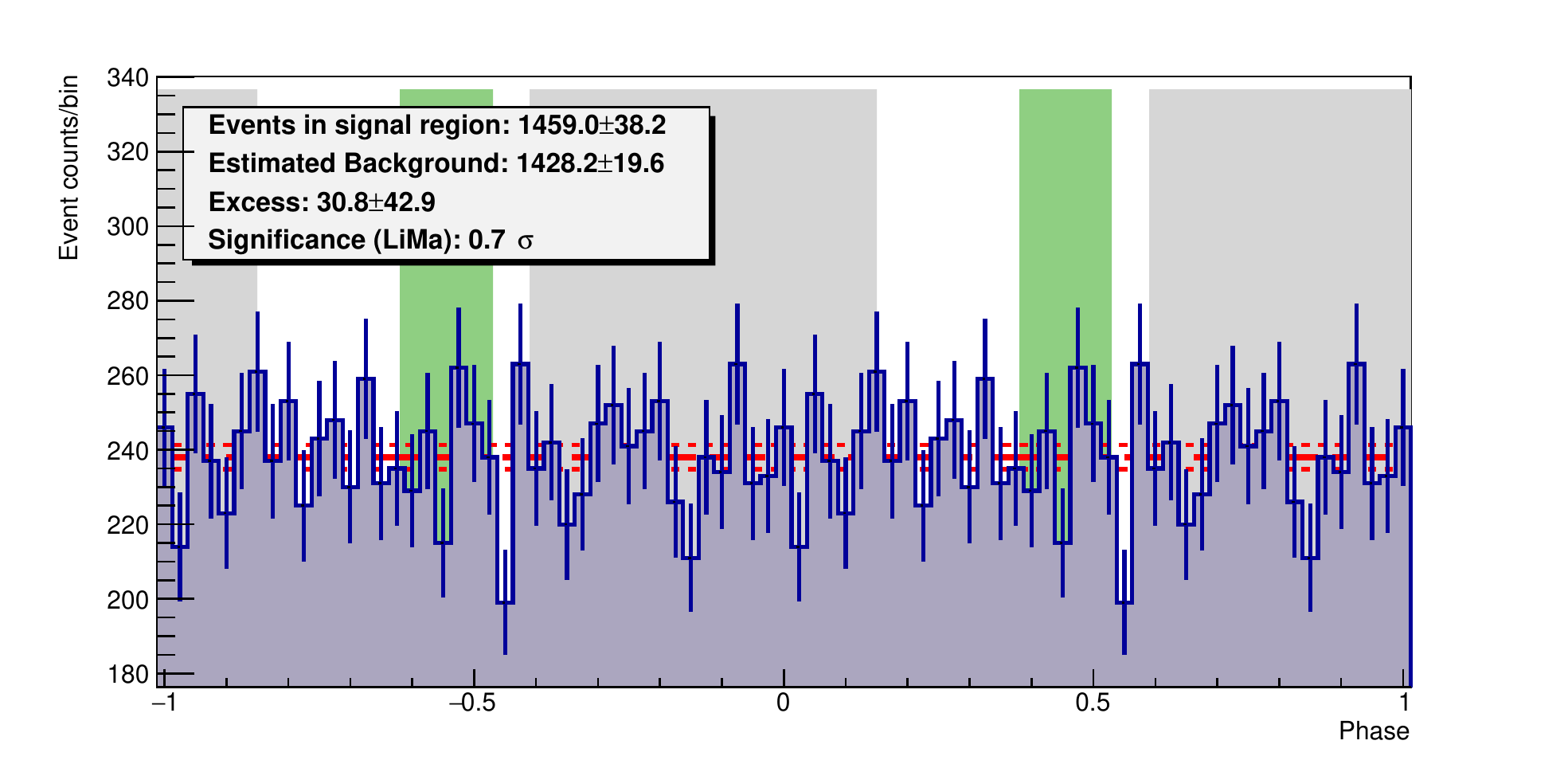}}   \\
  \subfloat{\includegraphics[scale=0.7]{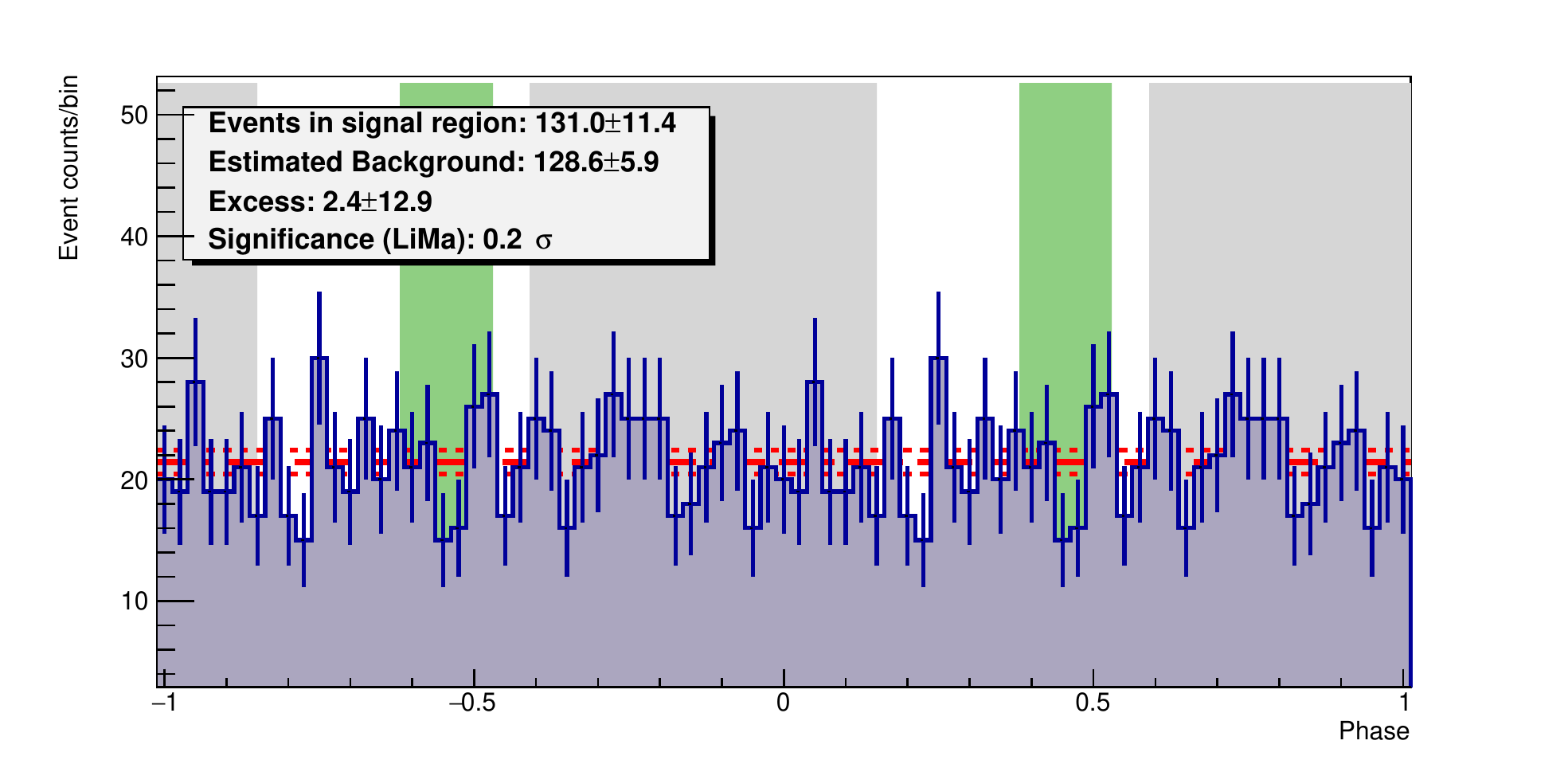}} \\
  \subfloat{\includegraphics[scale=0.7]{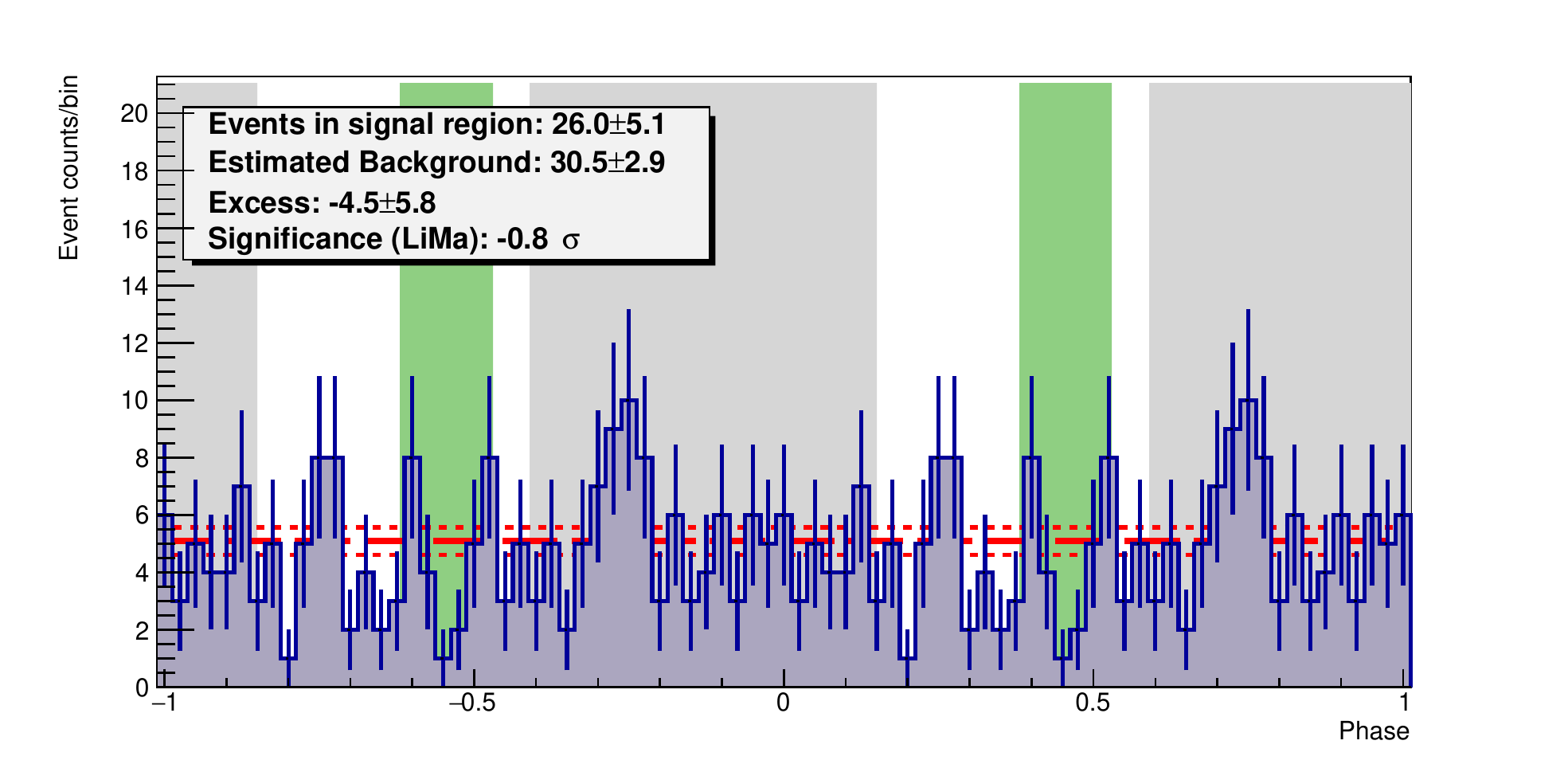}}   \\
 \caption{Pulse profiles of \textbf{PSR J2229+6114} from VERITAS data for soft cuts (top panel), moderate cuts (middle panel), and hard cuts (bottom panel).}
 \label{fig:PSRJ2229p6114_lcs}
\end{figure}

\pagebreak
\clearpage

\end{document}